\documentclass[11pt]{article}
\pdfoutput=1
\usepackage[margin=1in, centering]{geometry}
\usepackage{amsthm}
\makeatletter
\let\@fnsymbol\@arabic
\makeatother
\usepackage[table]{xcolor}
\usepackage[normalem]{ulem}
\definecolor{TSUYUKUSA}{RGB}{46, 169, 223}
\definecolor{KURENAI}{RGB}{203, 27, 69}
\usepackage[pdfstartview=FitH,pdfpagemode=UseNone,colorlinks=true,citecolor=KURENAI,linkcolor=TSUYUKUSA,backref=page,linktocpage=true]{hyperref}

\usepackage{hyperref}
\usepackage{fullpage}
\usepackage{amssymb}
\usepackage{mathtools}
\usepackage{amsthm}
\usepackage{eucal}
\usepackage{dsfont}
\usepackage{libertine}
\usepackage{newtxmath}
\usepackage{makecell}
\usepackage[capitalise]{cleveref}
\usepackage{tikz}
\usepackage{thm-restate}
\usepackage{subfig}
\usepackage{framed}
\usepackage{environ}

\usepackage{enumitem}
\usepackage{algorithmic}
\usepackage{float}

\NewEnviron{Anonymous}{\ifx\AnonymousSwitch\undefined\BODY\fi}

\makeatletter
\renewcommand{\section}{\@startsection {section}{1}{\z@}%
             {-3.5ex \@plus -1ex \@minus -.2ex}%
             {2.3ex \@plus .2ex}%
             {\normalfont\Large\scshape\bfseries}}
\renewcommand{\subsection}{\@startsection{subsection}{2}{\z@}%
             {-3.25ex\@plus -1ex \@minus -.2ex}%
             {1.5ex \@plus .2ex}%
             {\normalfont\large\scshape\bfseries}}
\renewcommand{\subsubsection}{\@startsection{subsubsection}{2}{\z@}%
             {-3.25ex\@plus -1ex \@minus -.2ex}%
             {1.5ex \@plus .2ex}%
             {\normalfont\normalsize\scshape\bfseries}}
\makeatother

\allowdisplaybreaks[1]

\def\A{\CMcal{A}}
\def\B{\CMcal{B}}
\def\C{\CMcal{C}}
\def\D{\CMcal{D}}

\def\F{\CMcal{F}}
\def\G{\CMcal{G}}
\def\H{\CMcal{H}}

\def\M{\CMcal{M}}

\def\P{\CMcal{P}}

\def\R{\CMcal{R}}

\def\V{\CMcal{V}}

\def\X{\CMcal{X}}
\def\Y{\CMcal{Y}}

\theoremstyle{plain}
\newtheorem{theorem}{Theorem}[section]
\newtheorem{lemma}[theorem]{Lemma}
\newtheorem{prop}[theorem]{Proposition}
\newtheorem{cor}[theorem]{Corollary}

\theoremstyle{definition}
\newtheorem{definition}[theorem]{Definition}
\newtheorem{claim}[theorem]{Claim}

\newtheorem{remark}[theorem]{Remark}
\newtheorem{fact}[theorem]{Fact}

\newcommand {\minusspace} {\: \! \!}

\newcommand {\Fn} [2] {\ensuremath{ #1 \minusspace \Br{ #2 } }}
\newcommand{\reals}{{\mathbb R}}
\newcommand {\set} [1] {\ensuremath{ \left\lbrace #1 \right\rbrace }}

\newcommand {\innerproduct} [2] {\ensuremath{\left \langle #1 , #2 \right \rangle}}

\newcommand{\normthree}[1]{{\left\vert\kern-0.25ex\left\vert\kern-0.25ex\left\vert #1 \right\vert\kern-0.25ex\right\vert\kern-0.25ex\right\vert}}
\newcommand {\br} [1] {\ensuremath{ \left( #1 \right) }}
\newcommand {\Br} [1] {\ensuremath{ \left[ #1 \right] }}

\DeclarePairedDelimiter{\norm}{\lVert}{\rVert}
\newcommand {\normsub} [2] {\ensuremath{ \norm{#1}_{#2} }}

\newcommand {\twonorm} [1] {\normsub{#1}{2}}
\newcommand {\ab} [1] {\ensuremath{ \left| #1 \right| }}
\DeclarePairedDelimiter{\abs}{\lvert}{\rvert}

\newcommand {\bra} [1] {\ensuremath{ \left\langle #1 \right| }}
\newcommand {\ket} [1] {\ensuremath{ \left| #1 \right\rangle }}
\newcommand {\ketbratwo} [2] {\ensuremath{ \left| #1 \middle\rangle \middle\langle #2 \right| }}
\newcommand {\ketbra} [1] {\ketbratwo{#1}{#1}}

\DeclareMathOperator*{\bigE}{\mathbb{E}}
\newcommand {\expec} [2] {\Fn{\bigE_{\substack{#1}}}{#2}}
\newcommand{\nnorm}[1]{{\left\vert\kern-0.25ex\left\vert\kern-0.25ex\left\vert #1
		\right\vert\kern-0.25ex\right\vert\kern-0.25ex\right\vert}}

\newcommand {\Tr} {\ensuremath{ \mathrm{Tr} }}

\newcommand {\id} {\ensuremath{\mathds{1}}}

\usetikzlibrary{calc}
\tikzset{meter/.append style={draw, inner sep=10, rectangle, font=\vphantom{A}, minimum width=30, line width=.8,
 path picture={\draw[black] ([shift={(.1,.3)}]path picture bounding box.south west) to[bend left=50] ([shift={(-.1,.3)}]path picture bounding box.south east);\draw[black,-latex] ([shift={(0,.1)}]path picture bounding box.south) -- ([shift={(.3,-.1)}]path picture bounding box.north);}}}

\newenvironment{Algorithm}{\begin{framed}}{\end{framed}}

\crefname{prop}{Proposition}{Propositions}
  
\renewcommand{\vec}[1]{\stackrel{\longrightarrow}{#1}}
\newcommand{\posint}{{\mathbb Z}_{>0}}
\newcommand{\e}{\mathrm{e}}
\renewcommand{\d}{\mathrm{d}}

\newcommand{\randP}{\ensuremath{\mathbf{P}}}
\newcommand{\randb}{\ensuremath{\mathbf{b}}}
\newcommand{\randg}{\ensuremath{\mathbf{g}}}
\newcommand{\randh}{\ensuremath{\mathbf{h}}}
\newcommand{\randz}{\ensuremath{\mathbf{z}}}
\newcommand{\randx}{\ensuremath{\mathbf{x}}}
\newcommand{\randy}{\ensuremath{\mathbf{y}}}
\newcommand{\randu}{\ensuremath{\mathbf{u}}}
\newcommand{\randv}{\ensuremath{\mathbf{v}}}

\newcommand{\randf}{\ensuremath{\mathbf{f}}}
\newcommand{\randA}{\ensuremath{\mathbf{A}}}
\newcommand{\randB}{\ensuremath{\mathbf{B}}}
\newcommand{\randC}{\ensuremath{\mathbf{C}}}
\newcommand{\randD}{\ensuremath{\mathbf{D}}}
\newcommand{\randM}{\ensuremath{\mathbf{M}}}
\newcommand{\randX}{\ensuremath{\mathbf{X}}}
\newcommand{\randa}{\ensuremath{\mathbf{a}}}
\newcommand{\randPi}{\ensuremath{\mathbf{\Pi}}}
\newcommand{\herspace}[1]{\ensuremath{\H_m^{\otimes #1}}}

\newcommand{\infi}[1]{\ensuremath{\mathrm{Inf}_i}\br{#1}}
\newcommand{\varinfi}[1]{\ensuremath{\mathrm{VarInf}_i}\br{#1}}
\newcommand{\inffj}[1]{\ensuremath{\mathrm{VarInf}_{f,j}}\br{#1}}
\newcommand{\infrfj}[1]{\ensuremath{\mathrm{VarInf}_{\randf,j}}\br{#1}}
\newcommand{\invf}[1]{\ensuremath{f^{-1}\br{#1}}}
\newcommand{\bits}[1]{\ensuremath{\set{-1,1}^{#1}}}
\newcommand{\usim}{\ensuremath{\sim_{\mathrm{u}}}}
\newcommand {\var} [1] {\mathrm{Var}\Br{#1}}
\newcommand{\gval}[3]{\ensuremath{\mathrm{val}_{#1}\br{\set{#2},\set{#3}}}}
\newcommand{\fourtuples}{\ensuremath{\br{x,y,a,b}\in\calX\times\calY\times\calA\times\calB}}
\newcommand{\sigmarange}[1]{\ensuremath{\sigma\in[m^2]_{\geq0}^{#1}}}
\newtheorem*{theorem*}{Theorem}

\newcommand{\poly} {\ensuremath{\operatorname{poly}}}
\newcommand{\polylog} {\ensuremath{\operatorname{polylog}}}
\newcommand{\polyloglog} {\ensuremath{\operatorname{polyloglog}}}
\newcommand{\ep} {\ensuremath{\varepsilon}}

\newcommand{\ccsm} {\ensuremath{C_{sm}}}
\newcommand{\dmcc} {\ensuremath{C_{pt}}}
\newcommand{\dlbo} {\ensuremath{{D}}}
\newcommand{\dldd} {\ensuremath{{d}}}
\newcommand{\wireal} {\ensuremath{{s_{w}}}}

\newcommand{\epreg} {\ensuremath{{\delta}}}
\newcommand{\epdr} {\ensuremath{\delta}}

\newcommand{\eprd} {\ensuremath{\ep_{rd}}}
\newcommand{\pseudostrategy} {pseudo-strategy}

\newcommand{\Pss} {\ensuremath{P^{\,x}_a}}
\newcommand{\Qss} {\ensuremath{Q^{\,y}_b}}
\newcommand{\Pwss} {\ensuremath{\tilde{P}^{\,x}_a}}
\newcommand{\Qwss} {\ensuremath{\tilde{Q}^{\,y}_b}}
\newcommand{\Pnn}[1] {\ensuremath{{P^{\,x, (#1)}_a}}}
\newcommand{\Qnn}[1] {\ensuremath{{Q^{\,y, (#1)}_b}}}
\newcommand{\Pff} {\ensuremath{\widehat{P}^{\,x}_a(\sigma)}}
\newcommand{\Qff} {\ensuremath{\widehat{Q}^{\,y}_b(\sigma)}}
\newcommand{\Pfn}[1] {\ensuremath{{{\widehat{P}}^{\,x, {(#1)}}_a}(\sigma)}}

\newcommand{\x}{\otimes}

\newcommand{\ct}{^{\dagger}}

\newcommand{\calH}{\mathcal{H}}
\newcommand{\calX}{\mathcal{X}}
\newcommand{\calY}{\mathcal{Y}}
\newcommand{\calA}{\mathcal{A}}
\newcommand{\calB}{\mathcal{B}}

\newcommand{\calL}{\mathcal{L}}

\newcommand{\calG}{\mathcal{G}}
\newcommand{\calR}{\mathcal{R}}

\newcommand{\appd}[1]{\approx_{#1}}

\newcommand{\bF}{\mathbb{F}}
\newcommand{\bZ}{\mathbb{Z}}

\newcommand{\ba}{\pmb{a}}
\newcommand{\bb}{\pmb{b}}
\newcommand{\bc}{\pmb{c}}
\newcommand{\bd}{\pmb{d}}
\newcommand{\be}{\pmb{e}}
\newcommand{\bw}{\pmb{w}}
\newcommand{\bx}{\pmb{x}}
\newcommand{\by}{\pmb{y}}

\newcommand{\bI}{\pmb{I}}
\newcommand{\bJ}{\pmb{J}}
\newcommand{\bT}{\pmb{T}}

\newcommand{\bU}{\pmb{U}}
\newcommand{\pmbF}{\pmb{F}}

\newcommand{\tN}{\tilde{N}}

\newcommand{\tG}{\tilde{G}}
\newcommand{\tH}{\tilde{H}}

\newcommand{\tpsi}{\tilde{\psi}}

\newcommand{\se}{\sqrt{\epsilon}}

\newcommand{\PCPP}{\mathrm{PCPP}}
\newcommand{\Unif}{\mathsf{Unif}}
\newcommand{\ipt}{\mathsf{input}}
\newcommand{\Enc}{\mathsf{Enc}}
\newcommand{\Dec}{\mathsf{Dec}}
\newcommand{\Alg}{\mathsf{Alg}}
\newcommand{\MIP}{\mathrm{MIP}}
\newcommand{\RE}{\mathrm{RE}}
\newcommand{\val}{\mathrm{val}}
\newcommand{\NEXP}{\mathrm{NEXP}}
\newcommand{\NP}{\mathrm{NP}}

\newcommand{\mlqin}[1]{\textcolor{blue}{mlqin: {#1}}}

\allowdisplaybreaks
\begin{document}

\title{The Computational Advantage of $\mathrm{MIP}^\ast$ Vanishes in the Presence of Noise\footnote{A preliminary  version has been accepted by the 39th Computational Complexity Conference (CCC 2024).}\\[2ex]}
\date{}

\ifx\AnonymousSwitch\undefined
\author{
    \and Yangjing Dong\thanks{\scriptsize State Key Laboratory for Novel Software Technology, New Cornerstone Science Laboratory, Nanjing University. Email: dongmassimo@gmail.com.}
    \and Honghao Fu \thanks{\scriptsize Massachusetts Institute of Technology. Email: honghaof@mit.edu.}
    \and Anand Natarajan \thanks{\scriptsize Massachusetts Institute of Technology. Email: anandn@mit.edu.}
    \and Minglong Qin\thanks{\scriptsize State Key Laboratory for Novel Software Technology, New Cornerstone Science Laboratory, Nanjing University. Email: mlqin@smail.nju.edu.cn.}
    \and Haochen Xu\thanks{\scriptsize Key Laboratory of System Software (Chinese Academy of Sciences) and State Key Laboratory of Computer Science, Institute of Software, Chinese Academy of Sciences. Email: xuhc@ios.ac.cn}~\thanks{\scriptsize University of Chinese Academy of Sciences, Beijing, China}
    \and Penghui Yao\thanks{\scriptsize State Key Laboratory for Novel Software Technology, New Cornerstone Science Laboratory, Nanjing University. Email: phyao1985@gmail.com.}~\thanks{\scriptsize Hefei National Laboratory, Hefei 230088, China.}
}
\fi

\clearpage\maketitle

\begin{abstract}
The class $\MIP^*$ of quantum multiprover interactive proof systems with entanglement is much more powerful than its classical counterpart $\mathrm{MIP}$~\cite{Babai1991, JNVWYuen'20, JNVWY'20}: while $\MIP = \mathrm{NEXP}$, the quantum class $\MIP^*$ is equal to $\RE$, a class including the halting problem. This is because the provers in $\MIP^*$ can share unbounded quantum entanglement. However, recent works \cite{qin2021nonlocal,qin_et_al:LIPIcs.ICALP.2023.97} have shown that this advantage is significantly reduced if the provers' shared state contains noise. This paper attempts to exactly characterize the effect of noise on the computational power of quantum multiprover interactive proof systems. We investigate the quantum two-prover one-round interactive system $\mathrm{MIP}^*\Br{\poly,O(1)}$, where the verifier sends polynomially many bits to the provers and the provers send back constantly many bits. We show that noise completely destroys the computational advantage given by shared entanglement in this model. Specifically, we show that if the provers are allowed to share arbitrarily many EPR states, where each EPR state is affected by an arbitrarily small constant amount of noise, the resulting complexity class is equivalent to $\mathrm{NEXP} = \mathrm{MIP}$. This improves significantly on the previous best-known bound of $\mathrm{NEEEXP}$ (nondeterministic triply exponential time) \cite{qin2021nonlocal}. We also show that this collapse in power is due to noise, rather than the $O(1)$ answer size, by showing that allowing for noiseless EPR states gives the class the full power of $\RE = \MIP^*\Br{\poly, \poly}$. Along the way, we develop two technical tools of independent interest. First, we give a new, deterministic tester for the positivity of an exponentially large matrix, provided that it has a low-degree Fourier decomposition in terms of Pauli matrices. Secondly, we develop a new invariance principle for smooth matrix functions having bounded third-order Fr\'echet derivatives or which are Lipschitz continuous.
\end{abstract}

\newpage
\tableofcontents

 \section{Introduction}

The power of entanglement in computation has been a central topic in the theory of quantum computing. In particular, the effect of entanglement in multiprover interactive proof systems has been studied for decades~\cite{doi:10.1137/090772885,doi:10.1137/090751293,Ito:2009:OTO:1602931.1603187,Ji:2017:CQM:3055399.3055441,Slo:2020,slofstra:2019}  leading to the seminal result $\mathrm{MIP}^*=\mathrm{RE}$~\cite{JNVWYuen'20,JNVWY'20} due to Ji, Natarajan, Vidick, Wright, and Yuen, which states that all recursively enumerable languages can be decided by multiprover interactive proof systems empowered by quantum entanglement. More precisely, the system only has two provers, one round of interaction between the provers and the verifier, and the provers share arbitrarily many copies of the EPR state.

Given the appearance of intractable complexity classes like $\RE$ in the previous result, a natural question is to what extent the body of results on $\MIP^*$ are relevant to the physical world. Of course, in reality, devices do not have access to unbounded numbers of perfect EPR pairs; in a sense, what $\MIP^* = \RE$ means is that the power of two entangled provers grows unboundedly as the number of shared EPR pairs increases, even when the message size is constrained to be polynomial. In fact, using a finite number of iterations of the ``compression" procedure from $\MIP^* = \RE$, one can show that the class $\mathrm{NTIME}[T(n)]$ for $T(n)$ any finite tower of exponentials has an $\MIP^*$ protocol, where the provers need only share a finite number of perfect EPR pairs scaling roughly with $\log T(n)$. However, the requirement that the EPR pairs be perfect seems essential to these protocols. The question naturally arises whether similar complexity results can be obtained even when the provers have access to \emph{imperfect} entanglement only.

To isolate the role played by noise, in this work we ask the following question: what is the power of $\MIP^*$ when the provers are given access to an \emph{unbounded} number of \emph{imperfect} EPR pairs, where each EPR pair is independently perturbed by a constant amount of depolarizing noise? (We choose this noise model for illustration, while it is mathematically elegant and also physically relevant, as recent experiments suggest that the dominating noise is the localized depolarizing noise in the neutral atom platform \cite{bluvstein2023logical}. In this paper, we are able to handle a more general noise model, see \Cref{sec:quantum}.) On the one hand, known $\MIP^*$ protocols all break down with states of this form. On the other hand, according to standard measures of entanglement such as distillable entanglement and entanglement of formation, such states have entanglement that grows unboundedly as the number of copies goes to infinity. Thus, it seems \emph{a priori} reasonable that the corresponding $\MIP^*$ class may also have unbounded power.

It is worth noting that this question is orthogonal to fault tolerance in quantum devices. As usual in $\MIP^*$, we assume that the provers are computationally unbounded, and may perform any quantum operation of their choice with no error. Nevertheless, this does not mean they can use techniques from fault tolerance to simulate provers with noiseless entangled states. This is because the provers cannot jointly correct their shared entangled state, since they are not allowed to communicate in this model.

This question is closely related to the quantum information primitive of self-testing. Self-tests are essentially $\MIP^*$ protocols that certify physical properties of quantum states, rather than computational statements. The protocols in $\MIP^* = \RE$ all rely on highly efficient self-tests for EPR pairs, but these tests are not at all tolerant of noise. Designing self-tests that \emph{are} tolerant to noise, and certify some useful measure of entanglement, is a current research question \cite{arnonFYen2018,arnon2019device}, and studying the power of $\MIP^*$ in the presence of noise gives us insight on this question from a different angle. In particular, for an entangled state $\rho$, one can think of the power of the complexity class $\MIP^*[\rho]$ where the provers are restricted to sharing copies of $\rho$, as a particular operational measure of the amount of useful entanglement in $\rho$. In passing, we remark that recent work of Vidick, Arnon-Friedman and Brakerski has studied ``computationally efficient" measures of entanglement from somewhat different perspective \cite{arnon2023computational}.

The first partial answer to this question was given by Qin and Yao~\cite{qin2021nonlocal}. They investigated two-player nonlocal games\footnote{An $\MIP^*$ protocol is essentially a uniform family of two-player nonlocal games, with efficient algorithms for sampling pairs of questions and for evaluating the game decision predicate.} when the states shared between the players are arbitrarily many copies of a maximally entangled state (MES) with an arbitrarily small but \emph{constant} amount of noise on each copy, which is termed as {\em noisy MES} in their paper. 
The noise will cause the quantum maximal correlation, as defined in \Cref{def:maximalcorrelation}, to be less than 1, and the marginal state to be a completely mixed state. For instance, applying a depolarizing channel to an MES results in a noisy MES.
They showed that the supremum winning probability over all strategies using these states can be computably approximated to any finite precision. In fact, they showed that for any $\varepsilon$, there is a number of copies of the noisy MES, which is a computable function of only $\varepsilon$ and the size of the nonlocal game, that is sufficient to achieve winning probability within $\varepsilon$ of this supremum. This implies that any language in $\MIP^*$ restricted to such states is decidable, meaning that this class is strictly smaller than $\RE$.

This result was later generalized to nonlocal games that allow quantum questions and quantum answers~\cite{qin_et_al:LIPIcs.ICALP.2023.97}. 
To put these results in the language of complexity classes, let $\mathrm{MIP}^*\Br{q,a,\psi}$ be the set of languages that are decidable in the model of two-prover, one-round quantum multiprover interactive proof systems, where the provers share arbitrarily many copies of $\psi$, the messages from the verifier are classical and $q$-bits long, and the messages from the provers are also classical and $a$-bits long. 
\cite{Reichardt2013,JNVWYuen'20,JNVWY'20}, while both the complexity classes $\mathrm{MIP}^*\Br{\mathrm{poly},\mathrm{poly},\psi}$ and $\mathrm{QMIP}\Br{\mathrm{poly},\mathrm{poly},\psi}$ are computable if $\psi$ is a noisy MES state~\cite{qin2021nonlocal,qin_et_al:LIPIcs.ICALP.2023.97}. Moreover, \cite{qin2021nonlocal,qin_et_al:LIPIcs.ICALP.2023.97} showed explicit, though very large, time bounds for computing approximations to the game value for noisy states.  

Although these results show that the full power of $\mathrm{MIP}^*$ is not robust against noise in the shared entanglement, it is still possible that multiprover interactive proof systems gain a finite but very large computational advantage by sharing noisy maximally entangled states, since the time bounds from the previous work are much larger than for the classes with no entanglement. Thus, it was consistent with prior work that $\mathrm{MIP}^*\Br{\poly, \poly, \psi}$ is contained in  nondeterministic quadruply exponential time complexity class for noisy $\psi$~\cite{qin2021nonlocal}, which is much more powerful than $\mathrm{MIP}\Br{\poly, \poly} = \mathrm{NEXP}$. 
This paper attempts to answer this question by investigating the complexity classes $\mathrm{MIP}^*\Br{\poly,O(1),\psi}$ (i.e. protocols with constant-size answers) when $\psi$ is a noisy MES, whose local dimension is a constant. 
Classically, it is known that $\mathrm{MIP}\Br{\poly, \poly} = \mathrm{MIP}\Br{\poly, O(1)} = \mathrm{NEXP}$ \cite{Babai1991,mie2009short}\footnote{$\mathrm{MIP}\Br{\poly, \poly} =\mathrm{NEXP}$ was proved in \cite{Babai1991}. $\mathrm{MIP}\Br{\poly, O(1)} = \mathrm{NEXP}$ can be proved using a scaled-up version of PCP theorem~\cite{mie2009short}.}. Our main result, stated in the language of nonlocal games, is the following.

\begin{theorem}[Informal]
    \label{thm:informal_1}
	Given a nonlocal game in which the players share arbitrarily many copies of a noisy MES $\psi$, and the size of the answer sets is constant, then approximating the value of the game up to any sufficiently small constant precision is  $\mathrm{NP}$-complete. 
\end{theorem}

The runtime in \cref{thm:informal_1} is measured in terms of the size of a description of the nonlocal game as a table containing the distribution over question pairs and the verifier's predicate for every tuple of questions and answers. Translating this result to the $\MIP^*$ world requires parametrizing the runtime in terms of the \emph{number of bits} in the questions and answers. Thus, \cref{thm:informal_1} shows that noisy $\MIP^*$ with $O(\log(n))$-bit questions and $O(1)$-bit answers is $\mathrm{NP}$-complete.
Scaling our result up to $\MIP^\ast$ protocols with $O(\poly(n))$-bit questions and $O(1)$-bit answers, we get the following.
\begin{cor}
    $\MIP^\ast[\poly, O(1),\psi]=\mathrm{NEXP}$, where $\psi$ is a noisy MES.
\end{cor}
Intuitively, \cref{thm:informal_1} says that for any nonlocal game, if the shared MES has constant noise, the players' optimal strategy has a concise classical description which is also easy to verify. 
It is interesting to compare such nonlocal games with their classical counterparts. H\aa stad in his seminal work~\cite{hastad} proved that it is $\mathrm{NP}$-hard to approximate the value of a classical nonlocal game to a constant precision even if the size of the answer set is a constant.  It is also worth noting that sharing entanglement does not always strengthen the hardness of nonlocal games. It may weaken the hardness of certain games as well. For example, the quantum XOR games and quantum unique games are easy~\cite{1313847,doi:10.1137/090772885}, while the classical XOR games are $\mathrm{NP}$-hard, and the classical unique games are conjectured to be $\mathrm{NP}$-hard as well~\cite{10.1145/509907.510017}. 
Thus introducing noisy quantum states doesn't introduce any quantum effect to the hardness at all.

One may wonder whether this surprising collapse in complexity is caused by the restriction to noisy states or the restriction to $O(1)$-size answers. We give strong evidence that it is the former, by showing that $\MIP^*$ with \emph{noiseless} states and $O(1)$-sized answers is still equal to $\RE$.
\begin{theorem}[\Cref{thm:re}]
    \label{thm:informal2}
    $\RE$ is equal to $\MIP^\ast[\poly, O(1),\ket{\mathrm{EPR}}]$ with completeness $1$ and constant soundness.
\end{theorem}
To put this in context, the original work~\cite{JNVWYuen'20,JNVWY'20} proves that nonlocal games with noiseless EPR states are $\mathrm{RE}$-complete to approximate if both the question set and answer set are of \emph{polynomial} size. Recently, Natarajan and Zhang~\cite{10.1145/3564246.3585208} proved, by repeatedly applying the ``question reduction" technique from~\cite{JNVWY'20}, that it is still $\mathrm{RE}$-complete if the question length is $O(1)$ and the answer length is $\mathrm{polylog}(n)$. 
Here, we achieve constant \emph{answer} length by combining a tightened version of the previous answer reduction technique with a new answer reduction transformation, obtained by instantiating the error-correcting code-based scheme of~\cite{neexp} with the Hadamard code. We also show how to alternately achieve constant answer length by iterative application of the tightened standard answer reduction, similarly to how \cite{10.1145/3564246.3585208} obtained constant question length.

\Cref{thm:informal_1,thm:informal2} give us strong evidence that the computational power of $\MIP^*$ will vanish in the presence of noise. 
So for any complexity class slightly larger than $\NEXP$, we cannot hope for an $\MIP^*$ protocol robust against noise.
They also suggest that the key resource behind the computational power of $\MIP^*$ is specifically copies of the MES state, not just entanglement. This is because as we remarked above,
as $n$ tends to infinity, $n$ copies of a noisy MES contain an amount of entanglement going to infinity under standard entanglement measures.\footnote{Note that quantum states from which MES can be obtained through local operations without any communication are considered equivalent to MES in this model. This is because two non-communicating provers can transform any such state to an MES.}
Alternately, using the power of $\MIP^*[\psi]$ as a measure of entanglement for $\psi$, we show that an MES and an $\varepsilon$-noisy MES are sharply separated by this measure for any constant $\varepsilon$.

Since efficient self-tests for large entangled states are the key technique behind the proof of $\MIP^* = \RE$, our result puts some limitations on the design of self-tests robust against noise. 
More specifically, our result suggests that to noise-robustly self-test larger entangled states, the numbers of questions and answers must grow with the dimension of the tested state.
For comparison, if we don't need a self-test to be noise-robust, this is not necessary \cite{fu2019}.

\subsection{Proof Overview}

The harder part is to show that there is an $\NP$-algorithm for this problem. 
To illustrate our algorithm, we adapt the framework of Fourier analysis on matrix spaces. This framework was initiated in~\cite{MO10,PhysRevA.84.052328} and views the set of $n$-qubit operators as a Hilbert space obtained by tensoring $n$ copies of 2-dimensional Hilbert spaces. Furthermore, we extend the results in the analysis of Boolean functions~\cite{Odonnell08} to such a space.
Readers may refer to~\cite{qin2021nonlocal} for a thorough treatment.

\subsubsection{Approximating the Values of Noisy Games is \texorpdfstring{$\NP$}{NP}-Complete.}

Given a nonlocal game sharing arbitrary copies of a noisy MES $\psi$, 
 Qin and Yao \cite{qin2021nonlocal} showed that
it suffices for the players to share
 $D$ copies of $\psi$ to achieve the value of the game to an arbitrarily small precision, where $D$ only depends on the size of the game and the precision.

We first improve the upper bound $D$ to make it only depend exponentially on the length of the questions instead of doubly exponentially as in \cite{qin2021nonlocal}.
To prove this upper bound, we use ideas from Fourier analysis. For illustration, 
let's assume $\psi=\rho\ketbra{\mathrm{EPR}}+(1-\rho)\id_2/2\otimes\id_2/2$ is a depolarized noisy EPR state for simplicity. 
 Given a strategy $S$, let $P$ be a POVM element from the strategy, which acts on $n$ qubits.
 We are going to show the upper bound is independent of $n$, so in the rest of the section by ``constant" we mean independent of $n$.
 Let the Fourier expansion of $P$ be

\[P=\sum_{\sigma\in\set{0,1,2,3}^n}\widehat{P}\br{\sigma}\P_{\sigma},\]
where $\P_{\sigma}=\otimes_{i=1}^n\P_{\sigma_i}$ and $\P_0=I,\P_1=X,\P_2=Y,\P_3=Z$ are the single-qubit Pauli operators.
The degree of a term $\widehat{P}\br{\sigma}\P_{\sigma}$ is the number of nontrivial Pauli's in it, denoted by $\abs{\sigma}$.
First, we adapt the smoothing technique in~\cite{qin2021nonlocal}, which applies a depolarizing channel with small noise to $P$ and removes the high-degree part of $P$, i.e. terms with $\abs{\sigma} > d$ where $d$ is a constant.  
After smoothing, $S$ only contains degree-$d$ operators 
\[P^{(\textrm{Smooth})}=\sum_{\sigma:\abs{\sigma}\leq d}\widehat{P^{(\textrm{Smooth})}}\br{\sigma}\P_{\sigma},\]
so we denote the new strategy by $S^{(\textrm{Smooth})}$.
Using the argument in~\cite{qin2021nonlocal}, the probability of winning the game with this new strategy changes at most slightly, i.e.
\begin{align*}
    \val^*(G, S^{(\textrm{Smooth})}) \approx \val^*(G, S).
\end{align*}
Let $\tau$ be a small constant independent of $n$. Since the degree of $P^{(\textrm{Smooth})}$ is $d$, using a standard argument in the analysis of Boolean functions, the number of registers having influence that exceeds a given small $\tau$ is at most $d/\tau$. Notice that $d$ is independent of $n$, so is $d/\tau$. 
Assume without loss of generality that $H=\set{1,\ldots, \abs{H}}$ is the set of all registers whose influence exceeds $\tau$. 
We apply the invariance principle from \cite{qin2021nonlocal} to replace all the non-identity Pauli bases in the registers with low influence by Gaussian variables while maintaining the strategy value.
Let
\[\randP^{(\textrm{Apprx})}=\sum_{\sigma:\abs{\sigma}\leq d}\widehat{P^{(\textrm{Smooth})}}\br{\sigma}\P_{\sigma_1}\otimes\P_{\sigma_2}\otimes\ldots\P_{\sigma_{|H|}}\otimes \randz^{(\abs{H}+1)}_{\sigma_{\abs{H}+1}}\id_2\otimes\randz^{(\abs{H}+2)}_{\sigma_{\abs{H}+2}}\id_2\otimes\ldots\otimes\randz^{(n)}_{\sigma_n}\id_2,\]
where $\id_2$ is a $2\times 2$ identity matrix; $\set{\randz^{(i)}_j}_{|H|+1\leq i\leq n, 1\leq j\leq 3}$ are independent Gaussian variables and $\randz^{(|H|+1)}_0=\ldots\randz^{(n)}_0=1$.
Denote the new strategy by $S^{(\textrm{Apprx})}$, then  
\begin{align*}
    \val^*(G, S^{(\textrm{Apprx})}) \approx \val^*(G, S^{(\textrm{Smooth})}).
\end{align*}
Notice that this process significantly reduces the dimension of $\randP^{(\textrm{Apprx})}$ to a constant.
To round such a randomized strategy back to a valid POVM strategy, we first need to reduce the number of Gaussian variables from $O(n)$ to a constant, which is the most difficult step. 
In this paper, we avoid the use of a crude union bound as in~\cite{qin2021nonlocal}, by taking the distribution of the questions into account. 
Furthermore, we manage to ensure that the expectation of the distance from a random operator in the intermediate step to positive matrices after the Gaussian dimension reduction step is independent of the question size.
Then the inverse of the invariance principle allows us to round the randomized strategy back to a valid POVM strategy only acting on constantly many qubits. 
The improvements in the Gaussian dimension reduction step give us the improved bound.
 
 This upper bound has already yielded an $\NEXP$ algorithm, where the certificate is an {\em exponential-sized} description of the strategy. 
To design a more efficient nondeterministic algorithm, we need to further compress the certificate to polynomial length. 
To compress the certificate, we first smoothen again the strategy by introducing additional noise as in the proof of the upper bound of $D$ to remove all the high-degree terms.
Such a transformation exponentially reduces the length of the certificate. 
The smoothed strategy only contains a polynomial number of coefficients since the maximal degree is a constant. Nonetheless, the smoothed strategy is only a  {\em \pseudostrategy}, probably not a valid strategy because these smoothed operators may not be positive semidefinite and thus do not form valid POVMs. 
The prover sends the description of a \pseudostrategy~to the verifier, which is of polynomial length. The verifier performs a test on the given certificate to see if it is close to a valid strategy that gives a high winning probability with
the following steps:
\begin{enumerate}
  \item Check that the pseudo-POVM elements contained in the \pseudostrategy~still sum up to the identity.
  \item Compute and check the winning probability of the \pseudostrategy.
  \item Check that all the operators in the \pseudostrategy~are close to being positive semidefinite.
\end{enumerate}
Item 1 is straightforward. For item 2, notice that $\Tr\br{\P_i\otimes\P_j}\psi=\delta_{i,j} c_i$, where $c_0=1$ and $c_1=c_2=c_3=\rho$. Thus for any degree-$d$ operators $A,B$, we have 
\begin{equation}\label{eqn:QP}
\Tr\br{A\otimes B}\psi^{\otimes D}=\sum_{\sigma:\abs{\sigma}\leq d}\widehat{A}\br{\sigma}\widehat{B}\br{\sigma}c_{\sigma},
\end{equation}
where $c_{\sigma}=c_{\sigma_1}\cdots c_{\sigma_n}$. This computation can be done in polynomial time. The winning probability is simply a linear combination of a polynomial number of the terms in the form of Eq.\eqref{eqn:QP}, which, therefore, can also be computed in polynomial time. Item 3 is the most challenging. Notice that the dimension of each operator in the \pseudostrategy~is still exponential. Thus, the verifier cannot directly compute its eigenvalues and check its positivity. 
Instead, we need an efficient {\em positivity tester} for large matrices.

The key component of our efficient positivity tester is a {\em derandomized invariance principle}, which enables us to further reduce the dimension of the operators to a constant and maintain the distance between the operator and the set of positive operators.
To be more specific, let us define the real function $\zeta$ to be 
\begin{equation}\label{eqn:introzeta}
	\zeta\br{x}=\begin{cases}x^2~&\mbox{if $x\leq 0$}\\ 0~&\mbox{otherwise}\end{cases}.
\end{equation}
Then $\Tr~\zeta(P)$ is the distance from $P$ to its positive part.
As before, when the degree of an operator is bounded by a constant $d$, the number of quantum registers having influence that exceeds a given small constant $\tau$ is at most $d/\tau$, which is also a constant.
To further reduce the dimension of the operators,
we prove a more general invariance principle for all smooth functions compared with the one in~\cite{qin2021nonlocal}. 
It states that if all non-identity Pauli bases in the registers with low influence are substituted by Rademacher variables or Gaussian variables, the expectation of the distance to the set of positive semidefinite matrices is almost unchanged. 
We replace all such registers with Rademacher variables, which significantly reduces the dimension of a constant-degree operator to a constant, making it possible to compute its expected $\zeta$ function value efficiently. 
However, the invariance principle introduces $\poly\br{s}$-many random variables, where $s$ is the size of the question sets. 
This only leads to a randomized positivity tester. 
To reduce the randomness, we further apply the well-known Meka-Zuckerman pseudorandom generator~\cite{10.1145/1806689.1806749} to obtain a derandomized invariance principle, which only uses a logarithmic number of independent bits to simulate these variables\footnote{An alternate approach is using Gaussian variables and derandomizing the Gaussian variables as in~\cite{kane:LIPIcs.CCC.2015.567}, which discretizes the Gaussian variables via the Box-Muller transformation and further derandomizes the discrete random variables.}. This gives a deterministic algorithm to approximately compute the expected $\zeta$ function values of all the measurement operators .

To prove the approximation problem is $\NP$-hard, we can compile any $\MIP[\log, O(1)]$ protocol for $3$-SAT into a family of noisy nonlocal games one for each $3$-SAT instance such that if a $3$-SAT instance is satisfiable, the corresponding game has value $1$ and if not, the value of the corresponding game is below some constant. 
In the compiled nonlocal game, the verifier checks with equal probability, if the provers can give consistent answers for the same question or if the provers can give valid answers for queries of their assignment of the instance.
Using Fourier analysis, we show that when the provers share noisy MESs, winning the consistency checks with high probability implies that their strategy is essentially deterministic. Then we can relate the classical completeness and soundness 
of the $\MIP$ protocol to the values of the noisy nonlocal games.

\subsubsection{Hardness of Noiseless \texorpdfstring{$\MIP^*[\poly, O(1)]$}{MIP*[poly, O(1)]}}
To show hardness of $\MIP^*[\poly, O(1)]$, we start from the known result $\MIP^*[\poly, \poly] = \RE$~\cite{JNVWY'20}, and apply \emph{answer reduction} transformations to the protocol to get answer length $O(1)$. Answer reduction is essentially PCP composition adapted to the $\MIP^*$ setting, and was already an essential component in \cite{neexp} and \cite{JNVWY'20}. Intuitively, the idea of answer reduction is to ask the two provers in an $\MIP^*$ protocol to compute a PCP proof that their answers satisfy the verifier's predicate. The verifier will check this proof rather than checking the answers directly. In order to instantiate this, one requires a PCP of proximity (PCPP) that remains sound when implemented as a two-player quantum game. Showing this soundness condition is technically challenging and usually involves showing that the local tester for a locally testable code, when converted to a two-prover game, is sound against entangled provers. In~\cite{JNVWY'20}, the code that was used was the Reed-Muller code, which has superconstant alphabet size.
Moreover, the formulation in~\cite{JNVWY'20} was for the setting of reducing the answer length from exponential to polynomial, and in fact the specific theorem shown there is incapable of reducing the answer length below $\polylog(n)$. Our first contribution is to improve the parameters of this answer reduction transformation to make sure that in each application it can reduce answer size exponentially and can be recursively applied to reduce answer size below $\log(n)$.

To go all the way down to $O(1)$-sized answers, we combine this Reed-Muller-based answer reduction with a new answer reduction theorem based on the Hadamard code, which is a locally testable code over the binary alphabet. Fortunately for us, it is known that the local tester for this code is ``quantum sound"~\cite{ito2012,NV17}. Moreover, the answer-reduction protocol in \cite{neexp} is \emph{modular}: it was shown in that work that \emph{any} code with sufficiently good parameters and a quantum-sound tester can be combined with an off-the-shelf PCPP to achieve answer reduction. Our main challenge is to show that the Hadamard code (or a slight variant of it) has a tester meeting the conditions of this theorem. Our new tester for the Hadamard code allows us to reduce the answer length from 
$O(\log(n))$ to $O(1)$ directly.

\subsection{Technical Contributions}
\label{sec:technical_contribution}

\subsubsection{Invariance Principle and Derandomized Invariance Principle for Matrix Functions} The invariance principle \cite{mossel2005noise} is a generalization of the Berry-Esseen Theorem, which is a quantitative version of the Central Limit Theorem, to multilinear low-degree polynomials. 
Before illustrating the invariance principle, we need to introduce the notion of {\em influence}, a fundamental notion in the analysis of Boolean functions. Given a real function $f:\reals^n\rightarrow\reals$ and i.i.d. random variables $\randx_1,\ldots, \randx_n$, the influence of $i$-th coordinate is 
\[\infi{f}=\expec{}{\ab{f\br{\randx}-f\br{\randx^{(i)}}}^2},\]
where $\randx^{(i)}$ is obtained from $\randx$ by resampling the $i$-th variable.  Hence, it captures the effect of the $i$-th variable on the function on average. Given a multilinear low-degree polynomial $f$ in which all variables have low influence, the invariance principle states that the distributions of $f\br{X_1,\ldots, X_n}$ and $f(Y_1,\ldots, Y_n)$ are similar as long as the first and second moments of the random vectors $\br{X_1,\ldots, X_n}$ and $(Y_1,\ldots, Y_n)$ match, and the variables $X_i, Y_i$ behave nicely\footnote{To be more specific, $\randx_i, \randy_i$ need to be hypercontractive. Informally speaking, the $p$-norms  $\norm{\randx_i}_p=\expec{}{\ab{\randx_i}^p}^{1/p}$ $\norm{\randy_i}_p=\expec{}{\ab{\randy_i}^p}^{1/p}$ do not increase drastically with respect to $p$. Many basic random variables, such as uniformly random variables and Gaussian variables, are hypercontractive.}. The invariance principle is a versatile tool that allows us to connect the distribution of a function on complicated random variables to the distribution obtained by replacing these random variables with simpler ones, such as Gaussian variables or Rademacher random variables. The proof of the classical invariance principle in \cite{mossel2005noise} is via Lindeberg's hybrid argument, which is also a classic method to prove the Central Limit Theorem.

In \cite{qin2021nonlocal}, Qin and Yao started investigating the invariance principle on matrix spaces. Suppose that $P$ is a $m^n\times m^n$ matrix, viewed as an operator acting on $n$ registers, each of dimension $m$. 
Let $\xi:\reals\rightarrow\reals$ be a smooth real function. Suppose all registers have low influence in $P$, where the influence is a generalization of the influence for functions. When substituting all registers with independent standard Gaussians or Rademacher variables multiplied by an identity matrix, we expect that the change of $\Tr~\xi(P)$ is small in expectation. The most challenging part of extending Lindeberg's argument to matrix functions is computing the high-order Fr\'echet derivatives, which are complicated and difficult to analyze in general \cite{SENDOV2007240}. Qin and Yao \cite{qin2021nonlocal} established an invariance principle for a specific spectral function by directly computing the Fr\'echet derivatives and applying many complicated matrix-analytic techniques.
Hence, the first obstacle we face is to prove an invariance principle for more general functions.

To overcome it, we adapt the theory of multilinear operator integrals \cite{skripka2019multilinear}, which provides a unified way to compute and bound the Fr\'echet derivatives. With such a tool, we establish an invariance principle applicable to a broader class of functions, including those that are smooth with a bounded third derivative and those that are Lipschitz continuous.

The invariance principle reduces the dimension from $\poly$ to constant but introduces a $\poly$ number of independent random variables. Thus, the second obstacle is that the size of the overall probability space is exponential.  
To improve the computational efficiency of our invariance principle, we use the ideas of \cite{10.1145/1806689.1806749,10.1145/2395116.2395118,10.1145/3460532} to use a Pseudorandom generator (PRG) to reduce the number of independent random variables. We apply this derandomized invariance principle to our positivity tester introduced below.
Derandomized invariance principles build upon the crucial observation that the highest moment of variables involved in the proof is at most $2d$, where $d$ is the degree of the operator, which is a constant.
Thus, it suffices to use $4d$-wise uniform random variables instead of polynomially many independent random variables when we replace the Pauli basis elements in the low-influence registers, which saves the randomness exponentially. To this end, we employ the well-known Meka-Zuckerman pseudorandom generator~\cite{10.1145/1806689.1806749} to construct $4d$-wise uniform random variables.

As the invariance principle has found numerous applications, we anticipate that the invariance principle for spectral functions is interesting in its own right. The positivity testing for low-degree matrices introduced below is an example of its applications.

\subsubsection{Positivity Tester for Low-degree Matrices} A Hermitian matrix $A$ is said to be positive semidefinite (PSD) if all the eigenvalues of $A$ are non-negative. This testing problem has received increasing attention in the past couple of years \cite{10.5555/644108.644112,han2017approximating,bakshi2020testing,9996743}. In this work, we present an efficient PSD tester for low-degree matrices, where the input matrix is given in terms of its Fourier coefficients. Given an $m^n\times m^n$ matrix, viewed as an operator acting on $n$-qudits, each of which has dimension $m$, if the degree of the operator is $d$, then the number of Fourier coefficients is bounded by $\sum_{i\le d}\binom{n}{i}(m^2-1)^i=O(dn^dm^{2d})$. Hence, this allows for a compact description of a low-degree, exponential-dimension operator.
If $m, d$ are constants, the input is of size $\poly(n)$, and we work in this setting when we explain how the tester works below.

Given the Fourier coefficients of a matrix $P$,
our tester estimates the distance between $P$ and the set of positive semidefinite matrices measured by $\Tr\zeta\br{P}$, where $\zeta\br{\cdot}$ is defined in Eq.~\eqref{eqn:introzeta}. 
Estimating $\Tr\zeta\br{P}$ involves applying the derandomized invariance principle introduced above.
More specifically, our tester enumerates all the possible seeds of the Meka-Zuckerman PRG to estimate this distance.
For each seed, the computation time is $O(1)$ because the derandomized invariance principle has effectively reduced the dimension of $P$ to a constant.
Hence, our tester runs in time $\mathrm{poly}(n)$, because there are only $\poly(n)$ seeds. 
Its guarantees are summarized below.

\begin{theorem*}[informal]
  Given as input the Fourier coefficients of a degree-$d$ operator $P$ acting on $n$ qudits, each of dimension $m$, 
  and error parameters $\beta\ge\delta\ge0$,
there exists an algorithm that runs in time $\exp(m^d/\delta)\cdot \poly(n)$
  such that
  \begin{itemize}
    \item the algorithm accepts if there exists a PSD operator $Q$ such that $\norm{P-Q}_F^2<\br{\beta-\delta}m^n$;
\item the algorithm rejects if $\norm{P-Q}_F^2>(\beta+\delta)m^n$ for any PSD operator $Q$.
  \end{itemize}
\end{theorem*}

 This approach completely differs from all previous works on positivity testing \cite{9996743,han2017approximating,bakshi2020testing}, where they only consider polynomial-sized matrices and the testers are randomized. In contrast, our tester is deterministic, and the dimension of the testing matrix can be exponential in input size if the degree is constant.

\subsubsection{Answer Reduction with the Hadamard Code} 
As mentioned above, we obtain $O(1)$-sized answers in the noiseless setting by applying the code-based answer reduction of~\cite{neexp}, with the code chosen to be the Hadamard code. To implement this required two new technical components. First, we showed a \emph{quantum-sound subset tester} for the Hadamard code: essentially, an interactive protocol that forces the provers to respond with the values of a subset $F$ of the coordinates of a Hadamard codeword, where $F$ is sampled from some (not necessarily uniform) distribution. Our proof of this result is essentially a generalization of the Fourier-analytic proof of the quantum soundness of the BLR test~\cite{BLR,NV17}. Secondly, the answer reduction procedure in~\cite{neexp} only works if the code has a relative distance close to $1$ (i.e., distinct codewords differ on almost all locations), whereas the Hadamard code has a distance $1/2$. 
To overcome this, we slightly modified the answer-reduced verifier's protocol of~\cite{neexp} by querying a large constant number of ``dummy coordinates" from the provers. It is worth mentioning that the answer reduction procedure from~\cite{neexp} is different from the procedure used in~\cite{JNVWY'20}; the former works for any error-correcting code satisfying certain properties but does not yield protocols that can be recursively compressed, whereas the latter is specialized to the low-degree code but is compatible with recursive compression.

In addition to this new answer reduction based on the Hadamard code, we also required a tightened version of the Reed-Muller-based answer reduction of \cite{JNVWY'20}, as noted above. This is because, due to the low rate of the Hadamard code, we must first reduce the answer length to $O(\log n)$ before applying our new answer reduction. However, the answer reduction as stated in \cite{JNVWY'20} can never reduce the answer size to smaller than $\polylog(n)$, because the reduced answer size depends poly-logarithmically on the verification time, which can never be smaller than $\poly(n)$ since the verifier must read the entire input. Our improvement is based on the observation that the verifier's verification process can be broken into two phases. In the first phase, a predicate of the answers is calculated, and in the second phase, the predicate is applied to the answers. We observe that the new answer size only depends poly-logarithmically on the size of the Boolean circuit implementing the predicate, which can be much smaller than the total runtime of the verifier when the answers are short. This observation is standard in the classical PCP literature, but was not necessary for \cite{JNVWY'20} since they were not concerned with obtaining sub-polynomial answer length.

Using this observation, we show that in each application of the answer reduction transformation, both the answer size and the predicate size are reduced exponentially, which allows us to apply it recursively to reduce answer size to below $O(\log n)$, at which point the Hadamard-based answer reduction can take us to constant answer size. We remark that it is also possible to achieve constant answer size by iteratively applying the improved answer reduction. The analysis of this is slightly less clean, but we sketch it at the end of the proof of \Cref{thm:re}.

\subsection{Discussions and Open Problems}

Our result characterizes the effect of depolarizing noise on the computational complexity class $\MIP^\ast$.
To our knowledge, this is the first example of a quantum computational complexity class whose quantum advantage over its classical counterpart completely vanishes in the presence of noise.
For comparison, noise causes \emph{no} collapse in the $\mathrm{BQP}$ model, or in general, for $\mathrm{BQTIME}$ because the algorithms in these classes can be implemented fault-tolerantly.
Even for algorithms with bounded space, it seems that the same reasoning still applies because all the intermediate measurements to achieve fault tolerance can be eliminated without a large space overhead \cite{fefferman2021eliminating}.
Hence, our work raises the natural question of which quantum complexity classes are truly fault tolerant. 
In contrast, for complexiy classes like $\MIP^*$, the fault-tolerance theorem \cite{aharonov2008fault} cannot be applied as the model of computation disallows the operations needed to perform error correction. For the specific case of $\MIP^*$, our result further shows that no form of fault tolerance is possible. 

Our proof techniques can be applied to the depolarizing noise but not the
bit-flipping noise, phase-flipping noise, or phase-damping noise.
This is because those types of noise do not reduce the quantum maximal correlation. Similarly, our techniques cannot be applied to the amplitude-damping noise because under this noise the marginal state is not completely mixed.
Hence, the effect of these noise channels on $\MIP^*$ is not clear.
On the other hand, if Alice and Bob start with tilted EPR pairs, for example, caused by some unitary noise, they can produce maximally entangled states via local operations, which is called entanglement concentration in literature \cite{bennett1996concentrating}. Then they can execute the $\MIP^*$ protocol for $\RE$.

More broadly, we know other examples where constant noise destroys the quantum advantage.
Random circuit sampling has been proposed to demonstrate the quantum advantage offered by near-term quantum devices \cite{boixo2018characterizing}.
However, when the random circuits are subject to constant noise, this sampling task becomes classically easy \cite{aharonov2023polynomial}.
We have more of such examples in quantum query algorithms. For example, if the oracle is noisy or faulty, no
quantum algorithm can achieve any speed-up in the unstructured search problem \cite{regev2008impossibility}.
In a setting closer to the near-term devices, where each gate in the circuit is subject to independent noise but the oracle is perfect,
the authors of \cite{chen2022complexity} showed that no quantum algorithm could achieve any speed-up in the unstructured search problem either.
For a more detailed survey of the effect of noise on quantum query algorithms, we refer to \cite[Section 3]{chen2022complexity}.

In recent years, the study of noise has focused on its effect on quantum circuits. In the circuit model, the study is about how noise accumulates in quantum circuits where each gate is subject to some noise. Now we know that noise effectively truncates a quantum circuit to a logarithmic depth \cite{mele2024noise}.
In our case, only the entangled states are subject to noise, and there is no accumulation of noise in the measurements. Our results show that the noise still limits the effective width of the circuit, but do not say anything about the effective depth, which means
that in our setting the prover could perform quantum circuits with arbitrary depths.

Our result also raises some natural but intriguing questions. We list some of them below.
\begin{enumerate}
    \item For $\MIP^\ast$ protocols with more rounds of interactions and larger answer sets, it is unclear how big the effect of noise is. The current answer reduction techniques do not work when the provers can only share noisy MES. Hence, we ask: Does the vanishing phenomenon for computational advantages occur for general $\MIP^\ast$ protocols?
    \item What non-computational capabilities of the $\MIP^*$ model remain in the noisy setting? Specifically, it is known that nonlocal games and correlations can be used to self-test entangled states. In the noisy setting, can we certify any properties of the provers' shared entanglement? Previous work on this question has studied entanglement of formation~\cite{arnonFYen2018} and one-shot distillable entanglement~\cite{arnon2019device}, but the general picture remains unclear.

    \item Classical invariance principle serves as a pivotal tool in the analysis of Boolean functions, which has found applications in designing various areas including pseudorandom generators and counting algorithms~\cite{10.1145/2395116.2395118,10.1145/3460532,10.1145/3357713.3384281,10.1145/3519935.3519965,10.4230/LIPIcs.CCC.2022.21}. Analysis on matrix spaces and the space of super-operators, a.k.a, Pauli analysis~\cite{10.1145/3618260.3649662} is receiving increasing attention\cite{pmlr-v195-bao23b, doi:10.1137/1.9781611977554.ch43,arunachalam_et_al:LIPIcs.ICALP.2024.13,10.1145/3618260.3649662,Rouze2024,klein_et_al:LIPIcs.ITCS.2024.69,slote2024noncommutativebohnenblusthilleinequalityqudit}.  Will our invariance principle lead to new applications?

    \item Testing whether a matrix is positive has played an important role in the study of algorithm designs for linear algebra problems, community structure detection, differential equations, etc (see~\cite{bakshi2020testing} and references therein). Multiple studies have been devoted to designing efficient algorithms for positivity testing~\cite{9996743,han2017approximating,bakshi2020testing}. Will our algorithm of positivity testing find new applications?
\end{enumerate}

\subsection*{Acknowledgment}
P.Y. would like to thank the discussion with Zhengfeng Ji. Part of the work was done when H.F. and H.X. visited Nanjing University. Y.D., M.Q. and P.Y. were supported by National Natural Science Foundation of China (Grant No. 62332009, 12347104), Innovation Program for Quantum Science and Technology (Grant No. 2021ZD0302901), NSFC/RGC Joint Research Scheme (Grant no. 12461160276) and Natural Science Foundation of Jiangsu Province (No. BK20243060). H.F. was supported by the US National Science Foundation QLCI program (grant OMA-2016245). H.X. was supported by Beijing Nova Program 20220484128.

\newcommand{\pinv}[1]{\ensuremath{#1^+}}
\newcommand {\influence} {\ensuremath{ \mathrm{Inf} }}

\section{Preliminary}

For $n\in\posint$, let $[n]$ and $[n]_{\geq 0}$ represent the sets $\set{1,\ldots, n}$ and $\set{0,\ldots, n-1}$, respectively. Given a finite set $\X$ and a natural number $k$, let $\X^k$ be the set $\X\times\cdots\times\X$, the Cartesian product of $\X$, $k$ times. For any $\sigma\in\mathbb{Z}_{\geq 0}^k$, we define $\abs{\sigma}=\abs{\set{i:\sigma_i\neq 0}}$.

In this paper, the lowercase letters in bold $\mathbf{x},\mathbf{y},\cdots$ are reserved for random variables.
The capital letters in bold, $\mathbf{A},\mathbf{B},\ldots$ are reserved for random operators.

\subsection{Quantum mechanics}\label{sec:quantum}

A quantum system is associated with a complex finite-dimensional Hilbert space, denoted by $A$. A quantum state in $A$ can be completely described by a density operator, a positive semidefinite operator with trace one. If the dimension of $A$ is $m$, we denote the set of Hermitian matrices in $A$ by $\H_m$. The identity matrix is denoted by $\id_m$ or $\id_A$. The state of a composite quantum system is the Kronecker product of the state spaces of the component systems. An important operation on a composite system $A\otimes B$ is the {\em partial trace} $\Tr_B\br{\cdot}$ which effectively derives the marginal state of the subsystem $A$ (denoted by $\psi_A$) from the quantum state $\psi_{AB}$.  The partial trace is given by \[\psi_A=\Tr_B\psi_{AB}=\sum_i\br{\id_A\otimes\bra{i}}\psi_{AB}\br{\id_A\otimes\ket{i}},\] where $\set{\ket{i}}$ is an orthonormal basis in $B$. A linear map from a system $A$ to a system $B$ is {\em unital} if it maps $\id_A$ to $\id_B$.  A {\em quantum measurement} is represented by a {\em positive operator-valued measure} (POVM), which is a set of positive semidefinite operators $\set{M_1,\ldots, M_n}$ satisfying $\sum_{i=1}^nM_i=\id$, where $n$ is the number of possible measurement outcomes. Suppose that the state of the quantum system is $\psi$, then the probability that it produces $i$ is $\Tr~ M_i\psi$. We use $\vec{M}=\br{M_1,\ldots, M_n}$ to represent an ordered set of operators.

The notion of {\em quantum maximal correlations} introduced by Beigi~\cite{Beigi:2013} is crucial to our analysis. 
\begin{definition}[Quantum maximal correlation]~\cite{Beigi:2013}\label{def:maximalcorrelation}
	Given quantum systems $A, B$ of dimension $m$ and a bipartite state $\psi_{AB}$ with $\psi_A=\psi_B=\frac{\id_{m}}{m}$, the quantum maximal correlation of $\psi_{AB}$ is defined to be
	\[\rho\br{\psi_{AB}}=\sup\set{\abs{\Tr\br{\br{P^{\dagger}\otimes Q}\psi_{AB}}}~:\genfrac{}{}{0pt}{}{P, Q\in\mathbb{C}^{m\times m},}{\Tr~P=\Tr~Q=0, \nnorm{P}_2=\nnorm{Q}_2=1.}}\]
\end{definition}

\begin{fact}~\cite{Beigi:2013}\label{fac:maximalcorrlationone}
	Given quantum systems $A, B$ and a bipartite quantum state $\psi_{AB}$ with $\psi_A=\id_{m_A}/m_A$ and $\psi_B=\id_{m_B}/m_B$, it holds that $\rho\br{\psi_{AB}}\leq 1$.
	\end{fact}
	
	\begin{definition}\label{def:noisyepr}
		Given quantum systems $A$ and $B$ with $\dim\br{A}=\dim\br{B}=m$, a bipartite state $\psi_{AB}\in\D\br{A\otimes B}$ is an $m$-dimensional \emph{noisy} maximally entangled state (MES) if $\psi_A=\psi_B=\id_m/m$ and its quantum maximal correlation $\rho=\rho\br{\psi_{AB}}<1$.
	\end{definition}

	An interesting class of noisy MESs is the isotropic states, which are the states obtained by depolarizing MESs with arbitrarily small noise.
	\begin{fact}\label{lem:noisyeprmaximalcorrelation}\cite[Lemma 3.9]{qin2021nonlocal}
		For any $0\leq\epsilon<1$ integer $m>1$, it holds that
		\[\rho\br{\br{1-\epsilon}\ketbra{\Psi}+\epsilon\frac{\id_{m}}{m}\otimes \frac{\id_{m}}{m}}=1-\epsilon,\]
		where $\ket{\Psi}=\frac{1}{\sqrt{m}}\sum_{i=0}^{m-1}|m,m\rangle$ is an $m$-dimensional MES.
	\end{fact}
 \begin{remark}
Fact~\ref{lem:noisyeprmaximalcorrelation} indicates the quantum maximal correlation of an isotropic state is strictly less than $1$. The class of noisy MES also contains other states. It is not hard to prove that any mixture of at least three out of the four orthogonal EPR states is a $2$-dimensional noisy MES.
 \end{remark}
	
	\begin{fact}\label{lem:normofM}\cite[Lemma 7.4]{qin2021nonlocal}
	Given $m\in\posint$, $m\geq2$, and a noisy $m$-dimensional MES $\psi_{AB}$. Then there exist standard orthonormal bases $\A=\set{\A_i}_{i=0}^{m^2-1}$ and $\B=\set{\B_i}_{i=0}^{m^2-1}$ in $\H_m$ such that
	\begin{equation}\label{eqn:corr}
\Tr\br{\br{\A_i\otimes\B_j}\psi_{AB}}=\begin{cases}c_i~&\mbox{if $i=j$}\\0~&\mbox{otherwise},\end{cases}
\end{equation}
where $c_0=1\geq c_1=\rho\br{\psi_{AB}}\geq c_2\geq\ldots c_{m^2-1}\geq0$ and $\rho\br{\psi_{AB}}$ is defined in \cref{def:maximalcorrelation}.
\end{fact}

\subsection{Matrix analysis}
\subsubsection{Matrix spaces}\label{}
Given $m\in\posint$ and $M\in\H_m$, we use $M_{i,j}$ to represent the $\br{i,j}$-th entry of $M$. For $1\leq p\leq\infty$, the $p$-norm of $M$ is defined to be \[\norm{M}_p=\br{\sum_{i=1}^{m}s_i\br{M}^p}^{1/p},\] where $\br{s_1\br{M},s_2\br{M},\ldots,s_m\br{M}}$ are the singular values of $M$ sorted in nonincreasing order. $\norm{M}=\norm{M}_{\infty}=s_1\br{M}$. 
The {\em normalized $p$-norm} of $M$ is defined as
\begin{equation}\label{eqn:nnormdef}
\nnorm{M}_p=\br{\frac{1}{m}\sum_{i=1}^ms_i\br{M}^p}^{1/p}
\end{equation}
 and $\nnorm{M}=\nnorm{M}_{\infty}=s_1\br{M}$.

	Given $P,Q\in\M_m$, we define
	\begin{equation}\label{eqn:innerproduct}	
	\innerproduct{P}{Q}=\frac{1}{m}\Tr~P^{\dagger}Q.
	\end{equation}
It is easy to verify that $\innerproduct{\cdot}{\cdot}$ is an inner product. $\br{\innerproduct{\cdot}{\cdot},\H_m}$ forms a Hilbert space. For any $M\in\H_m$, $\nnorm{M}_2^2=\innerproduct{M}{M}$.

	We say that $\set{\B_0,\ldots,\B_{m^2-1}}$ is a {\em standard orthonormal basis} in $\M_m$ if it is an orthonormal basis with all elements being Hermitian and $\B_0=\id_m$, which is an $m\times m$ identity matrix.
 \begin{fact}\cite[Lemma 2.10]{qin2021nonlocal}\label{fact:fourier-expansion}
     For any integer $m\geq 2$, a standard orthonormal basis exists in $\M_m$. 
 \end{fact}
	
\par{Given a standard orthonormal basis $\B=\set{\B_i}_{i=0}^{m^2-1}$ in $\H_m$, every matrix $M\in\H_m^{\otimes n}$ has a {\em Fourier expansion} with respect to the basis $\B$ given by
\[M=\sum_{\sigma\in[m^2]_{\geq 0}^{n}}\widehat{M}\br{\sigma}\B_{\sigma},\] where $\B_{\sigma}=\bigotimes_{i=1}^n\B_{\sigma_i}$.
}
\par{
\begin{definition}\label{def:inf}
		Let $\B=\set{\B_i}_{i=0}^{m^2-1}$ be a standard orthonormal basis in $\H_m$, $P\in\H_m^{\otimes n}$.
		\begin{enumerate}
			\item The {\em degree} of $P$ is defined to be \[\deg P=\max\set{\abs{\sigma}:\widehat{P}\br{\sigma}\neq 0}.\]
Recall that $\abs{\sigma}$ represents the number of nonzero entries of $\sigma$.
            \item For any $i\in[n]$, the {\em influence} of $i$-th coordinate is defined to be: \[\influence_i(P)=\nnorm{P-\id_m \otimes \Tr_iP}_2^2,\]
            where $\id_m$ is in the $i$'th quantum system, and the partial trace $\Tr_i$  is defined as the operator $\id\otimes\Tr$, with the trace operator $\Tr$ acting on the $i$'th quantum system.
			\item The total influence is defined by\[\influence\br{P}=\sum_i\influence_i\br{P}.\]
		\end{enumerate}
	\end{definition}
 }

\begin{fact}\label{lem:partialvariance}\cite[Lemma 2.16]{qin2021nonlocal}
Given $P\in\H_m^{\otimes n}$, a standard orthonormal basis $\B=\set{\B_i}_{i=0}^{m^2-1}$ in $\H_m$ and a subset $S\subseteq[n]$, it holds that
	\begin{enumerate}
		\item $\influence_i\br{P}=\sum_{\sigma:\sigma_i\neq0}\abs{\widehat{P}\br{\sigma}}^2$;
		\item $\influence\br{P}=\sum_{\sigma}\abs{\sigma}\abs{\widehat{P}\br{\sigma}}^2\leq\deg P\cdot\nnorm{P}^2_2$.
	\end{enumerate}
\end{fact}
The inequality in item 2 follows from Parseval's identity,
which is immediate by the Fourier expansion of $P$ (\cref{fact:fourier-expansion}). 
\begin{fact}[Parseval's identity]
    For any $P\in\H_m^{\otimes n}$,
    $$\nnorm{P}_2^2=\sum_{\sigma}\abs{\widehat{P}\br{\sigma}}^2.$$
\end{fact}

\begin{definition}\label{def:bonamibeckner}
	Given $m\in\posint$, $\rho\in[0,1]$, a noise operator $\Delta_{\rho}:\H_m\rightarrow\H_m$ is defined as follows. For any $P\in\H_m$,
	\[\Delta_{\rho}\br{P}=\rho P+\frac{1-\rho}{m}\br{\Tr~P}\cdot\id_m.\]
	With a slight abuse of notations, the noise operator $\Delta_{\rho}^{\otimes n}$ on the space $\herspace{n}$ is also denoted by $\Delta_{\rho}$.
\end{definition}
\begin{fact}\label{lem:bonamibecknerdef}\cite[Lemma 3.5]{qin2021nonlocal}
	Given integers $d,n,m>0$, $\rho\in[0,1]$, a standard orthonormal basis of $\H_m$: $\B=\set{\B_i}_{i=0}^{m^2-1}$, then for any $P\in\H_m^{\otimes n}$ with a Fourier expansion $P=\sum_{\sigma\in[m^2]_{\geq 0}^n}\widehat{P}\br{\sigma}\B_{\sigma}$, it holds that
		\[\Delta_{\rho}\br{P}=\sum_{\sigma\in[m^2]_{\geq 0}^n}\rho^{\abs{\sigma}}\widehat{P}\br{\sigma}\B_{\sigma}.\]
\end{fact}

\subsubsection{Random matrices.}
\mlqin{Check this subsubsection. To make the degree of functions well-defined, I add some preliminaries about Gaussian space.}

For integer $n\geq 1$, $\gamma_n$ represents the distribution of an $n$-dimensional  standard normal distribution. For any $0\leq \rho\leq 1$, $\G_{\rho}$ represents a  $\rho$-correlated Gaussian distribution, which is a  $2$-dimensional Gaussian distribution \[\br{X,Y}\sim N\br{\begin{pmatrix}
      0 \\
      0
    \end{pmatrix},\begin{pmatrix}
                    1 & \rho \\
                    \rho & 1
                  \end{pmatrix}}.\] Namely, the marginal distributions $X$ and $Y$ are distributed according to $\gamma_1$ and $\expec{}{XY}=\rho$.

We say a function $f:\reals^n\rightarrow\reals$ is in $L^2\br{\reals,\gamma_n}$ if
\[\int_{\reals^n}f(x)^2\gamma_n\br{\d x}<\infty.\]

We equip $L^2\br{\reals,\gamma_n}$ with an inner product
\[\innerproduct{f}{g}_{\gamma_n}=\expec{x\sim\gamma_n}{f(x)g(x)}.\]

Given $f\in L^2\br{\reals,\gamma_n}$, the 2-norm of $f$ is defined to be
\[\twonorm{f}=\sqrt{\innerproduct{f}{f}_{\gamma_n}}.\]

     The set of {\em Hermite polynomials} forms an orthonormal basis in $L^2\br{\reals,\gamma_1}$ with respect to the inner product $\innerproduct{\cdot}{\cdot}_{\gamma_1}$. The Hermite polynomials $H_r:\reals\rightarrow\reals$ for $r\in\mathbb{Z}_{\geq 0}$ are defined as
	\begin{equation*}
	H_0\br{x}=1; H_1\br{x}=x; H_r\br{x}=\frac{(-1)^r}{\sqrt{r!}}\e^{x^2/2}\frac{\d^r}{\d x^r}\e^{-x^2/2}.
	\end{equation*}
	For any $\sigma\in\br{\sigma_1,\ldots,\sigma_n}\in\mathbb{Z}_{\geq 0}^n$, define
	$H_{\sigma}:\reals^n\rightarrow\reals$ as \begin{equation*}
	H_{\sigma}\br{x}=\prod_{i=1}^nH_{\sigma_i}\br{x_i}.
	\end{equation*}
	The set $\set{H_{\sigma}:\sigma\in\mathbb{Z}_{\geq 0}^n}$ forms an orthonormal basis in $L^2\br{\reals,\gamma_n}$. Every function $f\in L^2\br{\reals,\gamma_n}$ has an {\em Hermite expansion}  as
$$f\br{x}=\sum_{\sigma\in\mathbb{Z}_{\geq 0}^n}\widehat{f}\br{\sigma}\cdot H_{\sigma}\br{x},$$
	where $\widehat{f}\br{\sigma}$'s are the {\em Hermite coefficients} of $f$, which can be obtained by $\widehat{f}\br{\sigma}=\innerproduct{H_{\sigma}}{f}_{\gamma_n}$. The degree of $f$ is defined to be \[\deg\br{f}=\max\set{\sum_{i=1}^n\sigma_i:~\widehat{f}\br{\sigma}\neq 0}.\]

We say $f\in L^2\br{\reals,\gamma_n}$ is {\em multilinear} if $\widehat{f}\br{\sigma}=0$ for $\sigma\notin\set{0,1}^n$.

Now we give the definition of random matrix.
\begin{definition}\label{def:randomoperators}
		Given $h, n, m\in\posint$, we say $P(\randg)$ is a random matrix if it can be expressed as
		\begin{equation}\label{eqn:randomoperatorexpansion}
		P(\randg)=\sum_{\sigma\in[m^2]_{\geq 0}^h}p_{\sigma}\br{\mathbf{g}}\B_{\sigma},
		\end{equation}
 		where $\set{\B_i}_{i=0}^{m^2-1}$ is a standard orthonormal basis in $\H_m$, $p_{\sigma}:\reals^n\rightarrow\reals$ for all $\sigma\in[m^2]_{\geq 0}^h$ and $\mathbf{g}\sim \gamma_n.$ Moreover, we say $P(\randg)\in L^2\br{\H_m^{\otimes h},\gamma_n}$ if $p_{\sigma}\in L^2\br{\reals,\gamma_n}$ for all $\sigma\in[m^2]_{\geq 0}^h$.
	\end{definition}

We define the degree of random operators:

	\begin{definition}\label{def:randop}
		Given integers $n,h>0, m>1$ and random operator $\mathbf{P}\in L^p\br{\H_m^{\otimes h},\gamma_n}$, the degree of $\mathbf{P}$, denoted by $\deg\br{\mathbf{P}}$, is \[\max_{\sigma\in[m^2]_{\geq 0}^h}\deg\br{p_{\sigma}}.\] We say $\mathbf{P}$ is multilinear if $p_{\sigma}\br{\cdot}$ is multilinear for all $\sigma\in[m^2]_{\geq 0}^h$.
	\end{definition}

\subsubsection{Fr\'echet derivatives and spectral functions.}

The Fr\'echet derivatives are derivatives on Banach spaces. In this paper, we only concern ourselves with Fr\'echet derivatives on matrix spaces. Readers may refer to \cite{Coleman} for a detailed treatment.

\begin{definition}\label{def:frechetderivative}
	Given a map $f:\H_m\rightarrow\H_m$ and $P, Q\in\H_m$, the Fr\'echet derivative of $f$ at $P$ with direction $Q$ is defined to be
	\[Df\br{P}\Br{Q}=\frac{d}{dt}f\br{P+tQ}|_{t=0}.\]
	The $k$-th order Fr\'echet derivative of $f$ at $P$ with direction $\br{Q_1,\ldots, Q_k}$ is defined to be
	\[D^kf\br{P}\Br{Q_1,\ldots, Q_k}=\frac{d}{dt}\br{D^{k-1}f\br{P+tQ_k}\Br{Q_1,\ldots, Q_{k-1}}}|_{t=0}.\]
To keep notations short, we use $D^kf\br{P}\Br{Q}$ to represent $D^kf\br{P}\Br{Q,\ldots,Q}$.
\end{definition}

In this paper, we are concerned with {\em spectral functions}, a special class of matrix functions. We say that the function $F:\H_m\rightarrow\H_m$ is a spectral function if there exists a function $f:\reals\rightarrow\reals$ such that $F\br{P}=\sum_if\br{\lambda_i}\ketbra{v_i},$ where $P=\sum_i\lambda_i\ketbra{v_i}$ is a spectral decomposition of $P$. With slight abuse of notations, we use the same notation $f$ to represent the function on $\reals$ and the corresponding spectral function, whenever it is clear from the context.

Given $n\in\posint$, we denote $\C^n$ to be the space of functions continuously differentiable $n$ times.

\begin{definition}
Let $\lambda_0,\dots,\lambda_n\in\reals$ and let $f\in\C^n$. The divided difference $f^{[n]}$ is defined recursively by
\[f^{[n]}(\lambda_0,\lambda_1,\tilde{\lambda})=\begin{cases}
\frac{f^{[n-1]}(\lambda_0,\tilde{\lambda})-f^{[n-1]}(\lambda_1,\tilde{\lambda})}{\lambda_0-\lambda_1}\quad\text{if }\lambda_0\ne\lambda_1,\\
\frac{\d}{\d\lambda_0}f^{[n-1]}(\lambda_0,\tilde{\lambda})\quad\text{if }\lambda_0=\lambda_1,
\end{cases}\]
where $\tilde{\lambda}=(\lambda_2,\dots,\lambda_n)$.
\end{definition}

It is well known that $f^{[n]}$ is a symmetric function.

\begin{fact}
 \cite[Theorem 5.3.2]{skripka2019multilinear}
 \cite[Theorem 6.1]{SENDOV2007240}\label{fac:frechetD}
Given $m,n\in\posint$, $P,Q\in\H_m$. Suppose that $P$ has a spectral decomposition
\begin{equation}\label{eqn:specdecom}
P=\sum_{i=1}^{m}\lambda_i\Pi_i,
\end{equation}
where $\lambda_1\geq\dots\geq\lambda_{m}$, $\set{\Pi_i}_{i\in[{m}]}$ are rank-one projectors satisfying that $\sum_{i=1}^{m}\Pi_i=\id$ and $\Pi_i\Pi_j=0$ for all $i\ne j$. Let $f\in\C^{n}$. Then
\[D^nf(P)\Br{Q}=\sum_{i_0,\dots,i_n\in[m]}f^{[n]}\br{\lambda_{i_0},\dots,\lambda_{i_n}}\Pi_{i_0}Q\Pi_{i_1}Q\dots Q\Pi_{i_n}.\]
\end{fact}

The following is one of the main results in the theory of multilinear operator integrals~\cite{skripka2019multilinear}. 
\begin{fact}\label{fact:remainder} \cite[Theorem 5.3.12]{skripka2019multilinear}
Given $m,n\in\posint$, $P,Q\in\H_m$. Let $f\in\C^{n}$. Denote
\[\Delta_{n,f}(P,Q)=f(P+Q)-\sum_{k=0}^{n-1}\frac{1}{k!}D^kf(P)\Br{Q},\]
then there exists a constant $c_n$ depending only on $n$ such that 
\[\ab{\Tr\Br{\Delta_{n,f}(P,Q)}}\leq c_n\norm{f^{(n)}}_\infty\norm{Q}_n^n,\]
where $\norm{f^{(n)}}_\infty$ denotes the supremum of $f^{(n)}$.
\end{fact}

\subsubsection{The distance from PSD matrices}

Define the function $\zeta:\reals\rightarrow\reals$ as follows.

\begin{eqnarray}
	&&\zeta\br{x}=\begin{cases}x^2~&\mbox{if $x\leq 0$}\\ 0~&\mbox{otherwise}\end{cases}.\label{eqn:zeta}
\end{eqnarray}

The function $\zeta$ measures the distance between a given matrix and its closest positive semi-definite matrix:
\begin{fact}\label{lem:closedelta1}\cite[Lemma 9.1]{qin2021nonlocal}
	Given an integer $m>0$, $M\in\H_m$, $\mathrm{Pos}=\set{X\in\H_m:X\geq0}$, let \[\R\br{M}=\arg\min\set{\twonorm{M-X}:X\in\mathrm{Pos}}\]
	 be a rounding map of $\mathrm{Pos}$ with respect to the distance $\twonorm{\cdot}$. It holds that
	\[\Tr~\zeta\br{M}=\twonorm{M-\R\br{M}}^2.\]
\end{fact}

\begin{fact}\label{lem:zetaadditivity}\cite[Lemma 10.4]{qin2021nonlocal}
	For any Hermitian matrices $P$ and $Q$, it holds that \[\abs{\Tr~\br{\zeta\br{P+Q}-\zeta\br{P}}}\leq2\br{\twonorm{P}\twonorm{Q}+\twonorm{Q}^2}.\]
\end{fact}

We will need to mollify\footnote{A mollified function $\zeta_\lambda$ is a smooth function that is close to the original function $\zeta$.}  $\zeta$ to get a smooth function:
\begin{fact}\label{fac:zetalambda}\cite[Lemma 3.21]{mossel2005noise}
Given $\lambda>0$, there exists a $C^\infty$ function $\zeta_\lambda$ satisfying
\begin{enumerate}
\item $\norm{\zeta_\lambda-\zeta}_\infty\leq2\lambda^2$,
\item For any integer $n\geq2$, there exists a constant $B_n$ independent of $\lambda$ such that \[\norm{\br{\zeta_\lambda}^{(n)}}_\infty\leq B_n\lambda^{2-n}.\]
\end{enumerate}
\end{fact}

\subsection{\texorpdfstring{$k$}{k}-wise uniform hash functions and random variables}\label{sec:k-uniform}

\begin{definition}
A family $\F=\set{f:[n]\to[p]}$ of hash functions is $k$-wise uniform if for any $y_1, \dots, y_k\in[p]$ and distinct $x_1, \dots, x_k\in[n]$:
\[\Pr_{f\in_u\F}\Br{f(x_i)=y_i\wedge\dots\wedge f(x_k)=y_k}=\frac{1}{p^k}.\]
\end{definition}
\begin{definition}
A random vector $\randz\in[p]^n$ is $k$-wise uniform if for any $y_1,\dots,y_k\in[p]$ and distinct $x_1,\dots,x_k\in[n]$:
\[\Pr_{\randz}\Br{\randz_{x_i}=y_i\wedge\dots\wedge \randz_{x_k}=y_k}=\frac{1}{p^k}.\]
\end{definition}

\begin{lemma}\label{lem:kwise-hash}
  Let $p$ be a power of $2$.
  There exists an efficient construction of $k$-wise uniform hash functions $\F=\set{f: [n]\to[p]}$ of size $\abs{\F}=O(\max(n, p)^k)$.
\end{lemma}
\begin{proof}
  For $k=2$, efficient constructions of size $\abs{\F}=O(np)$ are well known (see, e.g., \cite{10.1145/800105.803400}).
  For general $k$,
let $t$ be the minimal integer satisfying $2^t>\max(n, p)$ and consider the finite field $\mathbb{F}_{2^t}$.
  We can construct an irreducible polynomial in $\mathbb{F}_2$ of degree $t$ in polynomial time,
using, for example, the algorithms of Shoup~\cite{shoup1990new}.
  Thus, the basic operations in $\mathbb{F}_{2^t}$ can be carried out efficiently.
  Then the $k$-wise uniform hash functions $\tilde{\F}: \set{\tilde{f}: \mathbb{F}_{2^t}\to\mathbb{F}_{2^t}}$ can be efficiently constructed,
for example, using the construction in Section 3.5.5 in \cite{TCS-010},
  which has size $\abs{\mathbb{F}_{2^t}}^k=O(\max(n, p))^k$.
  Then $k$-wise uniform hash functions from $[n]$ to $\mathbb{F}_{2^t}$ can be constructed by restricting the input domain to $[n]$.
  $k$-wise uniform hash functions from $[n]$ to $[p]$ can be further constructed by cutting the output to $\log p$ bits.
\end{proof}
\begin{cor}\label{cor:kwise-z}
  There exists an efficient construction of $k$-wise uniform random variables $\randz\sim\set{-1,1}^n$,
which can be enumerated in $O(n^k)$ time.
\end{cor}
\begin{proof}
  Construct $k$-wise uniform hash functions $\F=\set{f: [n]\to\set{-1,1}}$, and then define $\randz=(f(1), \dots, f(n))$.
  By the definition of $k$-wise uniform hash functions,
  $\randz$ is $k$-wise uniform random variables.
  Moreover, the construction of $\F$ is efficient.
  Finally, the enumeration of $\randz$ takes time $O(n^k)$
  since we only need to enumerate the set $\F$.
\end{proof}

\subsection{Nonlocal games and \texorpdfstring{$\MIP^\ast$}{MIP*} protocols} 
Two-player one-round $\MIP^\ast$ protocols are also nonlocal games.
We follow the notations of \cite{JNVWY'20} for nonlocal games.
\begin{definition}[Two-player one-round games]
    A two-player one-round game $G$ is specified by a tuple $(\calX, \calY, \calA, \calB, \mu, V)$ where
    \begin{itemize}
        \item $\calX$ and $\calY$ are finite sets, called the \emph{question sets},
        \item $\calA$ and $\calB$ are finite sets, called the \emph{answer sets},
        \item $\mu$ is a probability distribution over $\calX \times \calY$, called the \emph{question distribution}, and
        \item $V: \calX \times \calY \times \calA \times \calB \to \set{0,1}$ is a function, called the \emph{decision predicate}.
    \end{itemize}
\end{definition}
\begin{definition}[Tensor-product strategies]
    A tensor-product strategy $S$ of a nonlocal game $G = (\calX, \calY, \calA, \calB, \mu, V)$ is a tuple 
    $(\psi, A, B)$ where
    \begin{itemize}
        \item a bipartite quantum state $\psi \in \calH_A \x \calH_B$ for finite dimensional complex Hilbert
spaces $\calH_A$ and $\calH_B$,
        \item $A$ is a set $\set{A^x}$ such that for every $x \in \calX$, $A^x = \set{A^x_a \mid a \in \calA}$ is 
        a POVM over $\calH_A$, and 
        \item $B$ is a set $\set{B^y}$ such that for every $y \in \calY$, $B^y = \set{B^y_b \mid b \in \calB}$ is 
        a POVM over $\calH_B$.
    \end{itemize}
\end{definition}
\begin{definition}[Tensor product value]
    The tensor product value of a tensor product strategy $S = (\psi, A ,B)$ for a nonlocal game
    $G = (\calX, \calY, \calA, \calB, \mu, V)$ is defined as
    \begin{align*}
        \val^\ast(G,S) = \sum_{x,y,a,b} \mu(x,y)V(x,y,a,b) \Tr\br{\br{A^x_a \x B^y_b}\psi}.
    \end{align*}
    For $v \in [0,1]$ we say that the strategy passes or wins $G$ with probability $v$ if $\val^\ast(G,S)\geq v$.
    The quantum value or tensor product value of $G$ is defined as
    \begin{align*}
        \val^*(G) = \sup_{S} \val^\ast(G,S)
    \end{align*}
    where the supremum is taken over all tensor product strategies $S$ for $G$.
\end{definition}
When we prove the quantum soundness of an $\MIP^\ast$ protocol, we focus on projective strategies, where the measurements
$A^x$ and $B^y$ are all projective, following Naimark's Dilation theorem \cite[Theorem 5.1]{JNVWYuen'20}.
\begin{definition}
    A game $G = (\calX, \calY, \calA, \calB, \mu, V)$ is symmetric if $\calX = \calY$ and $\calA = \calB$, the distribution $\mu$
    is symmetric (i.e. $\mu(x,y) = \mu(y,x)$ for all $x$ and $y$), and the predicate $V$ treats both players symmetrically (i.e. 
    $V(x,y,a,b) = V(y,x,b,a)$ for all $x,y,a,b$).

    We call a strategy $S = (\ket{\psi}, A, B)$ symmetric if $\ket{\psi}$ is a pure state in $\calH \x \calH$, for some Hilbert space $\calH$, that is invariant under permutation of the two factors, and the measurement operators of both players are identical. 
\end{definition}
A symmetric game is denoted by $(\calX, \calA, \mu, V)$, and a symmetric strategy is denoted by $(\ket{\psi}, M)$
where $M$ denotes the set of measurement operators for both players.
\begin{lemma}[Lemma 5.7 in \cite{JNVWY'20}]
    Let $G = (\calX, \calA, \mu, V)$ be a symmetric game with value $1 - \ep$ for some $\ep \geq 0$.
    Then there exists a symmetric and projective strategy $S = (\ket{\psi}, M)$ such that the $\val^\ast(G,S)\geq 1- \ep$.
\end{lemma}
Hence, for symmetric nonlocal games, it suffices to only consider symmetric strategies.

\subsection{Lemmas for the answer reduction of \texorpdfstring{$\MIP^\ast$}{MIP*}}
\label{sec:lemma_mip}
This section introduces several lemmas to prove the hardness of $\MIP^\ast[\poly,O(1)]$. 
We use the following notations for approximation in this section and \cref{sec:re}.
\begin{itemize}
\item For complex numbers $a$ and $b$, we write $a \appd{\delta} b$ if $\abs{a - b} \leq \delta$.
\item With respect to a distribution $D$ on $\calX$ and state $\ket{\psi}$, we write
\begin{align*}
	A^x_a \appd{\delta} B^x_a \quad \text{if} \quad \bigE_{x \sim D}\sum_{a \in \calA} \norm{ (A^x_a - B^x_a) \ket{\psi}}^2 \leq \delta.
\end{align*}
	\item With respect to a distribution $D$ on $\calX$ and state $\ket{\psi}$, we write
	\begin{align*}
		A^x_a \simeq_{\delta} B^x_a \quad \text{if} \quad \bigE_{x \sim D} \sum_{a \in \calA} \bra{\psi} A^x_a \x B^x_a \ket{\psi} \geq 1 - \delta.
	\end{align*}
\end{itemize}
In the rest of the section, the distribution on $\calX$ is implicit.
\begin{lemma}[Fact 4.13 of \cite{neexp}]
	\label{lm:consist_to_approx}
	Let $\set{A^x_a}$ and $\set{B^x_a}$ be POVM measurements.
	If $A^x_a \x \id \simeq_{\delta} \id \x B^x_a$, then $A^x_a \x \id \appd{2\delta} \id \x B^x_a$.
\end{lemma}
\begin{lemma} 
	\label{lm:close}
	Suppose $\set{A^x_a}$ and $\set{ B^x_a }$ are two measurements such that
	one of them is projective, and that 
	\begin{align*}
		A^x_a \x \id \appd{\delta} \id \x B^x_a
	\end{align*}
	with respect to some distribution $D$ of $x$ and the quantum state $\ket{\psi}$.
	Then
	\begin{align*}
		\ab{ \bigE_x \sum_a \bra{\psi} A^x_a \x \id - \id\x B^x_a \ket{\psi}} \leq  2 \sqrt{\delta}.
	\end{align*}
\end{lemma}
This proof is deferred to \cref{sec:ans}.
\begin{lemma}[Fact 4.14 of \cite{neexp}]
	\label{lm:convert}
	Suppose $\set{A^x_a}$ and $\set{ B^x_a }$ are two measurements such that 
	$A^x_a \x \id \appd{\delta} \id \x B^x_a$.
	Suppose that either $A$ or $B$ is a projective measurement and the other is a POVM measurement.
	Then $A^x_a \x \id \simeq_{\sqrt{\delta}} \id \x B^x_a$.
\end{lemma}
\begin{lemma}[Proposition 4.26 of \cite{JNVWYuen'20}]
\label{lm:prod_meas}
Let $\set{ C^x_{a,b} } \subseteq \calL(\calH)$ be a set of matrices such that $\sum_{b} (C^x_{a,b})\ct C^x_{a,b} \leq \id$
for all $x$ and $a$.
Then
\begin{align*}
	A^{x}_a \appd{\delta} B^{x}_a \quad \text{implies that } \quad C^x_{a,b}A^x_a \appd{\delta} C^x_{a,b}B^x_a. 
\end{align*}
\end{lemma}
\begin{lemma}[Proposition 4.28 of \cite{JNVWYuen'20}]
\label{lm:chain_rule}
	Suppose $A_i = \set{ (A_i)^x_a}$ be a set of matrices such that $(A_i)^x_a \appd{\delta_i} (A_{i+1})^x_a$
	for $i \in [k]$. Then 
	\begin{align*}
		(A_1)^x_ a \appd{k(\delta_1 + \ldots + \delta_k)} (A_{k+1})^x_a.
	\end{align*}
\end{lemma}
\begin{lemma}[Fact 4.33 of \cite{neexp}]
	\label{lm:k_prod}
	Let $k \geq 0$ be a constant.
	Let $\set{ A^x_{a_1, \dots, a_k}}$ be a projective measurement.
	For $1 \leq j \leq k$, let $\set{ (B_j)^x_{a_j} }$ be a projective measurement, and suppose that 
	\begin{align*}
		A^x_{a_j} \x \id \appd{\delta} \id \x (B_j)^{x}_{a_j}.
	\end{align*}
	Define the POVM measurement $\set{J^x_{a_1, \dots, a_k}}$ as
	\begin{align*}
		J^x_{a_1, \ldots, a_k} = (B_k)^x_{a_k} \dots (B_2)^{x}_{a_2} (B_1)^x_{a_1} (B_2)^x_{a_2} \dots (B_k)^x_{a_k}.
	\end{align*}
	Then
	\begin{align*}
		A^x_{a_1, \ldots, a_k} \x \id \appd{ (2k-1)^2 \delta }  \id \x J^x_{a_1, \ldots, a_k}.
	\end{align*}
\end{lemma}
This proof is also deferred to \cref{sec:ans}.

\begin{lemma}[Fact 4.35 of \cite{neexp}]
	\label{lm:k_prod_distr}
	Let $k \geq 0$ be a constant.
	Let $D$ be a distribution on questions $(x, y_1, \ldots, y_k)$, where
	each $y_i \in \calY_i$. For each $1 \leq i \leq k$, let $\calG_i$ be a set of functions $g_i : \calY_i \to \calR_i$,
	and let $\set{ (G_i)^x_g \mid g \in \calG_i}$ be a projective measurement.
	Suppose that the set $\calG_i$ has the following distance property: fix a question
	$z = (x, y_1,\ldots, y_{i-1}, y_{i+1},\ldots, y_k)$, and let $D_z$ be the distribution on $y_i$ conditioned on $z$.
	Then for any two nonequal $g_i, g_i' \in \calG_i$, the probability that $g_i(\by_i) = g_i'(\by_i)$, over a random 
	$\by_i \sim D_z$, is at most $\ep$.
	
	Let $\set{ A^{x,y_1,\ldots, y_k}_{a_1,\ldots, a_k}}$ be a projective measurement with outcomes $a_i \in \calR_i$.
	For each $1 \leq i \leq k$, suppose that 
	\begin{align}
		\label{eq:A_G_consist}
		A^{x,y_1, \ldots, y_k}_{a_i} \x \id \simeq_{\delta} \id \x (G_i)^x_{[g_i(y_i)=a_i]}  \\
		\label{eq:G_A_consist}
		(G_i)^x_{[g_i(y_i)=a_i]} \x \id \simeq_{\delta} \id \x A^{x,y_1, \ldots, y_k}_{a_i}.
	\end{align}
	Also suppose that 
	\begin{align}
		\label{eq:A_consist}
		A^{x,y_1, \ldots, y_k}_{a_i} \x \id \simeq_{\delta} \id \x A^{x,y_1, \ldots, y_k}_{a_i}. 
	\end{align}
	Define the POVM $\set{ J^x_{g_1, \ldots, g_k} }$ as
	\begin{align*}
		J^x_{g_1, \ldots, g_k}:= (G_k)^x_{g_k} \cdots (G_2)^x_{g_2} \cdot (G_1)^x_{g_1} \cdot (G_2)^x_{g_2}
		\cdots (G_k)^x_{g_k}.
	\end{align*}
	Then
	\begin{align*}
		A^{x,y_1, \ldots, y_k}_{a_1,\ldots, a_k} \x \id \appd{O(\exp(k) (\delta^{1/4^{k-1}} + \ep^{1/(2\cdot 4^{k-2})})}) \id \x J^x_{[g_1(y_1),\ldots, g_k(y_k) = a_1, \ldots, a_k]}.
	\end{align*}
\end{lemma}
This proof is the same as the original one, but we rewrite it to keep better track of the approximation errors.
We defer the proof to \cref{sec:ans}.

\section{Invariance principle for matrix spaces}\label{sec:invariance}
This section we will prove an invariance principle for general functions on matrix spaces. Hypercontractivity is crucial in the proofs of many invariance principles~\cite{mossel2005noise,isaksson2012maximally,10.1145/2395116.2395118,qin2021nonlocal,10.1145/3519935.3519965}. We also need to establish a new hypercontractive inequality before proving the invariance principle.

\subsection{Hypercontractivity}
In this subsection,  we adopt the concept of orthonormal ensembles as introduced in \cite{mossel2005noise}.
\begin{definition}
Given $m,n\in\posint$, a collection of $n$ real random variables $\set{\randz_1,\ldots,\randz_n}$ are orthonormal if $\expec{}{\randz_i\randz_j}=\delta_{i,j}$. We call a collection of $m$ orthonormal real random variables, the first of which is constant 1, an $m$-orthonormal ensemble. We call $\randx$ an $(m,n)$ ensemble if $\randx = (\randx_1,\dots ,\randx_n)$, where for all $i\in[n]$, $\randx_i = \set{\randx_{i,0} = 1,\randx_{i,1},\dots,\randx_{i,m-1}}$ is an $m$-orthonormal ensemble. 
\end{definition}

\begin{definition}\label{def:defofT}
Given $m,n\in\posint$, $\tau\in\Br{m}^n_{\geq0}$ and an $(m,n)$ ensemble $\randx$, denote $\randx_\tau=\prod_{i=1}^n\randx_{i,\tau_i}$. Define a multilinear polynomial over $\randx$ to be 
\[Q(\randx)=\sum_{\tau\in\Br{m}^n_{\geq0}}\widehat{Q}\br{\tau}\randx_\tau,\]
where the $\widehat{Q}\br{\tau}$'s are real constants. 

For $\gamma\in[0,1]$, we define the operator $T_\gamma$ acting on multilinear polynomial $Q(\randx)$ by 
\[T_\gamma Q(\randx)=\sum_{\tau\in\Br{m}^n_{\geq0}}\gamma^{\ab{\tau}}\widehat{Q}\br{\tau}\randx_\tau.\]

\end{definition}

\begin{definition}
For $1\leq r<\infty,$ let $\randy$ be a random variable with $\expec{}{\ab{\randy}^r}<\infty$. Define 
\[\norm{\randy}_r=\br{\expec{}{\ab{\randy}^r}}^{1/r}.\]

Given $1\leq p\leq q<\infty$, $0<\eta<1$, $m,n\in\posint$ and an $(m,n)$ ensemble  $\randx$, we say that $\randx$ is $(p,q,\eta)$ -hypercontractive if for any multilinear polynomial $Q$, it holds that 
\[\norm{\br{T_\eta Q}(\randx)}_q\leq\norm{Q(\randx)}_p.\]
\end{definition}

\begin{fact}\cite[Remark 3.10]{mossel2005noise}\label{fact:hc}
If $\randx$ is $(p,q,\eta)$-hypercontractive, then it is  $(p,q,\eta')$-hypercontractive for any $0<\eta' \leq\eta$.
\end{fact}

Consider an $(m,n)$ ensemble $\randx$. If for all $i\in[n]$, $j\in[m-1]$, $\randx_{i,j}$ are either independent standard Gaussians or independent Rademacher variables, then $\randx$ is $(2,q,(q-1)^{-1/2})$-hypercontractive. These two types are represented as significant examples of hypercontractive ensembles. Readers can refer to \cite{mossel2005noise} for an extensive treatment on hypercontractive ensembles.

We need the following lemma for technical reasons.
\begin{lemma}\label{lem:gausshyper}
	Given $m,n\in\posint$, $0<\eta<1$, a $(2,4,\eta)$-hypercontractive $(m,n)$ ensemble $\randx$, it holds that 
	\[\expec{ }{\br{\sum_{i=1}^k \br{T_{\eta}p_i}\br{\mathbf{x}}^2}^2}\leq\br{\expec{ }{\sum_{i=1}^kp_i\br{\mathbf{x}}^2}}^{2},\]
  for any multilinear polynomials $p_1,\ldots p_k$.
\end{lemma}

\begin{proof}
	Let $q_i= T_{\eta}p_i$. Then
	\begin{align*}
		\expec{ }{\br{\sum_{i=1}^k\br{T_{\eta}p_i}\br{\mathbf{x}}^2}^2}=~&\sum_{i, j}\expec{}{q_i\br{\mathbf{x}}^2q_j\br{\mathbf{x}}^2}\\
		\leq~&\sum_{i,j}\norm{q_i}_4^2\norm{q_j}_4^2\quad\quad\mbox{(Cauchy-Schwarz inequality)}\\
		\leq~&\sum_{i,j}\twonorm{p_i}^2\twonorm{p_j}^2\quad\quad\mbox{(\randx ~is $(2,4,\eta)$-hypercontractive)}\\
		=~&\br{\sum_i\twonorm{p_i}^2}^{2}\\
  =~&~\br{\expec{ }{\sum_{i=1}^kp_i\br{\mathbf{x}}^2}}^{2}.
	\end{align*}
\end{proof}

We then introduce the noise operator $\Gamma_{\gamma}$ for random matrices, which is a hybrid of $T_{\gamma}$ in \cref{def:defofT} and $\Delta_{\gamma}$ in \cref{def:bonamibeckner}.
	\begin{definition}\label{def:gamma}
	Given $0\leq\gamma\leq 1$, $h,n,m\in\posint,$ $m\geq 2$, an $(m^2,n)$ ensemble $\randx$, and a random matrix
 \begin{equation*}
	P(\randx)=\sum_{\sigma\in\Br{m^2}_{\geq0}^h}p_\sigma\br{\randx}\B_\sigma,
	\end{equation*}
	where $\set{\B_i}_{i=0}^{m^2-1}$ is a standard orthonormal basis and $p_\sigma$ is a real multilinear polynomial for all $\sigma\in\Br{m^2}_{\geq0}^h$, the noise operator $\Gamma_{\gamma}$ is defined to be	\[\Gamma_{\gamma}\br{P(\randx)}=\sum_{\sigma\in\Br{m^2}_{\geq0}^h}\br{T_\gamma p_\sigma}\br{\randx}\Delta_\gamma\br{\B_\sigma}.\]
\end{definition}

The lemma below follows directly from \cref{def:defofT} and \cref{lem:bonamibecknerdef}.

\begin{lemma}\label{lem:gammaoperator}
	Given $0\leq\gamma\leq 1$, $h,n,m\in\posint,$ $m\geq 2$, an $(m^2,n)$ ensemble $\randx$, and a random matrix
 \begin{equation*}
P(\randx)=\sum_{\sigma\in\Br{m^2}_{\geq0}^h}p_\sigma\br{\randx}\B_\sigma,
	\end{equation*}
	where $\set{\B_i}_{i=0}^{m^2-1}$ is a standard orthonormal basis and $p_\sigma$ is a real multilinear polynomial for all $\sigma\in\Br{m^2}_{\geq0}^h$, suppose that for all $\sigma\in\Br{m^2}_{\geq0}^h$, $p_\sigma$ has an expansion
	\[p_\sigma(\randx)=\sum_{\tau\in\Br{m^2}_{\geq0}^n}\widehat{p_\sigma}(\tau)\randx_\tau.\]
	 It holds that
	\begin{equation}\label{eqn:gamma}
	\Gamma_{\gamma}\br{P(\randx)}=\sum_{\sigma\in\Br{m^2}_{\geq0}^h}\sum_{\tau\in\Br{m^2}_{\geq0}^n}\gamma^{\ab{\sigma}+\ab{\tau}}\widehat{p_\sigma}(\tau)\randx_\tau\B_{\sigma}.
	\end{equation}
\end{lemma}

We need a hypercontractivity inequality for Hermitian matrices.
	\begin{fact}\label{lem:multihypercontractivity}\cite[Lemma 8.3]{qin2021nonlocal}
		Given $h,n,m\in\posint,$ $m\geq 2$, $0\leq\gamma\leq\br{9m}^{-1/4}$ and $P\in\H_m^{\otimes n}$, it holds that$$\nnorm{\Delta_\gamma^{\otimes n}\br{P}}_{4}\leq\nnorm{P}_2,$$
  where $\Delta_{\gamma}\br{\cdot}$ is defined in \cref{def:bonamibeckner}.
	\end{fact}
	
	The main result in this subsection is stated below.

	\begin{theorem}[Hypercontractivity for random matrices]\label{lem:hypercontractivity}
		Given $h,n,m\in\posint,$ $m\geq 2$, $0<\eta<1$, $0\leq\gamma\leq\min\set{\eta,\br{9m}^{-1/4}}$, a $(2,4,\eta)$-hypercontractive $(m^2,n)$ ensemble $\randx$ and a random matrix
 \begin{equation*}
P(\randx)=\sum_{\sigma\in[m^2]_{\geq0}^h}p_\sigma\br{\randx}\B_\sigma,
	\end{equation*}
	where $\set{\B_i}_{i=0}^{m^2-1}$ is a standard orthonormal basis, and $p_\sigma$ is a real multilinear polynomial for all $\sigma\in[m^2]_{\geq0}^h$, it holds that
		\[\expec{\randx}{\nnorm{\Gamma_{\gamma}\br{P(\randx)}}_4^4}\leq \br{\expec{\randx}{\nnorm{P(\randx)}_2^2}}^{2},\]
		where $\Gamma_\gamma$ is defined in \cref{def:gamma}.
	\end{theorem}	
	
	\begin{proof}

		Set $ Q(\randx)=\sum_{\sigma\in[m^2]_{\geq0}^h}\br{T_\gamma p_\sigma}\br{\randx}\B_\sigma$. Then by the definition of $\Gamma_{\gamma}$,
		\[\Gamma_{\gamma}\br{P(\randx)}=\Delta_{\gamma}\br{Q(\randx)}.\]
		\noindent Using \cref{lem:multihypercontractivity},
		\begin{equation}\label{eqn:np}
\expec{}{\nnorm{\Delta_{\gamma}\br{Q(\randx)}}_4^4}\leq\expec{}{\nnorm{Q(\randx)}_2^4}.
		\end{equation}

		Denote $q_\sigma=T_\gamma p_\sigma$. Notice that

\begin{align*}
\expec{}{\nnorm{Q(\randx)}_2^4}
=~&m^{-2h}\expec{}{\br{\sum_{\sigma\in[m^2]_{\geq0}^h}q_{\sigma}\br{\randx}^2}^2}\leq m^{-2h}\br{\expec{}{\sum_{\sigma\in[m^2]_{\geq0}^h}p_{\sigma}\br{\randx}^2}}^2
=\br{\expec{}{\nnorm{P(\randx)}_2^2}}^2,
		\end{align*}
		where the inequality follows from \cref{fact:hc} and \cref{lem:gausshyper}. We conclude the result by combining it with \cref{eqn:np}.
		
	\end{proof}
	
	The following is an application of \cref{lem:hypercontractivity}.
	\begin{theorem}\label{lem:hybridHC}
		Given $h,n,m,d\in\posint,$ $m\geq 2$, $0<\eta<1$, a $(2,4,\eta)$-hypercontractive  $(m^2,n)$ ensemble $\randx$, and a random matrix
 \begin{equation*}
P(\randx)=\sum_{\sigma\in\Br{m^2}_{\geq0}^h}p_\sigma\br{\randx}\B_\sigma,
	\end{equation*}
	where $\set{\B_i}_{i=0}^{m^2-1}$ is a standard orthonormal basis and for all $\sigma\in\Br{m^2}_{\geq0}^h$ and $p_\sigma$ is a real multilinear polynomial satisfying $\deg\br{p_\sigma}+\ab{\sigma}\leq d$, it holds that
   \[\expec{}{\nnorm{P(\randx)}_4^4}\leq\max\set{9m,1/\eta^4}^d\br{\expec{}{\nnorm{P(\randx)}_2^2}}^2.\]
	\end{theorem}
	\begin{proof}
	Suppose that for all $\sigma\in[m^2]_{\geq0}^h$, $p_\sigma$ has an expansion
\[p_\sigma(\randx)=\sum_{\tau\in\Br{m^2}_{\geq0}^n}\widehat{p_\sigma}(\tau)\randx_\tau.\]
		 Set $$P^{=i}(\randx)=\sum_{\genfrac{}{}{0pt}{}{\sigma\in[m^2]_{\geq0}^h,\tau\in\Br{m^2}_{\geq0}^n :}{\ab{\sigma}+\ab{\tau}=i}}\widehat{p_{\sigma}}\br{\tau}\randx_\tau\B_{\sigma}.$$
	 Set $\gamma=\min\set{\eta,\br{9m}^{-1/4}}$. Applying \cref{lem:gammaoperator} and \cref{lem:hypercontractivity},
		\begin{align*} \expec{}{\nnorm{P(\randx)}_4^4}&= \expec{}{\nnorm{\Gamma_{\gamma}\br{\sum_{i=1}^{d}\gamma^{-i} P^{=i}(\randx)}}_4^4}\leq \br{\expec{}{\nnorm{\sum_{i=1}^{d}\gamma^{-i} P^{=i}(\randx)}^2_2}}^{2}
		\end{align*}
		By the orthogonality of $\randx$ and $\B$, if $i\ne j$, we have
		\[\expec{}{\Tr~P^{=i}(\randx)P^{=j}(\randx)}=0.\]
		Therefore,
\begin{align*}
\expec{}{\nnorm{P(\randx)}_4^4}&\leq\br{\sum_{i=1}^{d}\gamma^{-2i}\expec{}{\nnorm{P^{=i}(\randx)}^2_2}}^{2}\leq\gamma^{-4d}\br{\sum_{i=1}^{d}\expec{}{\nnorm{P^{=i}(\randx)}^2_2}}^{2}=\gamma^{-4d}\br{\expec{}{\nnorm{P(\randx)}_2^2}}^2.
\end{align*}
		
	\end{proof}
	
\subsection{Invariance principle}

We are now prepared to introduce an invariance principle on matrix space applicable to general functions. Initially, we establish the proof for functions in $\C^{4}$.
\begin{theorem}\label{lem:generalinvariance}
	Given $0<\tau,\eta<1$, $d,h,m,n\in\posint$, $H\subseteq[n]$ of size $\ab{H}=h$, $\xi\in\C^{3}$ satisfying $\norm{\xi^{(3)}}_\infty\leq B$ where $B$ is a constant, and a $(2,4,\eta)$-hypercontractive  $(m^2,n)$ ensemble $\randx$, let $P\in\H_m^{\otimes n}$ be a degree-$d$ operator satisfying $\infi{P}\leq\tau$ for all $i\notin H$. Suppose that $P$ has a Fourier expansion
	\[P=\sum_{\sigmarange{n}}\widehat{P}\br{\sigma}\B_\sigma.\]
	Let 
	\[P^H(\randx)=\sum_{\sigmarange{n}}\widehat{P}\br{\sigma}\randx_{\sigma_{\overline{H}}}\B_{\sigma_H}.\]
	If $\sum_{\sigma\ne0}\widehat{P}\br{\sigma}^2\leq1$, we have
	\[\ab{m^{-n}\Tr~\xi\br{P}-m^{-h}\expec{}{\Tr~\xi\br{P^H(\randx)}}}\leq 2c_3B \max\set{9m,1/\eta^4}^d\sqrt{\tau}d\]
	for some absolute constant $c_3$.
\end{theorem}

\begin{proof}
Without loss of generality, we assume $\overline{H}=[n-h]$. We prove this by a hybrid argument. For any $0\leq i\leq n-h$, define the hybrid basis elements and the hybrid random operators as follows.
	\begin{align}
	&\X_{\sigma}^{\br{i}}=\randx_{\sigma_{\leq i}}\cdot\B_{\sigma_{>i}}~\mbox{for $\sigma\in[m^2]_{\geq 0}^n$};\label{eqn:hybridxi}\\
	&P^{\br{i}}\br{\randx}=\sum_{\sigma\in[m^2]^n_{\geq0}}\widehat{P}\br{\sigma}\X^{\br{i}}_{\sigma},\label{eqn:hybridmi}
	\end{align}
where $\randx_{\sigma_{\leq i}}=\randx_{\sigma_1}\cdots\randx_{\sigma_i}$
and $\B_{\sigma_{>i}}=\B_{\sigma_{i+1}}\otimes\ldots\otimes\B_{\sigma_n}$.
Then $P=P^{(0)}\br{\randx}$ and $P^{H}\br{\randx}=P^{(n-h)}\br{\randx}$. Note that

	\begin{align*}
	&P^{\br{i}}\br{\randx}=\sum_{\sigma:\sigma_{i+1}=0}\widehat{P}\br{\sigma}\X_{\sigma}^{\br{i}}+\sum_{\sigma:\sigma_{i+1}\neq 0}\widehat{P}\br{\sigma}\X_{\sigma}^{\br{i}},\\
	&P^{\br{i+1}}\br{\randx}=\sum_{\sigma:\sigma_{i+1}= 0}\widehat{P}\br{\sigma}\X_{\sigma}^{\br{i+1}}+\sum_{\sigma:\sigma_{i+1}\neq 0}\widehat{P}\br{\sigma}\X_{\sigma}^{\br{i+1}},
	\end{align*}

	Set
	\begin{align*}
	\mathbf{A}=\sum_{\sigma:\sigma_{i+1}=0}\widehat{P}\br{\sigma}\X_{\sigma}^{\br{i}}; \quad\quad\quad\quad
	&\mathbf{B}=\sum_{\sigma:\sigma_{i+1}\neq 0}\widehat{P}\br{\sigma}\X_{\sigma}^{\br{i}};\\
\mathbf{C}=\sum_{\sigma:\sigma_{i+1}=0}\widehat{P}\br{\sigma}\X_{\sigma}^{\br{i+1}};\quad\quad\quad\quad
	&\mathbf{D}=\sum_{\sigma:\sigma_{i+1}\neq 0}\widehat{P}\br{\sigma}\X_{\sigma}^{\br{i+1}}.
	\end{align*}
	Then we have
	\begin{align*}
	P^{\br{i}}\br{\randx}=\mathbf{A}+\mathbf{B};~P^{\br{i+1}}\br{\randx}=\mathbf{C}+\mathbf{D}.
	\end{align*}
Notice that $\mathbf{A}=\id_m\otimes\mathbf{C}$, where $\id_m$ is placed in the $(i+1)$-th register. Thus,

\begin{equation}\label{eqn:AC}
\Tr~\xi\br{\mathbf{A}}=m\cdot\Tr~\xi\br{\mathbf{C}}.
\end{equation}
From \cref{fact:remainder} and then \cref{eqn:AC},
	\begin{align*} &\ab{m^{i+1-n}\expec{}{\Tr~\xi\br{P^{\br{i+1}}\br{\randx}}}-m^{i-n}\expec{}{\Tr~\xi\br{P^{\br{i}}\br{\randx}}}}\\
 =~&\ab{\expec{}{\genfrac{}{}{0pt}{}{ m^{i+1-n}\br{\Tr~\xi\br{C}+\Tr~D\xi\br{\mathbf{C}}\Br{\mathbf{D}}+\frac{1}{2}\Tr~D^2\xi\br{\mathbf{C}}\Br{\mathbf{D}}+\Delta_{3,\xi}(\randC,\randD)}-}{ m^{i-n}\br{\Tr~\xi\br{A}+\Tr~D\xi\br{\mathbf{A}}\Br{\mathbf{B}}+\frac{1}{2}\Tr~D^2\xi\br{\mathbf{A}}\Br{\mathbf{B}}+\Delta_{3,\xi}(\randA,\randB)}}}}\\
	=~&\ab{\expec{}{\genfrac{}{}{0pt}{}{ m^{i+1-n}\br{\Tr~D\xi\br{\mathbf{C}}\Br{\mathbf{D}}+\frac{1}{2}\Tr~D^2\xi\br{\mathbf{C}}\Br{\mathbf{D}}+\Delta_{3,\xi}(\randC,\randD)}-}{ m^{i-n}\br{\Tr~D\xi\br{\mathbf{A}}\Br{\mathbf{B}}+\frac{1}{2}\Tr~D^2\xi\br{\mathbf{A}}\Br{\mathbf{B}}+\Delta_{3,\xi}(\randA,\randB)}}}}
\end{align*}
Both the first-order and second-order derivatives cancel out because of the following claim.
\begin{claim}\label{claim:1}
	It holds that
	\[\expec{}{\Tr~D\xi\br{\mathbf{A}}\Br{\mathbf{B}}}=m\expec{}{\Tr~D\xi\br{\mathbf{C}}\Br{\mathbf{D}}};\]	\[\expec{}{\Tr~D^2\xi\br{\mathbf{A}}\Br{\mathbf{B}}}=m\expec{}{\Tr~D^2\xi\br{\mathbf{C}}\Br{\mathbf{D}}}.\]		
\end{claim}
\bigskip
By \cref{fact:remainder}, there exists a universal constant $c_3>0$ such that

\begin{align*}
&\ab{\expec{}{m^{i+1-n}\Tr~\xi\br{P^{\br{i+1}}\br{\randx}}-m^{i-n}\Tr~\xi\br{P^{\br{i}}\br{\randx}}}}\\
\leq ~&c_3 B \br{\expec{}{\nnorm{\randB}_3^3}+\expec{}{\nnorm{\randD}_3^3}}\\
\leq ~&c_3 B \br{\expec{}{\nnorm{\randB}_2\nnorm{\randB}_4^2}+\expec{}{\nnorm{\randD}_2\nnorm{\randD}_4^2}}\quad\mbox{(H\"older's)}\\
\leq ~&c_3 B \br{\br{\expec{}{\nnorm{\randB}_2^2}\expec{}{\nnorm{\randB}_4^4}}^{1/2}+\br{\expec{}{\nnorm{\randD}_2^2}\expec{}{\nnorm{\randD}_4^4}}^{1/2}}\quad\mbox{(Cauchy-Schwartz)}\\
\leq ~&c_3 B \theta^d\br{\br{\expec{}{\nnorm{\randB}_2^2}}^{3/2}+\br{\expec{}{\nnorm{\randD}_2^2}}^{3/2}}\quad\mbox{(\cref{lem:hybridHC})},
\end{align*}
where $\theta=\max\set{9m,1/\eta^4}$. Notice that
	\[\expec{}{\nnorm{\randB}_2^2}=\expec{}{\nnorm{\randD}_2^2}=\sum_{\sigma:\sigma_{i+1}\neq 0}\ab{\widehat{P}\br{\sigma}^2}=\mathrm{Inf}_{i+1}\br{P}.\]
	Therefore,	\[\ab{\expec{}{m^{i+1-n}\Tr~\xi\br{P^{\br{i+1}}\br{\randx}}-m^{i-n}\Tr~\xi\br{P^{\br{i}}\br{\randx}}}}\leq 2c_3 B \theta^d\mathrm{Inf}_{i+1}\br{P}^{3/2}.\]
	
	Summing over $i\in[n-h]_{\geq0}$, we have 
	
	\begin{align*}
&\ab{m^{-n}\Tr~\xi\br{P}-m^{-h}\expec{}{\Tr~\xi\br{P^H(\randx)}}}\\
\leq~&2c_3 B \theta^d\sum_{i\notin H}\mathrm{Inf}_{i}\br{P}^{3/2}\\
\leq~&2c_3 B \theta^d\sqrt{\tau}\sum_{i\notin H}\mathrm{Inf}_{i}\br{P}\\
\leq~&2c_3 B \theta^d\sqrt{\tau}d\sum_{\sigma\ne0}\widehat{P}\br{\sigma}^2\\
\leq~&2c_3 B \theta^d\sqrt{\tau}d.
\end{align*}

\end{proof}
It remains to prove \cref{claim:1}.
\begin{proof}[Proof of \cref{claim:1}]
Note that $\mathbf{A},\mathbf{B},\mathbf{C}$ and $\randD$ can be expressed as
	\begin{align*}
	  \mathbf{A} =\id_m\otimes\mathbf{C}; \quad\quad\quad 
	  \mathbf{B} =\sum_{\sigma\in[m^2]_{\geq 0}:\sigma\neq 0}\B_{\sigma}\otimes\mathbf{X}_{\sigma}; \quad\quad\quad
	  \mathbf{D} = \sum_{\sigma\in[m^2]_{\geq 0}:\sigma\neq 0}\randx_{i+1,\sigma}\mathbf{X}_{\sigma}
	\end{align*}
for some random matrices $\mathbf{X}_{\sigma}$'s which are independent of $\randx_{i+1,\sigma}$'s, where $\id_m$ and $\mathbf{B}_{\sigma}$'s are in the $\br{i+1}$-th register.

Suppose that $\randC$ has a spectral decomposition
\begin{equation*}
\randC=\sum_{j=1}^{m'}\randa_j\randPi_j,
\end{equation*}
where $m'$ is the dimension of $\randC$, $\randa_1\geq\dots\geq\randa_{m'}$, $\set{\Pi_j}_{j\in[m']}$ are rank-one projectors satisfying that $\sum_{j=1}^{m'}\Pi_j=\id$ and $\Pi_j\Pi_k=0$ for all $j\ne k$.

By \cref{fac:frechetD}, we have 
\begin{align*}
&\expec{}{\Tr~D\xi\br{\mathbf{A}}\Br{\mathbf{B}}}\\
=~&\sum_{j,k\in[m']}\expec{}{\xi^{[1]}\br{\randa_j,\randa_k}\Tr\br{\br{\id\otimes\randPi_{j}}\randB\br{\id\otimes\randPi_{k}}}}\\
=~&\sum_{j,k\in[m']}\expec{}{\xi^{[1]}\br{\randa_j,\randa_k}\Tr\br{\br{\id\otimes\randPi_{j}\randPi_k}\randB}}\\
=~&\sum_{j\in[m']}\expec{}{\xi'\br{\randa_j}\Tr\br{\br{\id\otimes\randPi_{j}}\randB}}\\
=~&\expec{}{\Tr~\xi'\br{\randA}\randB}\\
=~&\sum_{\sigma\in[m^2]_{\geq 0}:\sigma\neq 0}\expec{}{\Tr~\br{\id_m\otimes\xi'\br{\randC}}\br{\B_{\sigma}\otimes\mathbf{X}_{\sigma}}}\\
=~&\sum_{\sigma\in[m^2]_{\geq 0}:\sigma\neq 0}\expec{}{\Tr~\B_{\sigma}\cdot\Tr~\xi'\br{\randC}\mathbf{X}_{\sigma}}=0,
\end{align*}
where the last equality follows from the orthogonality of $\set{\B_i}_{i=0}^{m^2-1}$.

\begin{align*}
&\expec{}{\Tr~D\xi\br{\mathbf{C}}\Br{\mathbf{D}}}\\
=~&\expec{}{\Tr~\xi'\br{\randC}\randD}\\
=~&\sum_{\sigma\in[m^2]_{\geq 0}:\sigma\neq 0}\expec{}{\randx_{i+1,\sigma}\cdot\Tr~\xi'\br{\randC}\mathbf{X}_{\sigma}}\\
=~&\sum_{\sigma\in[m^2]_{\geq 0}:\sigma\neq 0}\expec{}{\randx_{i+1,\sigma}}\cdot\expec{}{\Tr~\xi'\br{\randC}\mathbf{X}_{\sigma}}\\
=~&0,
\end{align*}
where the last equality follows from the orthogonality of $\randx$.

By \cref{fac:frechetD}, we have 
\begin{align*}
&\expec{}{\Tr~D^2\xi\br{\mathbf{A}}\Br{\mathbf{B}}}\\
=~&\sum_{j,k,\ell\in[m']}\expec{}{\xi^{[2]}\br{\randa_j,\randa_k,\randa_{\ell}}\Tr\br{\br{\id\otimes\randPi_{j}}\randB\br{\id\otimes\randPi_{k}}\randB\br{\id\otimes\randPi_{\ell}}}}\\
=~&\sum_{\sigma,\tau\ne0}\sum_{j,k,\ell\in[m']}\expec{}{\xi^{[2]}\br{\randa_j,\randa_k,\randa_{\ell}}\Tr\br{\B_{\sigma}\B_{\tau}}\cdot\Tr\br{\randPi_{j}\mathbf{X}_{\sigma}\randPi_{k}\mathbf{X}_{\tau}\randPi_{\ell}}}\\
=~&\sum_{\sigma\ne0}\sum_{j,k,\ell\in[m']}\expec{}{\xi^{[2]}\br{\randa_j,\randa_k,\randa_{\ell}}\Tr\br{\randPi_{j}\mathbf{X}_{\sigma}\randPi_{k}\mathbf{X}_{\sigma}\randPi_{\ell}}},\\
\end{align*}
where the last equality follows from the orthogonality of $\set{\B_i}_{i=0}^{m^2-1}$.

\begin{align*}
&\expec{}{\Tr~D^2\xi\br{\mathbf{C}}\Br{\mathbf{D}}}\\
=~&\sum_{j,k,\ell\in[m']}\expec{}{\xi^{[2]}\br{\randa_j,\randa_k,\randa_{\ell}}\Tr\br{\randPi_{j}\randD\randPi_{k}\randD\randPi_{\ell}}}\\
=~&\sum_{\sigma,\tau\ne0}\sum_{j,k,\ell\in[m']}\expec{}{\xi^{[2]}\br{\randa_j,\randa_k,\randa_{\ell}}\randx_{i+1,\sigma}\randx_{i+1,\tau}\cdot\Tr\br{\randPi_{j}\mathbf{X}_{\sigma}\randPi_{k}\mathbf{X}_{\tau}\randPi_{\ell}}}\\
=~&\sum_{\sigma,\tau\ne0}\sum_{j,k,\ell\in[m']}\expec{}{\randx_{i+1,\sigma}\randx_{i+1,\tau}}\expec{}{\xi^{[2]}\br{\randa_j,\randa_k,\randa_{\ell}}\cdot\Tr\br{\randPi_{j}\mathbf{X}_{\sigma}\randPi_{k}\mathbf{X}_{\tau}\randPi_{\ell}}}\\
=~&\sum_{\sigma\ne0}\sum_{j,k,\ell\in[m']}\expec{}{\xi^{[2]}\br{\randa_j,\randa_k,\randa_{\ell}}\Tr\br{\randPi_{j}\mathbf{X}_{\sigma}\randPi_{k}\mathbf{X}_{\sigma}\randPi_{\ell}}},\\
\end{align*}
where the last equality follows from the orthogonality of $\randx$.
\end{proof}

For those functions that are not sufficiently smooth, if they have a mollifier, which is a smooth approximator with a bounded third derivative, then the invariance principle still holds. The following lemma proves an invariance principle for $\zeta\br{\cdot}$ defined in \cref{eqn:zeta}, which has a mollifier $\zeta_{\lambda}\br{\cdot}$ guaranteed by \cref{fac:zetalambda}.

\begin{lemma}\label{lem:zetainvariance}
	Given $0<\tau,\eta<1$, $d,h,m,n\in\posint$, $H\subseteq[n]$ of size $\ab{H}=h$, a $(2,4,\eta)$-hypercontractive  $(m^2,n)$ ensemble $\randx$ and a degree-$d$ $P\in\H_m^{\otimes n}$ satisfying $\infi{P}\leq\tau$ for all $i\notin H$. Suppose that $P$ has a Fourier expansion
	\[P=\sum_{\sigmarange{n}}\widehat{P}\br{\sigma}\B_\sigma.\]
	Let
	\[P^H(\randx)=\sum_{\sigmarange{n}}\widehat{P}\br{\sigma}\randx_{\sigma_{\overline{H}}}\B_{\sigma_H}.\]
	If $\sum_{\sigma\ne0}\widehat{P}\br{\sigma}^2\leq1$, we have
	\[\ab{m^{-n}\Tr~\zeta\br{P}-m^{-h}\expec{}{\Tr~\zeta\br{P^H(\randx)}}}\leq C\br{\max\set{9m,1/\eta^4}^d\sqrt{\tau}d}^{2/3}\]
	for some universal constants $C$.
\end{lemma}

\begin{proof}
Let $\lambda>0$ be determined later, and $\zeta_\lambda$ be defined as in \cref{fac:zetalambda}. By \cref{lem:generalinvariance} and \cref{fac:zetalambda},

\[\ab{m^{-n}\Tr~\zeta_\lambda\br{P}-m^{-h}\expec{}{\Tr~\zeta_\lambda\br{P^H(\randx)}}}\leq 2c_3 B_3 \max\set{9m,1/\eta^4}^d\sqrt{\tau}d/\lambda,\]
where $c_3,B_3$ are universal constants. By \cref{fac:zetalambda} we also have
\[\ab{m^{-n}\Tr~\zeta\br{P}-m^{-n}\Tr~\zeta_\lambda\br{P}}\leq 2\lambda^2\]
and
\[\ab{m^{-h}\expec{}{\Tr~\zeta\br{P^H(\randx)}}-m^{-h}\expec{}{\Tr~\zeta_\lambda\br{P^H(\randx)}}}\leq 2\lambda^2.\]

By the triangle inequality, we have
\[\ab{m^{-n}\Tr~\zeta\br{P}-m^{-h}\expec{}{\Tr~\zeta\br{P^H(\randx)}}}\leq4\lambda^2+2c_3 B_3 \max\set{9m,1/\eta^4}^d\sqrt{\tau}d/\lambda.\]

Choosing $\lambda=\br{2c_3 B_3 \max\set{9m,1/\eta^4}^d\sqrt{\tau}d/8}^{1/3},$ we have
\[\ab{m^{-n}\Tr~\zeta\br{P}-m^{-h}\expec{}{\Tr~\zeta\br{P^H(\randx)}}}\leq3\br{2c_3 B_3 \max\set{9m,1/\eta^4}^d\sqrt{\tau}d}^{2/3}.\]

Let $C=3\br{2c_3B_3}^{2/3}$, we conclude the result.
\end{proof}

\begin{remark}
It is possible to prove an invariance principle for a broader class of functions. For example, we can prove it for Lipschitz continuous functions using the argument in \cite[Lemma 3.5]{isaksson2012maximally}. However, it is out of the focus of this paper. We will leave it for further research.
\end{remark}

\subsection{Derandomized invariance principle}\label{sec:derandomize}

From \cref{lem:generalinvariance}, it is not hard to see that the non-identity basis elements can be substituted by independent Rademacher variables. In this section, we will replace those Rademacher variables with pseudorandom variables to save the randomness.  It is worth noting that there is a large body of research on derandomization through invariance principles (readers may refer to\cite{10.1145/3460532} and the references therein). We adopt the pseudorandom generator (PRG) introduced in \cite{10.1145/1806689.1806749}. The PRG is constructed by pairwise uniform hash functions as follows.

  For $\F=\set{f: [n]\to[p]}$,
  define $G:\F\times\br{\bits{n}}^p\to\bits{n}$ by 
  \begin{equation}\label{eq:derandomize-G-def}
  G\br{f,z^1,\dots,z^p}=x\text{, where }x_{i}=z^{f(i)}_{i}\text{ for }i\in[n].
  \end{equation}

We define the influence of a random variable in a random matrix using the notation $\mathrm{VarInf}\br{\cdot}$ to distinguish from the notation for the influence of a register in \cref{def:inf}.
\begin{definition}
Given $m,n,p\in\posint$, let $P(\randb)=\sum_{S\subseteq[n]}\randb_SP_S$ be a random matrix with $\randb$ drawn uniformly from $\set{\pm 1}^n$, where $P_S\in\H_m$ and $\randb_S=\prod_{i\in S}\randb_i$ for all $S\subseteq[n]$. Then the influence of $i$'th coordinate of $\randb$ is defined to be
\[\varinfi{P(\randb)}=\sum_{S\ni i}\nnorm{P_S}_2^2.\]
We also define the influence of a block of coordinates.
Let $j\in[p]$ and $f:[n]\to[p]$ be a function, define the influence on the block $f^{-1}(j)\subseteq[n]$ to be
\[\inffj{P(\randb)}=\sum_{S:S\cap \invf{j}\ne\emptyset}\nnorm{P_S}_2^2.\]
\end{definition}

The following is the main theorem in this section.
\begin{theorem}[Derandomized invariance principle for $\zeta$]\label{thm:derandomize}
Given $d,h,m,n\in\posint$, $m>1$, and a random matrix
\[P(\randb)=\sum_{S\subseteq[n]}\randb_SP_S,\]
where $\randb\usim\bits{n}$, $\expec{\randb}{\nnorm{P(\randb)}_2^2}\leq1$, $\randb_S=\prod_{i\in S}\randb_i$ and $P_S\in\herspace{h}$, they satisfy  $\ab{S}+\deg\br{P_S}\leq d$ and $\varinfi{P(\randb)}\leq\tau$ for all $i\in[n]$. 

Let $p$ be the smallest power of $2$ satisfying $p\ge d/\tau$; $\F=\set{f:[n]\to[p]}$ be a family of pairwise uniform hash functions. For any $i \in [p]$, define $\randz^i$ to be a $4d$-wise uniform random vector drawn from $\set{\pm 1}^n$, and $\randz^i$ are independent across $i\in[p]$. Given $f\in\F$, denote $\randx_f=G\br{f,\randz^1,\dots,\randz^p}$ as in \cref{eq:derandomize-G-def}. Then we have
\[\ab{\frac{1}{m^h}\expec{\randb}{\Tr~\zeta\br{P(\randb)}}-\frac{1}{m^h}\expec{\randf,\randx_\randf}{\Tr~\zeta\br{\randP(\randx_\randf)}}}\leq C_2\sqrt{(9m)^dd\tau},\]
where $\randf$ is drawn uniformly from $\F$ and $C_2$ is a universal constant.
\end{theorem}

We first prove a derandomized invariance principle for the functions with bounded fourth derivative.
\begin{theorem}[Derandomized invariance principle]\label{thm:smoothderandomize}
Given $d,h,m,n\in\posint$, $m>1$, and a random matrix
\[P(\randb)=\sum_{S\subseteq[n]}\randb_SP_S,\]
where $\randb\usim\bits{n}$, $\expec{\randb}{\nnorm{P(\randb)}_2^2}\leq1$, $\randb_S=\prod_{i\in S}\randb_i$ and $P_S\in\herspace{h}$, they satisfy that  $\ab{S}+\deg\br{P_S}\leq d$ and $\varinfi{P(\randb)}\leq\tau$ for all $i\in[n]$. 

Let $p$ be the smallest power of $2$ satisfying $p\ge d/\tau$; $\F=\set{f:[n]\to[p]}$ be a family of pairwise uniform hash functions. For any $i\in[p]$, define $\randz^i$ to be a $4d$-wise uniform random vector drawn from $\set{\pm 1}^n$, and $\randz^i$ are independent across $i\in[p]$. Given $f\in\F$, denote $\randx_f=G\br{f,\randz^1,\dots,\randz^p}$ as in \cref{eq:derandomize-G-def}. Then for any $\xi\in\C^4$ with $\norm{\xi^{(4)}}_\infty\leq C_0$ where $C_0$ is a constant, it holds that 
\[\ab{\frac{1}{m^h}\expec{\randb}{\Tr~\xi\br{P(\randb)}}-\frac{1}{m^h}\expec{\randf,\randx_\randf}{\Tr~\xi\br{\randP(\randx_\randf)}}}\leq 4C_1C_0(9m)^dd\tau,\]
where $\randf$ is drawn uniformly from $\F$ and $C_1$ is a universal constant.
\end{theorem}

Assuming \cref{thm:smoothderandomize}, \cref{thm:derandomize} is straightforward: 

\begin{proof}[Proof of \cref{thm:derandomize}]
Let $\lambda>0$ be determined later and let $\zeta_\lambda$ be defined as in \cref{fac:zetalambda}. By \cref{thm:smoothderandomize} and \cref{fac:zetalambda},
\[\ab{\frac{1}{m^h}\expec{\randb}{\Tr~\zeta_\lambda\br{P(\randb)}}-\frac{1}{m^h}\expec{\randf,\randx_\randf}{\Tr~\zeta_\lambda\br{\randP(\randx_\randf)}}}\leq 4C_1B_4\lambda^{-2}(9m)^dd\tau,\]
where $C_1,B_4$ are universal constants. By \cref{fac:zetalambda} we also have
\[\ab{\frac{1}{m^h}\expec{\randb}{\Tr~\zeta\br{P(\randb)}}-\frac{1}{m^h}\expec{\randb}{\Tr~\zeta_\lambda\br{P(\randb)}}}\leq 2\lambda^2\]
and
\[\ab{\frac{1}{m^h}\expec{\randf,\randx_\randf}{\Tr~\zeta_\lambda\br{\randP(\randx_\randf)}}-\frac{1}{m^h}\expec{\randf,\randx_\randf}{\Tr~\zeta\br{\randP(\randx_\randf)}}}\leq 2\lambda^2.\]

By the triangle inequality, we have
\[\ab{\frac{1}{m^h}\expec{\randb}{\Tr~\zeta\br{P(\randb)}}-\frac{1}{m^h}\expec{\randf,\randx_\randf}{\Tr~\zeta\br{\randP(\randx_\randf)}}}\leq4\lambda^2+4C_1B_4\lambda^{-2}(9m)^dd\tau.\]

Choosing $\lambda=\br{C_1B_4(9m)^dd\tau}^{1/4}$, we have
\[\ab{\frac{1}{m^h}\expec{\randb}{\Tr~\zeta\br{P(\randb)}}-\frac{1}{m^h}\expec{\randf,\randx_\randf}{\Tr~\zeta\br{\randP(\randx_\randf)}}}\leq8\br{C_1B_4(9m)^dd\tau}^{1/2}.\]

Let $C_2=8\sqrt{C_1B_4}$, we conclude the result.
\end{proof}

\begin{remark}
It is also possible to generalize \cref{thm:derandomize} to Lipschitz continuous functions using the argument in~\cite[Lemma 3.5]{isaksson2012maximally}.
\end{remark}

\begin{lemma}\label{lem:simpletotalinf}
Given $d,n\in\posint$, and a random matrix
\[P(\randb)=\sum_{S\subseteq[n]:\ab{S}\leq d}\randb_SP_S,\]
where $\randb$ is a $2d$-wise uniform random vector from $\set{\pm 1}^n$ and $\expec{\randb}{\nnorm{P(\randb)}_2^2}\leq1$, it holds that  
\[\sum_{i=1}^n\varinfi{P(\randb)}\leq d.\]
\end{lemma}

\begin{proof}
    \begin{align*} \sum_{i=1}^n\varinfi{P(\randb)}=~&\sum_{i=1}^n\sum_{S\ni i}\nnorm{P_S}_2^2\\
    =~&\sum_{S\subseteq[n]:\ab{S}\leq d}\ab{S}\nnorm{P_S}_2^2\\
    \leq~&d\sum_{S\subseteq[n]:\ab{S}\leq d}\nnorm{P_S}_2^2\\
    =~&d\expec{\randb}{\nnorm{P(\randb)}_2^2}\quad\leq d.
    \end{align*} 
\end{proof}

The following lemma is crucial to our proof. The proof follows closely to the proof of \cite[Lemma 5.4]{10.1145/1806689.1806749}.

\begin{lemma}\label{fact:totalinfsqr}
Given $d,n,p\in\posint$, and a random matrix
\[P(\randb)=\sum_{S\subseteq[n]:\ab{S}\leq d}\randb_SP_S,\]
satisfying $\expec{\randb}{\nnorm{P(\randb)}_2^2}\leq1$, 
where $\randb$ is a $2d$-wise uniform random vector drawn from $\set{\pm 1}^n$, let $\F=\set{f:[n]\to[p]}$ be a family of pairwise uniform hash functions. Then for $\randf\usim\F$,
\[\expec{\randf}{\sum_{j=1}^p\infrfj{P(\randb)}^2}\leq\sum_{i=1}^n\varinfi{P(\randb)}^2+\frac{d^2}{p}.\]
\end{lemma}

\begin{proof}
    Fix $j\in[p]$ and for $1\leq i\leq n$, let $\randX_i$ be the indicator variable that is 1 if $f(i)=j$ and 0 otherwise. For brevity, let $\tau_i=\varinfi{P(\randb)}$ for $i\in[n]$. Now,
    \begin{align*}
        \infrfj{P(\randb)}=~&\sum_{S:S\cap\invf{j}\ne\emptyset}\nnorm{P_S}_2^2
\leq\sum_{S}\nnorm{P_S}_2^2\br{\sum_{i\in S}\randX_i}
=\sum_{i\in [n]}\randX_i\sum_{S\ni i}\nnorm{P_S}_2^2
=\sum_{i\in [n]}\randX_i\tau_i
    \end{align*}
    Thus
\[\infrfj{P(\randb)}^2\leq\br{\sum_{i\in [n]}\randX_i\tau_i}^2=\sum_{i\in[n]}\randX_i^2\tau_i^2+\sum_{i\ne k}\randX_i\randX_k\tau_i\tau_k.\]

    Note that $\expec{}{\randX_i}=1/p$ and for $i\ne k$, $\expec{}{\randX_i\randX_k}=1/p^2$. Thus
    \begin{align*}
        \expec{}{\infrfj{P(\randb)}^2}\leq~&\frac{1}{p}\sum_i\tau_i^2+\sum_{i\ne k}\tau_i\tau_k\frac{1}{p^2}\leq\frac{1}{p}\sum_i\tau_i^2+\frac{1}{p^2}\br{\sum_i\tau_i}^2.\\
    \end{align*}
    The lemma follows by using \cref{lem:simpletotalinf} and summing all $j\in[p].$
\end{proof}

We are ready to prove \cref{thm:smoothderandomize}.

\begin{proof}[Proof of \cref{thm:smoothderandomize}]
We prove this by a hybrid argument. 
Denote $\randb^{(0)}=\randb=G\br{f,\randb,\dots,\randb}$.
For $j\in[p]$, define $\randb^{(j)}=G\br{f,\randz^1,\dots,\randz^j,\randb,\dots,\randb}$,
i.e., substituting $\randb^{(j-1)}|_{\invf{j}}$ with $\randz^j_{\invf{j}}$. Then $\randb^{(p)}=\randx_\randf$, and

\begin{align*}
\randP(\randb^{(j-1)})&=\sum_{S:S\cap \invf{j}=\emptyset}\randb^{(j-1)}_SP_S+\sum_{S:S\cap \invf{j}\ne\emptyset}\randb^{(j-1)}_SP_S\\
\randP(\randb^{(j)})&=\sum_{S:S\cap \invf{j}=\emptyset}\randb^{(j)}_SP_S+\sum_{S:S\cap \invf{j}\ne\emptyset}\randb^{(j)}_SP_S.
\end{align*}
Note that for $S\cap \invf{j}=\emptyset$, $\randb^{(j-1)}_S=\randb^{(j)}_S$. Denote 
\begin{align*}
\randA=\sum_{S:S\cap \invf{j}=\emptyset}\randb^{(j)}_SP_S,\quad\quad\quad
\randB=\sum_{S:S\cap \invf{j}\ne\emptyset}\randb^{(j-1)}_SP_S,\quad\quad\quad
\randC=\sum_{S:S\cap \invf{j}\ne\emptyset}\randb^{(j)}_SP_S.
\end{align*}
We have
\begin{align*}
&\ab{\frac{1}{m^h}\expec{\randf,\randb^{(j-1)}}{\Tr~\xi\br{\randP(\randb^{(j-1)})}}-\frac{1}{m^h}\expec{\randf,\randb^{(j)}}{\Tr~\xi\br{\randP(\randb^{(j)})}}}\\
=~&\ab{\frac{1}{m^h}\expec{\randf,\randb^{(j-1)}}{\Tr~\xi\br{\randA+\randB}}-\frac{1}{m^h}\expec{\randf,\randb^{(j)}}{\Tr~\xi\br{\randA+\randC}}}\\
=~&\ab{\frac{1}{m^h}\expec{\randf,\randb^{(j-1)}}{\sum_{k=0}^{3}\frac{1}{k!}\Tr~D^k\xi(\randA)[\randB]+\Tr~\Delta_{4,\xi}(\randA,\randB)}-\frac{1}{m^h}\expec{\randf,\randb^{(j)}}{\sum_{k=0}^{3}\frac{1}{k!}\Tr~D^k\xi(\randA)[\randC]+\Tr~\Delta_{4,\xi}(\randA,\randC)}}
\end{align*}

By \cref{fac:frechetD} and the fact that $\randz_j$ is $4d$-wise uniform, we have for $k=0,1,2,3$, 
\[\expec{\randb^{(j-1)}}{\Tr~D^k\xi(\randA)\Br{\randB}}=\expec{\randb^{(j)}}{\Tr~D^k\xi(\randA)\Br{\randC}}.\]
Thus, 
\begin{align*}
&\ab{\frac{1}{m^h}\expec{\randf,\randb^{(j-1)}}{\Tr~\xi\br{\randP(\randb^{(j-1)})}}-\frac{1}{m^h}\expec{\randf,\randb^{(j)}}{\Tr~\xi\br{\randP(\randb^{(j)})}}}\\
\leq~&\frac{1}{m^h}\expec{\randf,\randb^{(j-1)}}{\ab{\Tr~\Delta_{4,\xi}(\randA,\randB)}}+\frac{1}{m^h}\expec{\randf,\randb^{(j)}}{\ab{\Tr~\Delta_{4,\xi}(\randA,\randC)}}\\
\leq~&C_1C_0\br{\expec{\randf,\randb^{(j-1)}}{\nnorm{\randB}_4^4}+\expec{\randf,\randb^{(j)}}{\nnorm{\randC}_4^4}},
\end{align*}
where the last inequality is from \cref{fact:remainder}, and $C_1$ is a universal constant. Because $\randz_j$ is $4d$-wise uniform, we have $\expec{\randb^{(j-1)}}{\nnorm{\randB}_4^4}=\expec{\randb^{(j)}}{\nnorm{\randC}_4^4}$. Using \cref{lem:hybridHC} with $\eta\gets1/\sqrt{3}$ Recall that $\randb$ is $(2,4,1/\sqrt{3})$-hypercontractive, \[\expec{\randb^{(j-1)}}{\nnorm{\randB}_4^4}\leq(9m)^d\br{\expec{\randb^{(j)}}{\nnorm{\randB}_2^2}}^2.\]
So we have 

\begin{align*}
&\ab{\frac{1}{m^h}\expec{\randf,\randb^{(j-1)}}{\Tr~\xi\br{\randP(\randb^{(j-1)})}}-\frac{1}{m^h}\expec{\randf,\randb^{(j)}}{\Tr~\xi\br{\randP(\randb^{(j)})}}}\\
\leq~&2C_1C_0(9m)^d\expec{\randf}{\br{\expec{\randb^{(j-1)}}{\nnorm{\randB}_2^2}}^2}\\
=~&2C_1C_0(9m)^d\expec{\randf}{\infrfj{P(\randb)}^2}.
\end{align*}
Summing over $j\in[p]$ and by \cref{fact:totalinfsqr}, we have
\begin{align*}
&\ab{\frac{1}{m^h}\expec{\randb}{\Tr~\zeta\br{P(\randb)}}-\frac{1}{m^h}\expec{\randf,\randx_\randf}{\Tr~\zeta\br{\randP(\randx_\randf)}}}\\\leq~&2C_1C_0(9m)^d\br{\sum_{i=1}^n\varinfi{P(\randb)}^2+\frac{d^2}{p}}\\
\leq~&2C_1C_0(9m)^d\br{\tau\sum_{i=1}^n\varinfi{P(\randb)}+\frac{d^2}{p}}\\
\leq~&4C_1C_0(9m)^dd\tau,\\
\end{align*}
where the last inequality is by \cref{lem:simpletotalinf} and $p\ge d/\tau$.
\end{proof}

\section{Positivity tester for low degree operators}\label{sec:positivitytesting}
In this section, we will present an algorithm deciding whether a low-degree operator is $\br{\beta-\delta}$-close to a positive semidefinite operator or $\br{\beta+\delta}$-far from all positive semidefinite operators,
for error parameters $\beta>\delta>0$.
The input operator is given in the form of a Fourier expansion. 
\begin{definition}[Positivity testing problem]\label{def:postest}
Given $d, \dlbo, m\in\posint,$ $m>1$, and real numbers $\beta>\delta> 0$, the input is a degree-$d$ operator in $\herspace{\dlbo}$ given in the form of Fourier expansion
$$P=\sum_{\substack{\sigma\in[m^2]_{\ge 0}^\dlbo\\\sigma: \abs{\sigma}\le d}}\widehat{P}(\sigma)\B_\sigma.$$
Distinguish the following two cases.
\begin{itemize}
    \item Yes: if $m^{-\dlbo}\,\Tr~\zeta(P)<\beta-\delta$.
    \item No: if $m^{-\dlbo}\,\Tr~\zeta(P)>\beta+\delta$.
\end{itemize}
\end{definition}
Notice that the number of Fourier coefficients is $\sum_{i=0}^d{\binom{\dlbo}{i} }\br{m^2-1}^i$. If we are concerned with constant-degree operators, then the dimension of the operator is exponential in the input size.

\begin{theorem}\label{thm:psd-test}
  Given $d, \dlbo, m\in\posint,$ $m>1$, and real numbers $\beta>\delta> 0$,
there exists a deterministic algorithm for the positivity testing problem that runs in time
  $$\exp\br{\poly\br{m^d, 1/\delta}}\cdot \dlbo^{O(d)}.$$
  In particular, if $m,d,\delta$ are constants, then the algorithm runs in time $\poly(\dlbo)$.
\end{theorem}

\subsection{Algorithm}
The algorithm is shown in \cref{fig:ptalg}, which applies the invariance principle \cref{lem:zetainvariance} to reduce the dimension of the matrices and then \cref{thm:derandomize} to derandomize, while the distance to positive operators is approximately preserved.
\begin{figure}
\begin{Algorithm}
\begin{description}
    \item[Input:] Parameters given in \cref{def:postest}.
    \item[Algorithm:] Perform the following steps
    
        \begin{enumerate}
          \item \textbf{Regularization}: 
            Let
            \begin{equation}\label{eq:chooseoftau}
                \tau = \frac{\delta^3}{C^\prime\cdot(9m)^{2\dldd}\cdot\dldd^2},
            \end{equation}
            where $C'=\max\{8C^3,4C_2^2\}$, with $C$ and $C_2$ originating from \cref{lem:zetainvariance} and \cref{thm:derandomize}, respectively. 
            
                For each $i$, compute the influence
                $\infi{P}=\sum_{\sigma: \sigma_i\neq 0}\widehat{P}(\sigma)^2$.
                Let $H=\set{i: \infi{P}>\tau}$.
          \item \textbf{Derandomized invariance principle}:
                Let $p$ be the smallest power of $2$ satisfying $p\ge\dldd/\tau$.
                Let $n=(m^2-1)\br{\dlbo-\abs{H}}$ and $\F=\set{f: [n]\to[p]}$ be a family of pairwise uniform hash functions.
                For any $i\in[p]$, let $\randz^i$ be $4d$-wise uniform random variables of length $n$ and $\br{\randz^i}$'s be independent across $i\in[p]$. 
                For any $f\in\F$, set $\randx_f=G\br{f,\randz^1,\dots,\randz^p}$ as defined in \cref{thm:derandomize}. Define the random operator
                \begin{equation}\label{eq:operator-after-derandomization}
                 P^\prime(f, \randz) = \sum_{\sigma\in[m^2]_{\ge 0}^{\dlbo}:\ab{\sigma}\le d}\widehat{P}(\sigma)\randx_{f,\sigma_{\bar{H}}}\B_{\sigma_H},
                \end{equation}
                where $\randx_{f,\sigma_{\bar{H}}} = \prod_{i\notin H}\br{\randx_f}_{(m^2-1)(i-1)+\sigma_i}$ and $\B_{\sigma_H}=\bigotimes_{i\in H}\B_{\sigma_i}$.
          \item Compute the distance to PSD: For each $f, \randz$, compute
                $$\delta_{f, \randz} = {m^{-\abs{H}}}\,\Tr~\zeta(P^\prime(f, \randz)).$$
          \item Accept if
                $$\expec{f, \randz}{\delta_{f, \randz}}<\beta.$$
        \end{enumerate}
\end{description}
\end{Algorithm}
\caption{Positivity testing algorithm}\label{fig:ptalg}
\end{figure}

\subsection{Time complexity}
\begin{enumerate}
  \item 
  Given that each computation of $\infi{P}$ entails calculating a sum of products of Fourier coefficients, the time required can be expressed as $\sum_{i=0}^d{\binom{\dlbo}{i}}\br{m^2-1}^i \le dm^{2d}\dlbo^d$. In addition, the time needed to determine the set $H$ is at most $\dlbo$.

  \item 
        When fixing $f$ and $\randz$, computing $\delta_{f, \randz}$ takes time
        $$\exp\br{\ab{H}} = \exp\br{\ab{d/\tau}} =\exp\br{\poly\br{m^d, 1/\delta}}.$$
        
  \item By \cref{lem:kwise-hash} and \cref{cor:kwise-z}, the enumeration over $\F$ and $\randz$ takes time polynomial in $\dlbo$, thus computing the expectation of $\delta_{f, \randz}$ also takes time polynomial in $\dlbo$.
\end{enumerate}

\subsection{Correctness}

Now we proceed to the correctness proof.
The first step in our algorithm is to use \cref{lem:zetainvariance} to reduce the dimension by introducing 
Rademacher random variables.
Let $\randb\in\bits{n}$ be uniformly distributed.
Consider the operator $P^{(1)}$ obtained by replacing the basis outside of $H$ with random bits.
That is,
\[ P^{(1)}(\randb) = \sum_{\sigma\in[m^2]_{\ge 0}^{\dlbo}:\ab{\sigma}\le d}\widehat{P}(\sigma)\randb_{\sigma_{\bar{H}}}\B_{\sigma_H},\]
        where $\randb_{\sigma_{\bar{H}}} = \prod_{i\notin H}\randb_{(m^2-1)(i-1)+\sigma_i}$ and $\B_{\sigma_H}=\bigotimes_{i\in H}\B_{\sigma_i}$.
Recall that $\randb$ is $(2,4,1/\sqrt{3})$-hypercontractive,
so in \cref{lem:zetainvariance} we have $1/\eta^4 = 9 \le 9m$. By our choice of $\tau$, the right hand side of the bound in \cref{lem:zetainvariance} can be upper bounded as
\begin{equation}\label{eqn:bound1}
C\br{(9m)^d\sqrt{\tau}d}^{2/3}\le\epreg/2.
\end{equation}
This implies
$$\ab{\frac{1}{m^{\abs{H}}}\expec{\randb}{\Tr~\zeta(P^{(1)}(\randb))}-\frac{1}{m^\dlbo}\Tr~\zeta(P)}\le\delta/2.$$

The second step is derandomization by \cref{thm:derandomize}.
We define $P^{(2)}$ to be the operator obtained by replacing $\randb$ with $\randx_{f, \randz}$,
which is the operator in \cref{eq:operator-after-derandomization}.
Then the right hand side of the bound in \cref{thm:derandomize} can be upper bounded as
\begin{equation}\label{eqn:bound2}
C_2\sqrt{(9m)^\dldd\dldd\tau}\le\epdr/2.
\end{equation}
By \cref{thm:derandomize} this implies
$$\ab{\frac{1}{m^{\abs{H}}}\expec{\randb}{\Tr~\zeta(P^{(2)}(\randx_{f, \randz}))}-\frac{1}{m^{\abs{H}}}\expec{f, \randz}{\Tr~\zeta(P^{(1)}(\randb))}}\le\delta/2.$$
Thus by triangle inequality, we have
\begin{equation}\label{eqn:bound3}
\ab{\frac{1}{m^{\abs{H}}}\expec{f, \randz}{\Tr~\zeta(P^{(2)}(\randx_{f, \randz}))}-\frac{1}{m^\dlbo}\Tr~\zeta(P)}\le\delta.
\end{equation}
The algorithm computes $m^{-\abs{H}}\expec{f, \randz}{\Tr~\zeta(P^{(2)}(\randx_{f, \randz}))}$. By Eq.\eqref{eqn:bound3}, the value is smaller than $\beta$  if $m^{-\dlbo}\,\Tr~\zeta(P)<\beta-\delta$; or greater than $\beta$ if $m^{-\dlbo}\,\Tr~\zeta(P)>\beta+\delta$. Therefore, the algorithm distinguishes the two cases correctly.

\section{Noisy nonlocal games are NP-complete}

\begin{definition}[Noisy Nonlocal Game Value Problem]\label{def:nonlocalgameproblem}
The input consists of the description of a nonlocal game,
which is a tuple $\mathfrak{G}=\br{\calX, \calY, \calA, \calB, \mu, V}$,
and real values $\rho, \beta$ and $\ep$.
$\calX$ and $\calY$ are question sets and assume $\abs{\calX} = \abs{\calY} = s$.
$\calA$ and $\calB$ are answer sets and assume $\abs{\calA} = \abs{\calB} = t$.
Let $\mu$ be a distribution on $\calX\times\calY$ and $V: \calX\times\calY\times\calA\times\calB\to\set{0,1}$ be the predicate.

Let $v=\val^\ast(\mathfrak{G}, \psi_{AB})$ be the value of the nonlocal game,
where Alice and Bob share arbitrarily many copies of a noisy MES $\psi_{AB}$ with the quantum maximal correlation $\rho$.
Let $1>\beta> \ep>0$. The task is to distinguish the following two cases.
\begin{itemize}
    \item Yes: $v>\beta+\ep$.
    \item No: $v<\beta-\ep$.
\end{itemize}
\end{definition}
In this section, we show: 
\begin{theorem}
    The noisy nonlocal game value problem is $\NP$-complete.
\end{theorem}
It follows from the two propositions below.

\begin{prop}\label{thm:nonlocal-game-is-NP}
  There exists a nondeterministic algorithm that runs in time
  $$\poly\br{s, \operatorname{eexp}\br{t, \log\br{\frac{1}{\rho}}, \frac{1}{\ep}}}$$
  that solves the noisy nonlocal game value problem.
  Here $\operatorname{eexp}(\cdot)$ means doubly exponential.
  In particular, if $t, \rho, \ep$ are constants, then the problem is in $\mathrm{NP}$.
\end{prop}
\begin{prop}
    \label{prop:noisy-np-hard}
    For each $3$-SAT instance $\phi$, there is a nonlocal game $G(\phi)$ such that
    its noisy game value is $1$ if $\phi$ is satisfiable, and below some constant $c$ if
    $\phi$ is not satisfiable.
\end{prop}
\Cref{thm:nonlocal-game-is-NP,prop:noisy-np-hard} are proved in \cref{sec:noisy_alg,sec:np-hard} respectively.

\subsection{The nondeterministic algorithm}
\label{sec:noisy_alg}

We first present an upper bound on the number of noisy MES sufficient to approximate the value of a nonlocal game to an arbitrary precision. 
The upper bound from \cite{qin2021nonlocal} is $D=\exp(\poly(s)$, $\exp\br{\poly(t)})$.
The follow-up work~\cite{qin_et_al:LIPIcs.ICALP.2023.97} studied fully quantum games in which both questions and answers are quantum and proved a better upper bound  $D=\exp\br{\poly(s),\poly(t)}$ using a refined Gaussian dimension reduction. We observe that this upper bound can be further improved to $D=\poly\br{s,\exp\br{\poly(t)}}$ for nonlocal games.
\begin{theorem}\label{thm:upperbound}
		Given parameters $0<\epsilon,\rho<1$, $n,m\in\posint$, $m\geq 2$, a noisy MES state $\psi_{AB}$, i.e.,  $\psi_A=\psi_B=\frac{\id_m}{m}$ with the quantum maximal correlation $\rho=\rho\br{\psi_{AB}}<1$
		as defined in \cref{def:maximalcorrelation}, let $\mathfrak{G}$  be a nonlocal game with the question sets $\calX,\calY$ and the answer sets $\calA,\calB$. Suppose the players share arbitrarily many copies of $\psi_{AB}$. Let $\omega_n(\mathfrak{G},\psi_{AB})$ be the highest winning probability that the players can achieve when sharing $n$ copies of $\psi_{AB}$. Then there exists an explicitly computable bound $D=D\br{\abs{\calX},\abs{\calY},\abs{\calA},\abs{\calB},m,\epsilon,\rho}$, such that 
		for any $n>D$, $\omega_n(\mathfrak{G},\psi_{AB})-\omega_D(\mathfrak{G},\psi_{AB})\leq\epsilon.$
		In particular, one may choose
		$$D=\mathrm{poly}\br{\abs{\calX},\abs{\calY}, \exp\br{\mathrm{poly}\br{\abs{\calA},\abs{\calB},\frac{1}{\epsilon},\frac{1}{1-\rho}},\log m}}.$$
	\end{theorem}
 The proof largely follows the framework in \cite{qin2021nonlocal} with several refinements. We include it in \cref{sec:upperbound}. \footnote{One may wonder why the upper bound in \cite{qin_et_al:LIPIcs.ICALP.2023.97} is still exponential in the size of the question set with the refined Gaussian dimension reduction. This is because of the different treatment of the questions. When the questions are classical, we take into account the distribution of the questions. However, if the questions are quantum as considered in \cite{qin_et_al:LIPIcs.ICALP.2023.97}, the question registers are expressed as a linear combination of matrix basis elements, where an extra factor on the size of the question sets is introduced.}.

Next we present the algorithm, shown in \cref{fig:valuealg}, which is deterministic provided with a certificate. 
By \cref{thm:upperbound} we know that sharing $\dlbo$ copies of $\psi_{AB}$ is sufficient to approximate the game value. However, outlining a strategy that shares $\dlbo$ copies of $\psi_{AB}$ requires $\exp\br{\dlbo}$ bits, rendering it excessively costly. Despite this, we've devised a more affordable certificate. Interpreted as a degree-$\dldd$ \pseudostrategy, this certificate is presented through its Fourier coefficients. By \pseudostrategy\ we mean two sets of operators $\set{\Pss}$ and $\set{\Qss}$ that may not be a valid quantum strategy. However, we can still define the winning probability on a 
\pseudostrategy, mathematically.

\begin{definition}\label{def:params-np}
We summarize the parameters we use for the algorithm in the table below.
\begin{itemize}
  \item $\dmcc=300$.
  \item $\eprd=\ep^2/(4t^3)$.
  \item $\delta = \frac{\eprd^2}{\dmcc t(t+1)}$.
  \item $\dldd=\frac{\ccsm\log^2\frac{1}{\delta}}{\delta(1-\rho)}$ as in \cref{lem:smoothing of strategies}.
  \item $\wireal = D\log m + \log\br{\frac{2}{\delta}}$ as in \cref{lem:truncating_low_degrees}.
  \item $\dlbo$ is the polynomial specified in \cref{thm:upperbound} with $\ep\gets\ep/2$.
\end{itemize}
\end{definition}

\begin{figure}
\begin{Algorithm}
\begin{description}
    \item[Input:]
        Parameters in \cref{def:nonlocalgameproblem}.
    \item[Certificate:]
        Let $\set{\br{\calA_i,\calB_i}}_{i=0}^{m^2-1}$ be a pair of standard orthonormal bases satisfying \cref{lem:normofM}.
        A tuple of real numbers of width $s_w$, which are non-zero Fourier coefficients of
        a degree-$\dldd$ \pseudostrategy\ on $\dlbo$ copies of $\psi_{AB}$.
        For each $x\in\calX, a\in\calA$ and $\sigma\in[m^2]_{\ge0}^\dlbo$ satisfying $\abs{\sigma}\le\dldd$,
        the certificate should contain the coefficient $\Pff$.
        Similarly, for $y\in\calY, b\in\calB$ and $\sigma$, the certificate should contain the coefficient $\Qff$.
        Then the degree-$\dldd$ \pseudostrategy\ can be written as $\Pss$ and $\Qss$ satisfying
        $$\Pss = \sum_{\abs{\sigma}\le\dldd}\Pff\calA_\sigma~\mbox{and}~\Qss = \sum_{\abs{\sigma}\le\dldd}\Qff\calB_\sigma.$$
    \item[Algorithm:] Perform the following steps
        \begin{enumerate}
          \item Compute the winning probability on the pseudo-strategy, which is 
                $$\gval{\dlbo}{\Pss}{\Qss} = \sum_{x, y, a, b}\mu(x, y)\cdot V(x, y, a, b)\sum_{\sigma\in[m^2]_{\geq 0}^D}c_{\sigma}\Pff\cdot \Qff,$$
                where $c_{\sigma}=c_{\sigma_1}\cdots c_{\sigma_D}$, and $\set{c_i}_{i=0}^{m^2-1}$ is given in \cref{lem:normofM}.
                Reject if $$\gval{\dlbo}{\Pss}{\Qss} < \beta.$$
          \item Check if the operators sum up to the identity by checking
                \begin{itemize}
                  \item For all $x, y$ and $\sigma\neq0^\dlbo$, it should hold that
                        $$\sum_{a}\Pff = \sum_b\Qff=0.$$
                  \item For all $x, y$, and $\sigma=0^\dlbo$, it should hold that
                        $$\sum_{a}\Pff=\sum_{b}\Qff=1.$$
                \end{itemize}
                Reject if any of the above equalities fails.
          \item For each $x, y, a, b$,
                run the positivity testing algorithm described in \cref{sec:positivitytesting} on $\Pss$ and $\Qss$ with parameters $\beta\gets4\delta$ and $\delta\gets2\delta$.
                Reject if any of the positivity tests fails.
          \item Accept.
        \end{enumerate}
\end{description}

\end{Algorithm}
\caption{Nondeterministic algorithm solving the noisy nonlocal game value problem}\label{fig:valuealg}
\end{figure}

\textbf{Time complexity}.
We upper bound the time complexity of each step.
\begin{enumerate}
  \item Certificate length: The certificate contains the non-zero Fourier coefficients of degree-$\dldd$ operators acting on $\dlbo$ qudits.
        Each degree-$\dldd$ operator consists of
        $$\sum_{d=0}^{\dldd}{\binom{\dlbo}{d} }\cdot (m^2-1)^{d} \le \dldd(m^2-1)^\dldd\dlbo^\dldd$$
coefficients, each $\wireal$ bits. Hence, the length of the certificate is $O(st\dldd m^{2\dldd}\dlbo^\dldd\wireal)$.
  \item To compute the game value, we need to enumerate over all $x, y, a, b, \sigma$ and compute a sum of products.
        This takes time
        $$s^2t^2(m^2-1)^dD^d.$$
  \item Check if the operators sum up to the identity takes linear time in            certificate length as it involves only summation over Fourier coefficients.
  \item Each positivity test takes time as specified in \cref{thm:psd-test}, which is
        $$\exp\br{\poly\br{m^d, 1/\delta}}\cdot D^{O(d)}.$$
\end{enumerate}
By the choices of parameters in \cref{def:params-np}, the overall running time is  upper bounded by
$$\poly\br{s, \operatorname{eexp}\br{t, \frac{1}{1-\rho}, \frac{1}{\ep}}}.$$

\textbf{Completeness}.
Suppose $\omega^*(G, \psi_{AB}) \ge \beta+\ep$.
Then by \cref{thm:upperbound},
there exists a strategy $\br{\Pss, \Qss}$ that uses $\dlbo$ copies of $\psi_{AB}$ with game value $\gval{\dlbo}{\Pss}{\Qss}\ge \beta+\ep/2$.
Let $f$ be the smoothing map in \cref{lem:smoothing of strategies}, and
let $\Pnn{1}=f(\Pss)$ and $\Qnn{1}=f(\Qss)$.
Then $\set{\Pnn{1}}, \set{\Qnn{1}}$ are of degree at most $\dldd$ and satisfy
\begin{enumerate}
    \item  For all $x, y$, we have $\sum_{a}=\Pnn{1}=\sum_{b}\Qnn{1}=\id$ (since $f$ is linear and unital)
	  \item For all $x, y, a, b$, $\nnorm{\Pnn{1}}_2\leq1~\mbox{and}~\nnorm{\Qnn{1}}_2\leq1$.

	  \item For all $x, y, a, b$, $\ab{\Tr\br{\br{\Pnn{1}\otimes \Qnn{1}}\psi^{\otimes n}_{AB}}-\Tr\br{\br{\Pss\otimes \Qss}\psi^{\otimes n}_{AB}}}\leq\delta$.
		\item For all $x, y, a, b$, $m^{-\dlbo}\,\Tr~\zeta\br{\Pnn{1}}\leq\delta$ and $m^{-\dlbo}\,\Tr~\zeta\br{\Qnn{1}}\leq\delta$.
\end{enumerate}
We observe that \cref{lem:smoothing of strategies} also guarantees
the Fourier coefficients of $\Pnn{1}$ and $\Qnn{1}$ have absolute values
bounded by $1$.
This allows us to truncate the strategy.
For each Fourier coefficient we preserve $\wireal$ digits and by \cref{lem:truncating_low_degrees} get $\set{\Pnn{2}}, \set{\Qnn{2}}$ satisfying
\begin{enumerate}
    \item For all $x, y$, $\sum_{a}\Pnn{2}=\sum_{b}\Qnn{2}=\id$.
	  \item For all $x, y, a, b$, $\nnorm{\Pnn{2}}_2\leq1~\mbox{and}~\nnorm{\Qnn{2}}_2\leq1;$
	  \item For all $x, y, a, b$, $\ab{\Tr\br{\br{\Pnn{2}\otimes \Qnn{2}}\psi^{\otimes n}_{AB}}-\Tr\br{\br{\Pnn{1}\otimes \Qnn{1}}\psi^{\otimes n}_{AB}}}\leq\delta,$
		\item For all $x, y, a, b$, $m^{-\dlbo}\,\Tr~\zeta\br{\Pnn{2}}\leq2\delta$ and $m^{-\dlbo}\,\Tr~\zeta\br{\Qnn{2}}\leq2\delta. $
\end{enumerate}
This \pseudostrategy\ is the certificate.
Specifically, by \cref{lem:maincauchy} the game value is
$$\gval{\dlbo}{\Pnn{2}}{\Qnn{2}}\ge\beta+\ep/2-2\delta t^2 = \beta+\ep/2-\frac{\ep^2}{2t\dmcc}\ge\beta,$$
and the first check is passed.
Also, by item 4, the positivity tests can also be passed.

\textbf{Soundness}.
Suppose that there exists a certificate that passes all tests,
then there exists a degree-$d$ \pseudostrategy\ $\set{\Pnn{1}}, \set{\Qnn{1}}$ satisfying
\begin{itemize}
  \item By the game value testing,
      $$\gval{\dlbo}{\Pnn{1}}{\Qnn{1}}\ge\beta.$$
  \item By ``summing up to the identity'' testings, for all $x, y$
      $$\sum_{a}\Pnn{1}=\id,~\mbox{and}~\sum_{b}\Qnn{1}=\id.$$

  \item By the positivity tests,
      for all $x, y, a, b$
      $$\frac{1}{m^\dlbo}\Tr~\zeta\br{\Pnn{1}}\leq 6\delta,~\mbox{and}~\frac{1}{m^\dlbo}\Tr~\zeta\br{\Qnn{1}}\leq 6\delta.$$
\end{itemize}
We then apply \cref{lem:closedelta} to get a strategy
$\set{\Pnn{2}}$ and $\set{\Qnn{2}}$.
It holds that for each $x\in\calX$
\begin{multline*}
  \sum_{a\in\calA}\nnorm{\Pnn{2}-\Pnn{1}}_2^2
  \le3(t+1)\br{\sum_{a\in A}\frac{1}{m^\dlbo}\Tr~\zeta\br{\Pnn{1}}}
    +6\sqrt{t}\br{\sum_{a\in A}\frac{1}{m^\dlbo}\Tr~\zeta\br{\Pnn{1}}}^{1/2} \\
  \le 18t(t+1)\delta + 6\sqrt{6}t\sqrt{\delta}
  \le \frac{18\eprd^2}{\dmcc} + \frac{6\sqrt{6}\eprd}{\sqrt{\dmcc}}\le\frac{18+6\sqrt{6\dmcc}}{\dmcc}\eprd\le\eprd.
\end{multline*}
Similarly, for each $y\in\calY$ we have
$$\sum_{b\in\calB}\nnorm{\Qnn{2}-\Qnn{1}}_2^2\le\eprd.$$
We get a strategy $\set{\Pnn{2}}$ and $\set{\Qnn{2}}$ with game value
\begin{align*}
&\ab{\gval{D}{\Pnn{2}}{\Qnn{2}}-\gval{D}{\Pnn{1}}{\Qnn{1}}}\\
\leq~&\ab{\gval{D}{\Pnn{2}-\Pnn{1}}{\Qnn{2}}}+\ab{\gval{D}{\Pnn{1}}{\Qnn{2}-\Qnn{1}}}\\
\leq~&\sum_{x,y,a,b}\mu(x,y)\br{\nnorm{\Pnn{2}-\Pnn{1}}_2\nnorm{\Qnn{2}}_2+\nnorm{\Pnn{1}}_2\nnorm{\Qnn{2}-\Qnn{1}}_2}\\
\leq~&\br{\sum_b\sum_{x,a}\mu_A(x)\nnorm{\Pnn{2}-\Pnn{1}}_2^2}^{1/2}\br{\sum_a\sum_{y,b}\mu_B(y)\nnorm{\Qnn{2}}_2^2}^{1/2}\\
&+\br{\sum_b\sum_{x,a}\mu_A(x)\nnorm{\Pnn{1}}_2^2}^{1/2}\br{\sum_a\sum_{y,b}\mu_B(y)\nnorm{\Qnn{2}-\Qnn{1}}_2^2}^{1/2}\quad\mbox{(Cauchy-Schwartz)}\\
\le~&2t\sqrt{t\eprd}.
\end{align*}

Thus there exists a strategy with game value
$$\gval{D}{\Pnn{2}}{\Qnn{2}} >\beta - 2t\sqrt{t\eprd}=\beta - \ep.$$

\subsection{NP-hardness}
\label{sec:np-hard}
In this subsection, we first show that if $L \in \MIP[s,t]$ with perfect completeness and constant soundness, then $L \in \MIP^*[s,t,\psi_{AB}]$ also with perfect completeness and constant soundness.

\begin{figure}
 \begin{Algorithm}
    \begin{description}
    	\item[Setup:] Flip two unbiased coins $\bc_1, \bc_2 \sim \set{0,1}$. 
             Sample questions $(\bx,\by) \sim \Alg_Q(\ipt)$.
             With probability $1/2$ each, perform one of the following two tests.
             \item[Verify:] Distribute the questions as follows
             \begin{itemize}
                 \item Player $\bc_1$: give $\bx$; receive $\ba$.
    		\item Player $\overline{\bc_1}$: give $\by$; receive $\bb$
    		\end{itemize}
    		Accept if $V(\ipt, \bx, \by)$ accepts on $\ba, \bb$.
            \item[Consistency:] Distribute the questions as follows:
                if $\bc_2 = 0$
                \begin{itemize}
                 \item Player $\bc_1$: give $\bx$; receive $\ba$,
    		      \item Player $\overline{\bc_1}$: give $\bx$; receive $\bb$,
                \end{itemize}
                otherwise 
                \begin{itemize}
                 \item Player $\bc_1$: give $\by$; receive $\ba$,
    		      \item Player $\overline{\bc_1}$: give $\by$; receive $\bb$,
    		\end{itemize}
    		Accept if $\ba = \bb$.
    \end{description}
    \end{Algorithm}
    \caption{The noisy $\MIP^*$ verifier $V^*$ from an $\MIP$ verifier $V = (\Alg_Q,\Alg_A)$}\label{fig:verifier}
    \end{figure}
\begin{theorem}\label{prop:NPC}
    Let $V= (\Alg_Q, \Alg_A)$ be an $\MIP$ protocol for a language $L$ with perfect completeness.
    Then the verifier $V^*$ described in \cref{fig:verifier} is an $\MIP^*$ verifier for $L$
    with the following conditions:
    \begin{description}
        \item[Completeness.] If $\ipt \in L$, there is a value-$1$ strategy for $V^*$.
        \item[Soundsness.] Given $\ipt$, if there is a strategy for $V^*$ with value $1-\epsilon$ where the provers share
        arbitrarily many copies of a noisy MES, then there is
        a classical strategy for $V$ with value $1 - 2\ep - \frac{32\epsilon}{1-\rho}$.
    \end{description}
\end{theorem}
\begin{proof}[Proof of \cref{prop:noisy-np-hard}]
Notice that $3$-SAT $\in \MIP[\log, 1]$ \cite{Babai1991} with perfect completeness and an arbitrarily small constant soundness. By \Cref{prop:NPC}, there exists an $\MIP^*[\log,1,\psi_{AB}]$ protocol for $3$-SAT with perfect completeness and a constant soundness. 
\end{proof}

\begin{proof}[Proof of \Cref{prop:NPC}]
    \textbf{Completeness.} If $\ipt$ is satisfiable, the value-$1$ strategy for $V$ is also a value-$1$ strategy for $V^*$.

    \textbf{Soundness.} In the consistency test, with probability $1/2$ both provers get Alice's question $x$. Suppose that Alice and Bob share $n$ copies of a noisy $m$-dimensional MES $\psi_{AB}$, and that they apply the measurements $\set{P^x_a}_{a\in\calA}$ and $\set{Q^x_a}_{a\in\calA}$, respectively. Hence the probability for the provers to pass the consistency test of $x$ is at least $1 - 4\epsilon$.
     It means that    
\begin{align*}
    \bigE_{x} \sum_{a \in \calA} \Tr\br{\br{P^x_a \otimes Q^x_a} \psi_{AB}^{\otimes n}} \geq 1 - 4\epsilon.
\end{align*}
    Let $\set{\br{\calA_i,\calB_i}}_{i=0}^{m^2-1}$ be a pair of standard orthonormal bases satisfying \cref{lem:normofM}. Using the Fourier expansions of $P^x_a = \sum_\sigma \widehat{P^x_a}\br{\sigma}\calA_{\sigma}$ and $Q^x_a = \sum_\sigma \widehat{Q^x_a}\br{\sigma}\calB_{\sigma}$, the condition above is equivalent to 
\begin{align*}
\bigE_x \sum_a \sum_\sigma c_{\sigma} \widehat{P^x_a}\br{\sigma}\widehat{Q^x_a}\br{\sigma} \geq 1 - 4\epsilon,
\end{align*}
where $c_{\sigma}=c_{\sigma_1}\cdots c_{\sigma_D}$, and $\set{c_i}_{i=0}^{m^2-1}$ is given in \cref{lem:normofM}. By the Cauchy-Schwartz inequality,
\begin{align*}
\br{\bigE_x \sum_a \sum_\sigma c_{\sigma} \widehat{P^x_a}\br{\sigma}^2}^{1/2}\br{\bigE_x \sum_a \sum_\sigma c_{\sigma} \widehat{Q^x_a}\br{\sigma}^2}^{1/2}\geq1-4\epsilon.
\end{align*}

Notice that $$\sum_a \sum_\sigma c_{\sigma} \widehat{Q^x_a}\br{\sigma}^2\leq\sum_a \sum_\sigma\widehat{Q^x_a}\br{\sigma}^2=\sum_a\nnorm{Q^x_a}^2_2\leq1$$ for all $x$, we have
\[\bigE_x \sum_a \sum_\sigma c_{\sigma} \widehat{P^x_a}\br{\sigma}^2\geq(1-4\epsilon)^2.\]
On the other hand, we have
\begin{align*}
    \bigE_x \sum_a \sum_\sigma c_{\sigma} \widehat{P^x_a}\br{\sigma}^2
    &\leq \bigE_x \sum_a 
    \left[\widehat{P^x_a}\br{0^n}^2 + \rho  \sum_{\sigma \neq 0^n} \widehat{P^x_a}\br{\sigma}^2\right] \\
    &\leq \bigE_x \sum_a
    \left[\widehat{P^x_a}\br{0^n}^2 + \rho(\nnorm{P^x_a}_2^2 - \widehat{P^x_a}\br{0^n}^2) \right] \\
    &\leq \rho + (1-\rho) \bigE_x \sum_a \widehat{P^x_a}\br{0^n}^2.
\end{align*}
Therefore,
\begin{align}
    \bigE_x \sum_a \widehat{P^x_a}\br{0^n}^2 \geq 1 - \frac{8\epsilon -16\epsilon^2}{1-\rho}\geq1-\frac{8\epsilon}{1-\rho}.
\end{align}
Note that for all $x$, $\sum_a \widehat{P^x_a}\br{0^n} =m^{-n}\sum_a\Tr~P^x_a=1$.
For each $x$, let $a_x$ be the answer that maximizes $\widehat{P^x_a}\br{0^n}$.
Then $\sum_a \widehat{P^x_a}\br{0^n}^2 \leq \widehat{P^x_{a_x}}\br{0^n}\sum_a \widehat{P^x_a}\br{0^n} = \widehat{P^x_{a_x}}\br{0^n}$, and
\begin{align*}
  \bigE_x \widehat{P^x_{a_x}}\br{0^n} \geq 1 - \frac{8\epsilon }{1-\rho}.
\end{align*}

Similarly, from the fact that they can pass the consistency test on Bob's questions, we can conclude that the measurements $\set{ Q^y_b }$ satisfy the conditions above.
In particular, let $b_y$ be the answer that maximizes $\widehat{Q^y_b}\br{0^n}$. Then
\begin{align*}
    \bigE_{y} \widehat{Q^y_{b_y}}\br{0^n} \geq 1 - \frac{8\epsilon}{1-\rho}.
\end{align*}

In the deterministic strategy, Alice answers $a_x$ for question $x$ and Bob answers $b_y$ for question $y$.
The difference in the probability of satisfying $V$ between the original strategy and the deterministic strategy is 
\begin{align*}
    &\left| \bigE_{x,y} \sum_{a,b} \Tr\Br{ \br{P^x_a \otimes Q^y_b} \psi_{AB}^{\otimes n}}V(x,y,a,b) -
    \bigE_{x,y} V(x,y,a_x,b_y)\right| \\
    &\leq \bigE_{x,y} \br{ 1 - \Tr\Br{ \br{P^x_{a_x} \otimes Q^y_{b_y}} \psi_{AB}^{\otimes n}}}V(x,y,a_x,b_y)
    + \bigE_{x,y} \sum_{\substack{a \neq a_x \text{ or}\\ b \neq b_y}}
    \Tr\Br{ \br{P^x_a \otimes Q^y_b} \psi_{AB}^{\otimes n}} V(x,y,a,b) \\
    &\leq \bigE_{x,y} \br{ 1 - \Tr\Br{ \br{P^x_{a_x} \otimes Q^y_{b_y}} \psi_{AB}^{\otimes n}}}
    + \bigE_{x,y} \sum_{\substack{a \neq a_x \text{ or}\\ b \neq b_y}}
    \Tr\Br{ \br{P^x_a \otimes Q^y_b} \psi_{AB}^{\otimes n}}
\end{align*}
where we use the fact that $V(x,y,a,b) \leq 1$ for all $x,y,a,b$.
Writing $1 = \sum_{a,b} \Tr\Br{ \br{P^x_{a} \otimes Q^y_{b}} \psi_{AB}^{\otimes n}}$, we get that the expression above equals
\begin{align*}
    &2 \bigE_{x,y}  \sum_{\substack{a \neq a_x \text{ or}\\ b \neq b_y}}
    \Tr\Br{ \br{P^x_a \otimes Q^y_b} \psi_{AB}^{\otimes n}} \\
    &\leq 2 \bigE_{x,y} \sum_{a \neq a_x,b} \Tr\Br{ \br{P^x_a \otimes Q^y_b} \psi_{AB}^{\otimes n}}
    + 2 \bigE_{x,y} \sum_{b \neq b_y}  \Tr\Br{ \br{P^x_{a_x} \otimes Q^y_b} \psi_{AB}^{\otimes n}} \\
    &\leq 2 \bigE_{x,y} \Br{\sum_{a \neq a_x} \widehat{P^x_a}(0^n) + \sum_{b\neq b_y} \widehat{Q^y_b}(0^n)} \\
    & \leq \frac{32\epsilon}{1-\rho}.
\end{align*}
The probability for the original strategy to satisfy $V$ is at least $1 - 2\ep$, 
so the probability for the deterministic strategy to satisfy $V$ is at least $1 - 2\ep- 32\ep/(1-\rho)$. 
\end{proof}

\section{\texorpdfstring{$\MIP^\ast$}{MIP*} protocol for \texorpdfstring{$\RE$}{RE} with \texorpdfstring{$O(1)$}{O(1)}-size answers}
\label{sec:re}

In this section, we prove that there is an $\MIP^\ast$ protocol for any language in $\RE$ with poly-size questions and constant-size answers.
The first step is to tighten the answer reduction techniques from \cite{JNVWY'20,10.1145/3564246.3585208} so that they can reduce the verifier's answer size sequetially from $\poly(n)$ to $\polylog(n)$, and then to $\polyloglog(n)$. 
The second step is to develop a new answer reduction technique that can reduce the answer size of an $\MIP^\ast$ protocol from $\polyloglog(n)$ to $O(1)$ while preserving other parameters of the protocol.

We achieve the second step by modifying the answer reduction technique from \cite{neexp}.
Natarajan and Wright's answer reduction follows a modular design with two major components: Probabilistically checkable proofs of proximity ($\PCPP$) and a tester of the low-degree code.
Hence, to achieve constant answer size, it suffices to change the code to be the Hadamard code, and derive a new tester for the Hadamard code
that allows a verifier to test multiple bits of a codeword at the same time.
Then in our final construction of the $\MIP^\ast[\poly, O(1)]$ protocol for $\RE$, we successively apply the tightened answer reduction technique, followed by the new technique with the Hadamard code, to the $\MIP^\ast[\poly, \poly]$ protocol for $\RE$ \cite{JNVWY'20}. 

Note that \cite{JNVWY'20} doesn't use the answer reduction technique of \cite{neexp}. 
The authors of \cite{JNVWY'20} use a specific $\PCPP$ tailored to the low individual-degree code in their answer reduction technique so that it fits the recursive compression framework.
However, the answer reduction technique of \cite{JNVWY'20} is more difficult to modify due to its less modular design.

\subsection{Tighter answer reduction}

For $\MIP^*$ protocols with short answers, it is useful to separate the part of the verifier that directly acts on the answer bits from the rest. Without loss of generality, we imagine that the decider $\Alg_A$ acts in two phases: first, given its internal randomness and the question pair $x,y$, it computes a Boolean circuit $C^{\ipt}_{x,y} $, and then it applies $C^{\ipt}_{x,y}$ to the answers and returns its output. The \emph{verification time} is the total runtime of this process. We define the \emph{decision complexity} denoted by $d_{V,A}(n)$ to be the maximal size of the circuit $C^{\ipt}_{x,y}$ over all $\ipt$ of size $n$ and question $x,y$ sampled by $\Alg_Q(\ipt)$.
This is always at most as large as the verification time, but it can be much smaller. 

In this section, we will observe that tighter bounds on the answer reduction theorem of~\cite{JNVWY'20,10.1145/3564246.3585208} can be given in terms of the decider complexity.
In particular, the next theorem is an improved version of \cite[Theorem 51]{10.1145/3564246.3585208}.
\begin{theorem}
    Let $V = (\Alg_Q, \Alg_A)$ be an $\MIP^*$ protocol for a language $L$, with question length $\ell_{V,Q}(n)$, 
    answer length $\ell_{V,A}(n)$,
    sampling time $t_{V,Q}(n)$, verification time $t_{V,A}(n)$, and decision complexity $d_{V,A}(n)$. 
    Suppose further that $V$ has the following property: for any $\ipt \in L$, the prover has a 
	real commuting symmetric EPR strategy of value $1$.
    Then there is an answer-reduced verifier $V^{AR} = (\Alg_Q', \Alg_A')$ with the following properties:
    \begin{description}
        \item[Question length.] The new question length is $2\ell_{V,Q}(n) + \poly(\log d_{V,A}(n))$.
        \item[Answer length.] The new answer length is $\polylog( d_{V,A}(n))$.
        \item[Sampling time.] The new sampling time is $t_{V,Q}(n) +\polylog (d_{V,A}(n))$.
        \item[Verification time.] The new verification time is $t_{V,Q}(n) + t_{V,A}(n)+ \poly(d_{V,A}(n))$. 
        \item[Decision complexity.] The new decision complexity is $\polylog(d_{V,A}(n))$.
        \item[Completeness.] If $\ipt \in L$, there is a value-$1$ real commuting symmetric EPR strategy for $V^{AR}$.
        \item[Soundness.] Given $\ipt$, if there is a strategy to $V^{AR}$ with value $1 -\epsilon$, then there is a strategy to $V$ with value $1 - 
    \delta(
    \epsilon, n)$, where $\delta(\epsilon, n) = a((\log d_{V,A}(n))^{a}\epsilon^b + (\log d_{V,A}(n))^{-100b})$, $a$ is a universal constant such that $a>0$, and $b$ is a universal constant such that $0 < b < 1$.
        \item[Efficient computability.] There exists an algorithm that takes the description of $V = (\Alg_Q, \Alg_A)$ as input and outputs the description of $V^{AR} = (\Alg_Q', \Alg_A')$ in time $O( \abs{\Alg_Q}+ \abs{\Alg_A})$,
        where $\abs{\Alg_Q}$ and $\abs{\Alg_A}$ denote the sizes of the descriptions of $\Alg_Q$ and $\Alg_A$ respectively.
        Moreover, $\abs{\Alg_Q'} = \abs{\Alg_Q} +O(1)$ and $\abs{\Alg_A'} = \abs{\Alg_A}+O(1)$.
    \end{description}
\end{theorem}
\begin{proof}
    The proof will follow \cite{10.1145/3564246.3585208} very closely. The main nontrivial thing to prove is the bound on the decision complexity of $V^{AR}$. 
    
    In more detail, the answer-reduced verifier first apply the Cook-Levin theorem to the circuit $C^{\ipt}_{x,y}$ computed in the first phase of $\Alg_A$ to
    get a 5SAT instance with size $s = \poly(d_{V,A}(n))$.
    Following the proof of \cite[Theorem 51]{10.1145/3564246.3585208}, we can choose the parameters used in the PCP part of the answer reduction techniques as
    \begin{align*}
        &m = O(\log s) = O(\log(d_{V,A}(n))),\\
        &m' = 5m + 5 = O(\log(d_{V,A}(n))),\\
        &q = 2^{O(n)} = \poly(d_{V,A}(n)),\\
        &d = O(m) = O(\log(d_{V,A}(n))),
    \end{align*}
    so that the proofs of the provers are $m'+6$ low-degree polynomials on $\bF_q^m$ and $\bF_q^{m'}$ of individual degree at most $d$ in each variable.

    \textbf{Question size.} The question of $V^{AR}$ has at most two questions sampled by $V$ and queries to the low-degree polynomials prepared by the provers. Hence, 
    \begin{align*}
        \ell_{V^{AR},Q}(n) = 2 \ell_{V,Q}(n) + O(m' \log q) = 2 \ell_{V,Q}(n) + \polylog(d_{V,A}(n)).
    \end{align*}

    \textbf{Answer size.} The answer size is at most $O((m'+6)(d+1)\log q)$, which is the number of bits to specify the coefficients in $\bF_q$ of $m'+6$ polynomials of degree at most $d$. Hence,
    \begin{align*}
        \ell_{V^{AR},A}(n) = O(m'd \log q) = \polylog(d_{V,A}(n)).
    \end{align*}

    \textbf{Sampling time.} The verifier will first sample the questions $x,y$ and then sample queries to the low-degree polynomials. Following the proof of \cite[Theorem 10.27]{JNVWY'20}, we have
    \begin{align*}
        t_{V^{AR}, Q}(n) = t_{V,Q}(n) + \poly(m',\log q) = t_{V,Q}(n) + \polylog(d_{V,A}(n)).
    \end{align*}

    \textbf{Verification time.} Following the description of the answer-reduced verifier in \cite[Figure 14]{JNVWY'20},
    we can see that the run time of the answer-reduced verifier is sum of the run time of each step.
    Step 1 and 2 are consistency checks, so the run time is $O(\ell_{V^{AR},A}(n))$.
    Step 3 and 4 are low-degree tests, so the run time is at most $\poly(m,d,m',\log q) = \polylog( d_{V,A}(n))$.
    In Step 5, $\Alg_A'$ needs to run the functions $L^A$ and $L^B$ of the original $\Alg_Q$ first, which takes time $t_{V,Q}(n)$ and then
    the PCP verification.
    The PCP verification takes time $\poly(s) + \poly(s, \log q) + \poly(m',\log q) + \poly(\log q)= \poly(d_{V,A}(n))$ according to the proof of \cite[Theorem 10.25]{JNVWY'20}. Hence, overall,
    \begin{align*}
        t_{V^{AR},A}(n) = t_{V,Q}(n) + t_{V,A}(n) + \poly(d_{V,A}(n)).
    \end{align*}

    \textbf{Decision complexity.} The checks performed by $V^{AR}$ in \cite{10.1145/3564246.3585208} are the same as those in \cite{JNVWY'20},  
    which are specified in \cite[Figure 14]{JNVWY'20}. 
    These checks are simple arithmetics over $\bF_q$ as explained below.
    \begin{enumerate}
        \item Step 1 and 2 of $V^{AR}$ as in \cite[Figure 14]{JNVWY'20} are consistency checks.
        \item Step 3 and 4 of $V^{AR}$ as in \cite[Figure 14]{JNVWY'20} are low-degree test, which is evaluating a univariate polynomial over $\bF_q$ of degree at most $m'd$, whose coefficients are specified by the prover, at a point chosen by the verifier.
        Note that the checks in these 4 steps do not depend on $C^{\ipt}_{x,y}$.
        \item In Step 5, the checks of the PCP verifier are executed:
        \begin{enumerate}
            \item In Step 4 of the PCP verifier as in \cite[Figure 13]{JNVWY'20}, the values of $\mathcal{F}_{arith}(x,o,w), o_1, \dots, o_5$ are precomputed, so the verifier only needs to evaluate an individual-degree-1 polynomial on the prover's answers.
            \item In Step 5 of the PCP veifier as in \cite[Figure 13]{JNVWY'20}, the values of $\mathrm{zero}(z_i)$ are precomputed, so again the verifier only needs to evaluate an individual-degree-1 polynomial on the prover's answers.
        \end{enumerate}
    \end{enumerate}
    Hence, $d_{V^{AR},A}(n) = \poly(\ell_{V^{AR},A}(n)) = \polylog(d_{V,A}(n))$.

    \textbf{Completeness, soundness and efficient computability} follow from the same argument in the proof of \cite[Theorem 51]{10.1145/3564246.3585208}.
    The soundness error is calculated using the values of $q, m, m'$ and $d$ in our setting.
\end{proof}

We will actually use a parallel-repeated version of this answer reduction, which obtains constant soundness. This is closely modeled on \cite[Theorem 52]{10.1145/3564246.3585208}.
\begin{theorem}
\label{thm:newar-pr}
    Let $V = (\Alg_Q, \Alg_A)$ be an $\MIP^*$ protocol for a language $L$, with question length $\ell_{V,Q}$, sampling time $t_{V,Q}$, verification time $t_{V,A}$, and decision complexity $d_{V,A}$. 
    Suppose further that $V$ has the following property: for any $\ipt \in L$, the prover has a 
	real commuting symmetric EPR strategy with a value $1$.
    Then there exists an efficiently computable function $k(n) = \poly(\log d_{V,A}(n))$ and an answer-reduced verifier $V^{AR} = (\Alg_Q', \Alg_A')$
    such that the following hold:
    \begin{description}
        \item[Question length.] The new question length is $k(n) \cdot (2\ell_{V,Q}(n) + \poly(\log d_{V,A}(n)))$.
        \item[Answer length.] The new answer length is $k(n) \cdot \polylog( d_{V,A}(n))$.
        \item[Sampling time.] The new sampling time is $k(n) \cdot (t_{V,Q}(n) +\polylog (d_{V,A}(n)))$.
        \item[Verification time.] The new verification time is $k(n) \cdot (t_{V,Q}(n) + t_{V,A}(n)+ \poly(d_{V,A}(n)))$. 
        \item[Decision complexity.] The new decision complexity is $k(n) \cdot \polylog(d_{V,A}(n))$.
        \item[Completeness.] If $\ipt \in L$, there is a value-$1$ strategy for $V^{AR}$.
        \item[Soundness.] Given $\ipt$, if the value of $\,V$ is at most $1/2$, then the value of $\,V^{AR}$ on $\ipt$ is also at most $1/2$.
        \item[Efficient computability.] There exists an algorithm that takes the description of $V = (\Alg_Q, \Alg_A)$ as input and outputs the description of $V^{AR} = (\Alg_Q', \Alg_A')$ in time $O( \abs{\Alg_Q}+ \abs{\Alg_A})$.
        Moreover, $\abs{\Alg_Q'} = \abs{\Alg_Q} +O(1)$ and $\abs{\Alg_A'} = \abs{\Alg_A}+O(1)$.
    \end{description}
\end{theorem}
The choice of $k(n)$ follows the same reasoning in the proof of \cite[Theorem 52]{10.1145/3564246.3585208}. 
The only difference is that here $s$, the size of the SAT formular, is $\poly(d_{V,A}(n))$, instead of $\poly(t_{V,A}(n), \abs{\Alg_A})$, so $k(n) = \polylog(s) = \polylog(d_{V,A}(n))$.

We note that, unlike in the answer reduction theorems of \cite{JNVWY'20}, we do not keep track of the \emph{level} of the sampler in our answer reduction theorems. To recall, the level is a measure of the complexity of the sampler distribution when expressed in terms of a ``conditional linear'' function of a uniform random seed. The level is important in the context of \cite{JNVWY'20} because it affects the complexity of the question reduction procedure, in which the sampling of questions is delegated to the provers. Thus, in that paper, it was important to keep track of the level to make sure that it remains bounded by a universal constant, so that question reduction can be recursively applied. In our setting, we will never apply question reduction directly, so we do not need to track the level. It can be checked that the answer reduction theorems as stated here hold for all levels, and all of the asymptotic bounds are independent of the number of levels.

\begin{remark}\label{rem:qr-decision-complexity}
The important conclusion from \Cref{thm:newar-pr} is that answer reduction shrinks the decision complexity of a protocol. Looking ahead, this will help us when we repeatedly apply answer reduction. One might ask how the decision complexity behaves under \emph{question reduction}, which is the other component of the compression procedure in~\cite{JNVWY'20}. It turns out that question reduction will in general always ``reset'' the decision complexity to be the same as the decider runtime prior to question reduction. This is because question reduction delegates sampling the questions to the provers, so the ``new'' decider, after question reduction has been applied, must wait for the ``new'' answers in order to simulate the computation of the old decider. Thus, even if the old decider could do most of its work as precomputation before the answers were received, the new decider may not be able to do any precomputation, so the decision complexity can only be bounded by the runtime of the old decider.
\end{remark}

\subsection{Subset tester for the Hadamard code}
\label{sec:subset_tester}
To use the \cite{neexp} answer reduction procedure with a particular error-correcting code, one must show that this code satisfies certain efficient testability properties. Here we show this for the Hadamard code. Specifically, we show that the Hadamard code has a \emph{subset tester} in the sense of \cite[Section 16]{neexp}, shown in \cref{fig:hadamardcode}, which ensures that the provers have a global Hadamard encoding of some bitstring.

First, we recall the definition and key properties of the Hadamard code.
\begin{definition}
    The Hadamard code encodes $x \in \bF_2^k$ as $\Enc_k(x) = (x\cdot y)_{y \in \bF_2^k}$. Moreover,
    \begin{itemize}
        \item For $x \neq y \in \bF_2^k$, $\Enc_k(x)$ and $\Enc_k(y)$ have normalized Hamming agreement at most $\eta_H = \frac{1}{2}$.
        \item There exists an embedding $\mu_k: [k] \to [2^k]$ such that for each $i \in [k]$, $\mu_k(i) = 2^{i-1}$ and $x_i = (\Enc(x))_{\mu_k(i)}$.
        \item There exists a decoding algorithm $\Dec_k$ such that $\Dec_k(\Enc_k(x)) = x$ and, for every $w$ not in the range of $\Enc_k$, $\Dec_k(w) = \perp$.
    \end{itemize}
\end{definition}
The decoding algorithm $\Dec_k$ on input $w$, first computes $x = (w_{\mu_k(k)},\ldots,w_{\mu_k(1)})$
outputs $x$ if $w = \Enc_k(x)$ and $\perp$ otherwise.
Hence the running time of the decoding algorithm is $t_{\Dec}(k) = O(2^k).$
Note that both $\Enc_k$ and $\Dec_k$ run in time exponential in $k$.

\begin{figure}    
\begin{Algorithm}
    Let $k \leq n$ and $D$ be a distribution on the subsets of $\bF_2^n$ with size $k$.
    Flip an unbiased coin $\bb \sim \set{0,1}$.
    Sample $\pmbF = \set{x_1, \ldots, x_k} \sim D$ and a uniformly random $\by \in \bF_2^n$,
    Perform one of the following three subtests with equal probability.
    \begin{description}
	\item[Subtest 1:] Perform one of the following checks with equal probability.
    \begin{description}
        \item[Check 1:] Distribute the question as follows:
        \begin{itemize}
            \item Player $\bb$: give $\pmbF$ and $\by$; receive $(\ba_1, \ldots, \ba_k, \bc, \ba_1', \ldots, \ba_k') \in \bF_2^{2k+1}$.
            \item Player $\overline{\bb}$: give $\pmbF$, receive $(\bd_1,\ldots, \bd_k) \in \bF_2^k$.
        \end{itemize}
        Accept if $\ba_i + \bc = \ba_i'$ and $\ba_i = \bd_i$ for all $i$.
        \item [Check 2:] Distribute the question as follows:
        \begin{itemize}
            \item Player $\bb$: give $\pmbF$ and $\by$; receive $(\ba_1, \ldots, \ba_k, \bc, \ba_1', \ldots, \ba_k') \in \bF_2^{2k+1}$.
            \item Player $\overline{\bb}$: give $\by$, receive $\be \in \bF_2$.
        \end{itemize}
        Accept if $\ba_i + \bc = \ba_i'$ for all $i$, and $\be = \bc$.
        \item[Check 3:]
        Distribute the question as follows:
        \begin{itemize}
            \item Player $\bb$: give $\pmbF$ and $\by$; receive $(\ba_1, \ldots, \ba_k, \bc, \ba_1', \ldots, \ba_k') \in \bF_2^{2k+1}$.
            \item Player $\overline{\bb}$: give $\pmbF+\by = \set{\bx_1+\by, \ldots, \bx_k+\by}$, receive $(\bd_1,\ldots, \bd_k) \in \bF_2^k$.
        \end{itemize}
        Accept if $\ba_i + \bc = \ba_i'$ and $\ba'_i = \bd_i$ for all $i$.
    \end{description}
    \item[Subtest 2:]
    Distribute the question as follows:
        \begin{itemize}
            \item Player $\bb$: give $\pmbF+\by = \set{\bx_1+\by, \ldots, \bx_k+\by}$; receive $(\ba_1, \ldots, \ba_k)$.
            \item Player $\overline{\bb}$: give $\bx_i + \by$ for a random $i$, receive $\bd$.
        \end{itemize}
        Accept if $\ba_i = \bd$.
	\item[Subtest 3:] Perform one of the following checks with equal probability
    \begin{description}
        \item[Check 1:] Distribute the question as follows:
        \begin{itemize}
            \item Player $\bb$: give $\pmbF$; receive $(\ba_1, \ldots, \ba_k)$.
            \item Player $\overline{\bb}$: give $\pmbF$; receive $(\bd_1, \ldots, \bd_k)$.
        \end{itemize}
        Accept if $\ba_i = \bd_i$ for all $i$.
        \item[Check 2:] Distribute the question as follows:
        \begin{itemize}
            \item Player $\bb$: give $\bx_i+\by$ for a random $i$; receive $\ba$.
            \item Player $\overline{\bb}$: give $\bx_i+\by$ for a random $i$; receive $\bd$.
        \end{itemize}
        Accept if $\ba = \bd$.
    \end{description}
\end{description}
\end{Algorithm}
\caption{Subset tester for the Hadamard code}
    \label{fig:hadamardcode}
\end{figure}

\begin{prop}
	\label{prop:subset_test}
    For the subset $F = \set{x_1,\ldots,x_k} \subseteq \bF_2^n$ sampled according to a distribution $D$ and a uniformly random $y \in \bF_2^n$, 
	if a quantum strategy with $\ket{\psi} \in \calH_A \x \calH_B$ and measurements
	\begin{align*}
	\set{ M^{F,y}_{a, c, a'} \mid a,a' \in \bF_2^k, c \in \bF_2}, \set{N^F_b \mid b \in \bF_2^k}, \set{ N^y_d \mid d \in \bF_2 }
	\end{align*}
	can pass the subset tester with probability $1 - \ep$, then there is a Hilbert space $\calH'_A \x \calH'_B$,
	a state $\ket{aux} = \ket{aux_A} \x \ket{aux_B} \in \calH'_A \x \calH'_B$ and a projective measurement $\set{ \hat{G}_u \mid u \in \bF_2^n }$
	on $\calH_B \x \calH_B'$
	such that if we write $\ket{\psi'} = \ket{\psi} \x \ket{aux}$
	\begin{align*}
		\bigE_{F \sim D} \sum_{a \in \bF_2^k} \norm{ N^F_a \x \id_{\calH'} \x \id_B \ket{\psi'} - \id_A \x \sum_{\substack{u: u\cdot x_i = a_i\\\forall i \in [k]}} \hat{G}_u \ket{\psi'}}^2 \leq  (2k-1)^2 (45+ 12\sqrt{k}) \sqrt{\ep}. 
	\end{align*}
\end{prop}
\begin{proof}
Let $F+y = (x_1+y, \ldots, x_k+y)$.
Let 
\begin{align*}
\Omega = \set{( a, c ,a') \mid a_i + c = a_i' \text{ for all } i \in [k]}.
\end{align*}
The set $\Omega$ is the set of valid answer tuples for Alice in \textbf{Subtest 1}; we also use $\Omega$ to denote the \emph{event} that Alice's answers are valid.
Winning the subset tester with probability $1 - \ep$ implies that 
winning each subtest with a probability of at least $1 - 3\ep$.
Furthermore,
winning \textbf{Subtest 1} with a probability of at least $1 - 3\ep$ implies that 
when Alice gets question $(F,y)$ and Bob gets Player $1$'s questions:
\begin{align*}
	&\bigE_{F \sim D} \bigE_{y \sim D_{\Unif}} \Pr[a_1 = b_1 \land \ldots \land a_k = b_k \land \Omega \mid q_A = (F,y), q_B = F] \geq 1 - 18\ep \\
	&\bigE_{F \sim D} \bigE_{y \sim D_{\Unif}} \Pr[ c = d \land \Omega \mid q_A = (F,y) ,q_B = y] \geq 1 - 18 \ep \\
	&\bigE_{F \sim D} \bigE_{y \sim D_{\Unif}} \Pr[ a'_1 = b_1 \land \ldots \land a_k' = b_k\land \Omega \mid q_A = (F,y) ,q_B = F+y] \geq 1 - 18 \ep \\
	&\bigE_{F \sim D} \bigE_{y \sim D_{\Unif}} \Pr[ \Omega \mid q_A = (F,y)] \geq 1 - 6\ep,
\end{align*}
for all $i \in [k]$;
winning \textbf{Subtest 2} with a probability of at least $1 - 3\ep$ implies that 
when Alice gets Player 0's question and Bob gets Player 1's question
\begin{align*}
	&\bigE_{F \sim D} \bigE_{y \sim D_{\Unif}} \Pr[ a_i = d \mid q_A = F+y, q_B = x_i + y] \geq 1 - 6k \ep;
\end{align*}
and winning \textbf{Subtest 3} with a probability of at least $1 - 3\ep$ implies that 
when Alice gets Player 0's question and Bob gets Player 1's question
\begin{align*}
	&\bigE_{F \sim D} \Pr[a_1 = b_1 \land \ldots \land a_k = b_k \mid q_A = q_B =F] \geq 1 - 12 \ep \\
	&\bigE_{F \sim D} \bigE_{y \sim D_{\Unif}} \Pr[ a= b \mid q_A = q_B = x_i + y] \geq 1 - 12k \ep \quad \text{ for all } i.
\end{align*}
In terms of the measurements and the state $\ket{\psi}$, these conditions are equivalent to 
\begin{align*}
	&\bigE_{F \sim D} \bigE_{y \in D_{\Unif}} \sum_{\substack{a,c,a': \\ a_i + c = a_i' \forall i}} \bra{\psi} M^{F,y}_{a, c, a'} \x N^F_a \ket{\psi} \geq 1 - 18 \ep \\
	&\bigE_{F \sim D} \bigE_{y \in D_{\Unif}}  \sum_{\substack{a,c,a': \\ a_i + c = a_i' \forall i}} \bra{\psi} M^{F,y}_{a, c, a'} \x N^y_c \ket{\psi }\geq 1 - 18 \ep \\
	&\bigE_{F \sim D} \bigE_{y \in D_{\Unif}} \sum_{\substack{a,c,a': \\ a_i + c = a_i' \forall i}}\bra{\psi} M^{F,y}_{a, c, a'} \x N^{F+y}_{a'} \ket{\psi }  \geq 1 - 18 \ep \\
	&\bigE_{F \sim D} \bigE_{y \in D_{\Unif}} \sum_{\substack{a,c,a': \\ a_i + c = a_i' \forall i}} \bra{\psi} M^{F,y}_{a, c, a'} \x \id_B \ket{\psi } \geq 1 - 6\ep \\
	&\bigE_{F \sim D} \bigE_{y \in D_{\Unif}} \sum_{a \in \bF_2^k} \bra{\psi} N^{F+y}_a \x N^{x_i + y}_{a_i} \ket{\psi} \geq 1 - 6k \ep \quad\text{ for all } i\\
	&\bigE_{F \sim D} \sum_{a \in \bF_2^k} \bra{\psi} N^F_a \x N^F_a \ket{\psi} \geq 1 - 12\ep \\
	&\bigE_{F \sim D} \bigE_{y \in D_{\Unif}} \sum_{a \in \bF_2} \bra{\psi} N^{x_i+y}_a \x N^{x_i+y}_a \ket{\psi} \geq 1 - 12k\ep \quad\text{ for all } i.
\end{align*}
We define binary observables 
\begin{align*}
	&M^{x_i|F,y} = \sum_{a, c, a'} (-1)^{a_i} M^{F,y}_{a,c,a'} &&
	M^{y|F,y} = \sum_{a, c, a'} (-1)^{c} M^{F,y}_{a,c,a'} &&
	M^{x_i + y | F, y} = \sum_{a,c,a'} (-1)^{a_i'} M^{F,y}_{a,c,a'} \\
	&N^{x_i | F } = \sum_{b} (-1)^{b_i} N^F_b && 
	N^y = N^y_0 - N^y_1 &&
	N^{x_i+y | F+y} = \sum_{b} (-1)^{b_i} N^{F+y}_b.
\end{align*}
We can prove
\begin{align*}
	&\quad \quad \bigE_{F \sim D} \bigE_{y \in D_{\Unif}} \bra{\psi} M^{x_i| F, y} \x N^{x_i | F} \ket{\psi} \\
	& =\bigE_{x \sim D} \bigE_{y \in D_{\Unif}} \Big[ \Pr[a_i = b_i \land \Omega \mid q_A = (F,y), q_B = F] \\
	&\quad\quad- (\Pr[ a_i \neq b_i \mid q_A = (F,y),  q_B = F] - \Pr[ a_i = b_i \land \overline{\Omega} \mid q_A= (F,y), q_B= F]) \Big]\\
	& \geq \bigE_{F \sim D} \bigE_{y \in D_{\Unif}} \Big[\Pr[a_i = b_i \land \Omega \mid q_A = (F,y), q_B = F] - ( 1 - \Pr[a_i = b_i \land \Omega \mid q_A =(F,y), q_B = F]) \Big]\\
	& =\bigE_{F \sim D} \bigE_{y \in D_{\Unif}} \Big[ 2 \Pr[a_i = b_i \land \Omega \mid q_A = (F,y), q_B = F] - 1\Big] \\
	& \geq \bigE_{F \sim D} \bigE_{y \in D_{\Unif}} \Big[2 \Pr[a_1 =  b_1 \land \ldots \land a_k = b_k \land \Omega \mid q_A = (F,y), q_B = F] - 1\Big]\\
 &\geq 1 - 36\ep,
\end{align*}
which implies that $\bigE_{F \sim D}  \bigE_{y \in D_{\Unif}} \norm{ M^{x_i| F, y}\x\id_B \ket{\psi} - \id_A \x N^{x_i | F} \ket{\psi}}^2 \leq 72 \ep$ by expanding the vector norm.
Similarly, from the two other checks of \textbf{Subtest 1},
\begin{align*}
	&\bigE_{F \sim D}  \bigE_{y \in D_{\Unif}} \norm{ M^{y | F, y} \x \id_B\ket{\psi} - \id_A \x N^{y } \ket{\psi}}^2 \leq 72 \ep \\
	&\bigE_{F \sim D}  \bigE_{y \in D_{\Unif}} \norm{ M^{x_i+ y | F, y} \x \id_B \ket{\psi} - \id_A \x N^{x_i + y | F + y } \ket{\psi}}^2 \leq 72 \ep.
\end{align*}
Applying a similar argument to the probability of the event $\Omega$, we can also show 
\begin{align*}
	&\bigE_{F \sim D}  \bigE_{y \in D_{\Unif}} \bra{\psi} M^{x_i | F, y} M^{y | F, y} M^{x_i+y| F, y} 
	\x \id_B \ket{\psi} \\
	&= \bigE_{F \sim D}  \bigE_{y \in D_{\Unif}} \sum_{a, c, a'} (-1)^{a_i + c + a_i'} \bra{\psi} M^{F,y}_{a,c,a'} \x \id_B \ket{\psi} \\
	&= \bigE_{F \sim D} \bigE_{y \in D_{\Unif}} 2\Pr[ a_i + c = a_i' \mid q_A = (F,y)] -1 \\
	&\geq \bigE_{F \sim D} \bigE_{y \in D_{\Unif}}  2 \Pr[ \Omega \mid q_A = (F, y)] - 1 \geq 1 - 12\ep.
\end{align*}

Next, we would like to replace $M^{x_i | F, y}$ by $N^{x_i | F}$, $M^{y | F,y}$ by $N^y$ and $M^{x_i +y | F,y}$ by $N^{x_i+y|F+y}$
and show 
\begin{align}
	\label{eq:N_xy}
	\abs{\bigE_{F \sim D}  \bigE_{y \in D_{\Unif}} \bra{\psi} \id_A \x N^{x_i+y| F+y} N^y N^{x_i  | F } \ket{\psi}-1} \leq 38\sqrt{\ep}.
\end{align}
In the first step
\begin{align*}
	&\quad \abs{ \bigE_{F \sim D}  \bigE_{y \in D_{\Unif}} \bra{\psi} M^{x_i | F, y} M^{y | F, y} ( M^{x_i+y| F, y}\x\id_B -\id_A \x N^{x_i+y | F + y}) \ket{\psi} }\\
	&\leq \bigE_{F \sim D}  \bigE_{y \in D_{\Unif}} \norm{ M^{y | F, y} M^{x_i | F, y} \x \id_B \ket{\psi}}
	\cdot \norm{ (M^{x_i+y| F, y} \x\id_B- \id_A \x N^{x_i+y | F + y}) \ket{\psi} } \quad\mbox{(Cauchy-Schwarz)}\\
	&= \bigE_{F \sim D}  \bigE_{y \in D_{\Unif}} \norm{ (M^{x_i+y| F, y}\x\id_B - \id_A \x N^{x_i+y | F + y}) \ket{\psi} } \\
	& \leq \sqrt{  \bigE_{F \sim D}  \bigE_{y \in D_{\Unif}} \norm{ (M^{x_i+y| F, y} \x \id_B- \id_A \x N^{x_i+y | F + y}) \ket{\psi} }^2 } \quad\mbox{(Jensen)}\\
	& \leq 6\sqrt{2\ep}.
\end{align*}
Similarly,
\begin{align*}
	&\abs{ \bigE_{F \sim D}  \bigE_{y \in D_{\Unif}} \bra{\psi} M^{x_i | F, y} \x N^{x_i+ y | F, y} \cdot ( M^{y | F, y} \x \id_B - \id_A \x N^y) \ket{\psi}}  \leq 6\sqrt{2\ep} \\
	&\abs{ \bigE_{F \sim D}  \bigE_{y \in D_{\Unif}} \bra{\psi} \id_A \x N^{x_i+y | F + y} N^y \cdot  (M^{x_i | F,y} \x \id_B - \id_A \x N^{x_i | F}) \ket{\psi}} \leq 6\sqrt{2\ep} .
\end{align*}
Hence 
\begin{align*}
	\abs{  \bigE_{F \sim D}  \bigE_{y \in D_{\Unif}} \bra{\psi}  \id_A \x N^{x_i+y | F + y} N^y N^{x_i | F} \ket{\psi} - 1} \leq 18\sqrt{2\ep} + 12 \ep \leq 38\sqrt{\ep}.
\end{align*}

On the other hand, from \textbf{Subtest 2}, we have that for all $i \in [k]$
\begin{align*}
	&\quad \bigE_{F \sim D}  \bigE_{y \in D_{\Unif}} \bra{\psi} N^{x_i+y | F + y} \x N^{x_i+y} \ket{\psi} \\
	&= 2 \bigE_{F \sim D}  \bigE_{y \in D_{\Unif}} \Pr[a_i = b \mid q_A=F+y, q_B =x_i + y] - 1
	\geq 1 - 12k\ep,
\end{align*}
which implies that 
\begin{align*}
	\bigE_{F \sim D}  \bigE_{y \in D_{\Unif}} \norm{ (N^{x_i + y | F + y} \x \id_B- \id_A \x N^{x_i + y}) \ket{\psi} }^2 \leq 24k\ep.
\end{align*}
From \textbf{Subtest 3}, with similar reasoning we know
\begin{align*}
	&\bigE_{F \sim D} \norm{ (N^{x_i | F} \x \id_B - \id_A \x N^{x_i |F})\ket{\psi}}^2 \leq 48\ep \\
	&\bigE_{F \sim D}  \bigE_{y \in D_{\Unif}} \norm{ (N^{x_i + y} \x \id_B- \id_A \x N^{x_i + y}) \ket{\psi} }^2 \leq 48k\ep \quad \text{ for all } i.
\end{align*}
Then 
\begin{align*}
	&\bigE_{F \sim D}  \bigE_{y \in D_{\Unif}}  \bra{\psi} \id_A \x N^{x_i+y | F + y} N^y N^{x_i | F} \ket{\psi} \\
	&\appd{\sqrt{24 k\ep} } \bigE_{F \sim D}  \bigE_{y \in D_{\Unif}}  \bra{\psi} N^{x_i + y} \x N^y N^{x_i | F} \ket{\psi} \\
	&\appd{\sqrt{48\ep}} \bigE_{F \sim D}  \bigE_{y \in D_{\Unif}}  \bra{\psi} N^{x_i + y}N^{x_i | F}  \x N^y  \ket{\psi} \\
	&\appd{\sqrt{48k\ep}} \bigE_{F \sim D} \bigE_{y \in D_{\Unif}}  \bra{\psi}N^{x_i | F}  \x  N^{x_i + y} N^y  \ket{\psi}
\end{align*}
Hence \cref{eq:N_xy} implies that 
\begin{align}
	\label{eq:N_xy_noxi+y}
	\abs{ \bigE_{F \sim D}  \bigE_{y \in D_{\Unif}} \bra{\psi} N^{x_i| F} \x N^{x_i +y }N^y  \ket{\psi} -1 } \leq (45 + 12\sqrt{k})\sqrt{\ep}.
\end{align}

Let $C_1 := 45 + 12\sqrt{k}$.
 Let $\tN_u = \frac{1}{2^{n}}\sum_{y \in \bF_2^{n}} (-1)^{u\cdot y} N^{y}$ and $G_u = (\tN_u)^2$. Since each $N^{y}$ is a binary observable, $\set{G_u}$ is a POVM.
	 It can be checked that $N^{y} = \sum_{u\in \bF_2^n} (-1)^{u \cdot y} \tN_u$.
	 Averaging over $F \sim D$, the consistency between $\set{ N^{x_i | F}_0, N^{x_i | F}_1}$ and $\set{ \sum_{u:u\cdot x_i = 0} G_u, \sum_{u:u\cdot x_i = 1} G_u}$,
     where $N^{x_i | F}_c = \sum_{b: b_i = c} N^F_b$ for $c = 0,1$,
	 is 
	 \begin{align*}
	 	&\bigE_{F \sim D} \frac{1}{2}(1 + \bra{\psi} \sum_u (-1)^{u \cdot x_i}N^{x_i | F} \x G_u \ket{\psi}) \\
		&\quad =\frac{1}{2} + \frac{1}{2} \bra{\psi} \bigE_{F \sim D} \bigE_{y,z \in D_{\Unif}} \sum_u (-1)^{u \cdot (x_i+y+z)} N^{x_i | F} \x N^{y} N^{z} \ket{\psi} \\
		&\quad =  \frac{1}{2} + \frac{1}{2} \bra{\psi} \bigE_{F \sim D} \bigE_{z\in D_{\Unif}}N^{x_i | F }\x N^{x_i + z} N^{z}  \ket{\psi} 
		\approx_{\frac{C_1}{2} \sqrt{\ep}} 1,
	 \end{align*}
	 which follows \cref{eq:N_xy_noxi+y}.
We consider the Naimark's dialation of $\set{ G_u }$ on $\calH \x \calH'$ denoted by $\set{ \hat{G}_u }$, which is a projective measurement.
There exists $\ket{aux} \in \calH'$ such that 
averaging over $F \sim D$, the consistency between $\set{ N^{x_i | F}_0 \x \id_{\calH'}, N^{x_i | F}_1\x \id_{\calH'}}$ and $\set{ \sum_{u:u\cdot x_i = 0} \hat{G}_u, \sum_{u:u\cdot x_i = 1} \hat{G}_u}$ with respect to $\ket{\psi'} = \ket{\psi} \x \ket{aux} \x \ket{aux}$ is 
\begin{align*}
	&\bigE_{F \sim D} \sum_{a =0,1} \bra{\psi'} (N^{x_i | F}_a \x \id_{\calH'}) \x  (\sum_{u: u\cdot x_i = a} \hat{G}_u) \ket{\psi'} \\
	&=  \bigE_{F \sim D} \sum_{a =0,1} \bra{\psi} N^{x_i | F}_a \x \left(\sum_{u: u\cdot x_i = a} (\id \x \bra{aux}) \hat{G}_u (\id \x \ket{aux} ) \right) \ket{\psi} \\
	&=   \bigE_{F \sim D} \sum_{a =0,1} \bra{\psi} N^{x_i | F}_a \x\left(\sum_{u: u\cdot x_i = a} G_u \right) \ket{\psi} \\
	&\appd{C_1/2\sqrt{\ep}} 1.
\end{align*}
Since both $\set{ N^{x_i | F}_a \x \id_{\calH'} }$ and $\set{ \sum_{u: u\cdot x_i = a} \hat{G}_u }$ are projective measurements, their consistency implies that 
\begin{align*}
	\bigE_{F \sim D} \sum_{d = 0, 1} \norm{ N^{x_i | F}_d \x \id_{\calH'} \ket{\psi'} - \sum_{u: u \cdot x_i = d} \hat{G}_u \ket{\psi'}}^2 \leq C_1 \sqrt{\ep}.
\end{align*}

Next, notice that 
\begin{align*}
	N^F_a = N^{x_k | F }_{a_k} \ldots N^{x_1 | F}_{a_1} \text{ and }  
	 \sum_{\substack{u: u\cdot x_i = a_i\\\forall i \in [k]}} \hat{G}_u = 
	 \left( \sum_{u: u\cdot x_k = a_k}\hat{G}_u \right)\ldots \left( \sum_{u: u\cdot x_1 = a_1}\hat{G}_u \right) \ldots \left( \sum_{u: u\cdot x_k = a_k}\hat{G}_u \right)
\end{align*}
Then by \cref{lm:k_prod}
\begin{align*}
	\bigE_{F \sim D} \sum_{a \in \bF_2^k} \norm{ N^F_a \x \id_{\calH'} \x \id_B \ket{\tpsi} - \id_A \x \sum_{\substack{u: u\cdot x_i = a_i\\\forall i \in [k]}} \hat{G}_u \ket{\tpsi}}^2 \leq  (2k-1)^2 C_1 \sqrt{\ep},
\end{align*}
which completes the proof.
\end{proof}

\subsection{Answer reduction protocol}

The subset tester of the Hadamard code implies that we can replace the low-degree code of the answer reduction technique
in \cite[Section 17.4]{neexp} by the Hadamard code. 
The other key ingredient of Natarajan and Wright's answer reduction is Probabilistically Checkable Proofs of Proximity (PCPP),
so we recall its definition and key properties that we will use later.

\begin{definition}[PCPP]
    \label{def:pcpp}
    For functions $r, q: \bZ_{>0} \to \bZ_{>0}$, $t: \bZ_{>0} \times \bZ_{>0} \to \bZ_{>0}$,
    an $(r,q,t)$-restricted $\PCPP$ verifier is a probabilistic machine that, given a string $x$
    (called the \emph{explicit} input) and a number $K$ (in binary) as well as oracle access to an implicit input $y \in \set{0,1}^K$ and to a \emph{proof oracle} $\pi \in \set{0,1}^*$,
    tosses $r(\abs{x}+K)$ coins, queries the oracles $(y,\pi)$ for a total of $q(\abs{x}+K)$ symbols,
    runs in time $t(\abs{x}, K)$, and outputs a Boolean verdict.
    
    For constants $s,\gamma \in [0,1]$, a pair language $L \subseteq \set{0,1}^\ast \times \set{0,1}^\ast$ is in $\PCPP_{s,\gamma}[r,q,t]$  if there exists an $(r,q,t)$-restricted $\PCPP$
    verifier $V$ with the following properties:
    \begin{description}
        \item[Completeness:] If $(x,y) \in L$, there exists a proof $\pi$ such that 
        $\Pr_R[V^{y,\pi}(x, \abs{y};R) = 1] = 1$ where $V^{y,\pi}(x, \abs{y};R)$ denotes the decision of $V$
        on input $(x, \abs{y})$, oracle access to $(y, \pi)$, and randomness $R$.
        \item[Soundness:] Let $L_x = \set{y \mid (x,y) \in L}$. If $(x,y)$ is such that $y$ is $\gamma$-far from $L_x \cap \set{0,1}^{\abs{y}}$, then for every $\pi$, $\Pr_R[V^{y,\pi}(x, \abs{y};R) = 1] \leq s$.
    \end{description}
\end{definition}
We work with the $\PCPP$ such that when $L$ is an $\mathsf{NTIME}(T)$ pair language,
\begin{description}
    \item[Randomness complexity:] $r(m) = \log_2T(m) + O(\log_2\log_2 T(m))$,
    \item[Query complexity:] $q(m) = O(1)$, and 
    \item[Verification time:] $t(n, K) = \poly(n , \log_2 K, \log_2 T(n+K))$.
\end{description}

We are going to apply the $\PCPP$ defined above to the following language.
\begin{definition}
    \label{def:L_enc}
    Let $V = (\Alg_Q, \Alg_A)$ be an $\MIP^\ast$ verifier, where $\Alg_Q$ is his algorithm to sample the questions and $\Alg_A$ is his algorithm to check the answers.
    Suppose on inputs of length $n$ it has question length $\ell_{V,Q}(n)$ and answer length $\ell_{V,A}(n)$.
    We define 
    \begin{align*}
        L_{\Enc} = \set{ (\ipt, x_0, x_1, \Enc_{\ell_{V,A}(\abs{\ipt})}(y_0), \Enc_{\ell_{V,A}(\abs{\ipt})}(y_1) ) 
        \mid C^{\ipt}_{x_0,x_1}(y_0,y_1) = 1
        },
    \end{align*}
    which are all the accepted tuples with the answers encoded by $\Enc_{\ell_A(\abs{\ipt})}$.
\end{definition}
Note that when $\abs{\ipt} = n$, the running time of the decider of $L_{\Enc}$ is the maximal of the running time of $\Alg_A$ and $\Dec_{\ell_{V,A}(n)}$ as pointed out in \cite[Proposition 17.7]{neexp}.
Suppose $\gamma \leq \eta_H/2 = 1/4$. 
Then by \cite[Proposition 17.8]{neexp}, if $(\ipt, x_0, x_1, z_0, z_1)$ does not correspond to 
the encoding of any assignment accepted by $\Alg_A$, for every proof $\pi$
\begin{align*}
    \Pr_R[V_{\PCPP}^{z_0,z_1,\pi}(\ipt, x_0,x_1, \abs{z_0} + \abs{z_1};R) = 1 ] \leq s
\end{align*}
where $s$ is the soundness of $V_{\PCPP}$.
\begin{definition}
    \label{def:params}
    We instantiate the answer-reduced $\MIP^\ast$ protocol with the following components and notations.
    \begin{itemize}
        \item Let $V = (\Alg_Q, \Alg_A)$ be an $\MIP^\ast$ verifier for a Language $L$.
        Suppose on inputs of size $n$, the verifier $V$ has question length $\ell_{V,Q}(n)$,
        answer length $\ell_{V,A}(n)$, question sampling time $t_{V,Q}(n)$ and answer verification time $t_{V,A}(n)$.
        \item Let $G_k(\bT)$ be the subset tester from \cref{sec:subset_tester} for the Hadamard code of $\bF_2^k$ with the embedding $\mu_k$, and for
        the subset $\bT$ sampled according to some distribution $D$.
        \item Let $L_{\Enc}$ be the language defined in \cref{def:L_enc}, which is in $\text{TIME}(T)$ with
        \begin{align*}
            T(n) = t_{V,A}(n) + t_{\Dec}(\ell_{V,A}(n)).
        \end{align*}
        and let $V_{\PCPP}$ be its $\PCPP$ verifier with 
        $\gamma \leq 1/4$ and constant soundness $s$. 
        The verification time is 
        \begin{align*}
            t_{\PCPP,A}(n) = \poly(n+\ell_{V,Q}(n), \log_2(2^{\ell_{V,A}(n)}), \log_2(T(n))).
        \end{align*}
        The sampling time is also upper bounded by the verification time of the PCPP, which includes both the sampling time and answer verification time, so
        \begin{align*}
            t_{\PCPP,Q}(n) = \poly(n+\ell_{V,Q}(n), \log_2(2^{\ell_{V,A}(n)}), \log_2(T(n))).
        \end{align*}
        Finally, on inputs of size $n$, the proof length is
        \begin{align*}
            \ell_\pi(n) = 2^{r(n)} = T(n)\cdot \polylog(T(n)),
        \end{align*}
        where $r(n)$ is the randomness complexity of the PCPP verifier.
        \item We write $\ell_1 \coloneqq \ell_{V,A}(n)$ and $\ell_2 \coloneqq \ell_{\pi}(n)$.
    \end{itemize}
\end{definition}
Next, we give the protocol of the answer reduced verifier $V^{AR}$, which requires the provers to encode their proof $\pi$ by
the Hadamard code of $\bF_2^{\ell_2}$.
The protocol is very similar to the protocol presented in \cite[Figure 15]{neexp}, but we include it for completeness.
\begin{figure}
    \begin{center}
    \scalebox{0.8}{
    \begin{minipage}{0.9\linewidth}
    \begin{Algorithm}
        \begin{center}{\large The answer reduced verifier $V^{AR}$}
        \end{center}
        \textbf{Setup:} Flip one unbiased coin $\bb \sim \set{0,1}$. Sample questions $(\bx_0,\bx_1) \sim \Alg_Q(\ipt)$.
    	Sample a view $\bI_0, \bI_1, \bJ \sim V_{\PCPP}(\ipt, \bx_0, \bx_1)$.
    	Set $\bJ' = \mu_{\ell_2}(\bJ)$.
    	Randomly select $\bI_0', \bI_1'  \subseteq [2^{\ell_1}]$ and $\bJ'' \subseteq [2^{\ell_2}]$ such that
            $\abs{\bI_0'} = \abs{\bI_1'} = \abs{\bJ''} = \kappa$, which is a sufficiently large constant.
            Details about how to choose $\kappa$ can be found in the proof below.
    	Set $\bT_0 = \bI_0 \cup \bI_0'$, $\bT_1 = \bI_1 \cup \bI_1'$ and $\bU = \bJ' \cup \bJ''$.
    	
            With probability $1/10$ each, perform one of the following ten tests
            \footnote{There are two answer cross-checks, one for $c=0$ and one for $c=1$, similarly for the 
            answer consistency checks and answer code checks.}.
    \begin{description}
    	\item[Verify]: Distribute the questions as follows:
    		\begin{itemize}
    		\item
    		Player $\bb$: give $(\bx_0, \bx_1), \bT_0, \bT_1, \bU$; receive $\ba_0, \ba_1, \ba_2$.
    		\end{itemize}
    	Accept if $V_{\PCPP}(\ipt, \bx_0, \bx_1)$ accepts on $\ba_0|_{\bI_0}$, $\ba_1|_{\bI_1}$ and $\ba_2|_{\bJ'}$.
    	\item[Cross check]: 
    	\begin{description}
    		\item[Consistency test:] Distribute the questions as follows:
    		\begin{itemize}
    		\item
    		Player $\bb$: give $(\bx_0, \bx_1), \bT_0, \bT_1, \bU$; receive $\ba_0, \ba_1, \ba_2$.
    		\item
    		Player $\overline{\bb}$: give $(\bx_0, \bx_1), \bT_0, \bT_1, \bU$; receive $\ba'_0, \ba'_1, \ba'_2$
    		\end{itemize}
    		Accept if $\ba_0 = \ba_0'$, $\ba_1 = \ba_1'$ and $\ba_2 = \ba_2'$.
    		\item[Answer cross-check:] For $c=0,1$, distributed the questions as follows:
    		\begin{itemize}
    		\item
    		Player $\bb$: give $(\bx_0, \bx_1), \bT_0, \bT_1, \bU$; receive $\ba_0, \ba_1, \ba_2$.
    		\item
    		Player $\overline{\bb}$: give $\bx_{c}, \bT_{c}$; receive $\ba'_{c}$
    		\end{itemize}
    		Accept if $\ba_{c} = \ba'_{c}$.
    		\item[Answer consistency check:] For $c = 0, 1$, distributed the questions as follows:
    		\begin{itemize}
    		\item
    		Player $\bb$: give $\bx_{c}, \bT_{c}$; receive $\ba_{c}$.
    		\item
    		Player $\overline{\bb}$: give $\bx_{c}, \bT_{c}$; receive $\ba'_{c}$
    		\end{itemize}
    		Accept if $\ba_{c} = \ba'_{c}$.
    		\item[Proof cross-check:] Distribute the questions as follows:
    		\begin{itemize}
    		\item
    		Player $\bb$: give $(\bx_0, \bx_1), \bT_0, \bT_1, \bU$; receive $\ba_0, \ba_1, \ba_2$.
    		\item
    		Player $\overline{\bb}$: give $(\bx_0, \bx_1), \bU$; receive $\ba'_2$
    		\end{itemize}
    		Accept if $\ba_2 = \ba_2'$.
    	\end{description}
    	\item[Code checks]:
    	\begin{description}
    		\item[Answer code check:] For $c=0,1$, 
    		sample questions $(\bw_0, \bw_1) \sim G_{\ell_1}(\bT_{c})$.
    		Distributed the questions as follows:
    		\begin{itemize}
    		\item
    		Player $\bb$: give $\bx_{c}, \bw_0$; receive $\ba_0$.
    		\item
    		Player $\overline{\bb}$: give $\bx_{c}, \bw_1$; receive $\ba_1$.
    		\end{itemize}
    		Accept if $G_{\ell_1}(\bT_{c})$ accepts on $\ba_0$ and $\ba_1$.
    		\item[Proof code check:] 
    		Sample questions $(\bw_0, \bw_1) \sim G_{\ell_2}(\bU)$.
    		Distribute the questions as follows:
    		\begin{itemize}
    		\item
    		Player $\bb$: give $(\bx_0, \bx_1), \bw_0$; receive $\ba_0$.
    		\item
    		Player $\overline{\bb}$: give $(\bx_0, \bx_1), \bw_1$; receive $\ba_1$.
    		\end{itemize}
    		Accept if $G_{\ell_2}(\bU)$ accepts on $\ba_0$ and $\ba_1$.
    	\end{description}
    \end{description}
    \end{Algorithm}
    \caption{The answer reduced verifier $V^{AR}$.}
    \end{minipage}
    }
     \end{center}
     \label{fig:VAR}
 \end{figure}

\begin{theorem}
    \label{thm:answer_reduce}
	Let $V = (\Alg_Q, \Alg_A)$ be an $\MIP^\ast$ protocol for a language $L$,
    with question length $\ell_{V,Q}$, answer length $\ell_{V,A}$, sampling time $t_{V,Q}$
    and verification time $t_{V,A}$.
	Suppose the $\PCPP$ verifier is chosen so that $\gamma \leq 1/4$.
	Suppose further that $V$ has the following property: for any $\ipt \in L$, the prover has a 
	real commuting symmetric EPR strategy with a value $1$.
	Then $V^{AR}$ obtained by applying the the answer reduction procedure to $V$ as shown in \cref{fig:VAR} is also an
	$\MIP^\ast$ verifier for $L$ with the following two conditions:
	\begin{description}
        \item[Question length.] The new question length is $\ell_{V^{AR}, Q}(n) = O(\ell_{V,Q}(n) + \ell_1(n) + \ell_2(n))$.
        \item[Answer length.] The new answer length is $\ell_{V^{AR}, A}(n) = O(1)$.
        \item[Sampling time.] The new question sampling time is 
        \begin{align*}
            t_{V^{AR},Q}(n) = t_{V,Q}(n) + \poly(n+\ell_{V,Q}(n),\log_2(2^{\ell_{V,A}(n)}), \log_2(T(n))) + O(\ell_1(n) + \ell_2(n)).
        \end{align*}
        \item[Verification time.] The new verification time is $t_{V^{AR},A} = \poly(n+\ell_{V,Q}(n), \ell_1(n), \log_2(T(n)))$.
		\item[Completeness.] If $\ipt \in L$, there is a value-$1$ strategy for $V^{AR}$.
		\item[Soundness.] Given $\ipt$, suppose there is a strategy for $V^{AR}$ with value $1 - \ep$.
		Then there exists constants $K_1$ and $K_2$ such that there is a strategy for $V$ on $\ipt$ with
		value $1 - K_1 - K_2 \ep^{1/128}$.
        \item[Efficient computability.] There exists an algorithm that takes the description of $V = (\Alg_Q, \Alg_A)$ as input and outputs the description of $V^{AR} = (\Alg_Q', \Alg_A')$ in time $O( \abs{\Alg_Q}+ \abs{\Alg_A})$.
        Moreover, $\abs{\Alg_Q'} = \abs{\Alg_Q} +O(1)$ and $\abs{\Alg_A'} = \abs{\Alg_A}+O(1)$.
	\end{description}
\end{theorem}

\begin{proof}
\textbf{Question length}. The question of $V^{AR}$ consists of a question from $V$ and queries to the encodings of the answers and the $\PCPP$ proof. Hence,
\begin{align*}
    \ell_{V^{AR}, Q}(n) \leq 2\ell_{V,Q}(n) + 2(\kappa+q(n)) \cdot \ell_1(n) + (\kappa + q(n)) \cdot \ell_2(n) = O(\ell_{V,Q}(n) +\ell_1(n)+\ell_2(n)),
\end{align*}
where $q(n)$ is the query complexity of the PCPP verifier.

\textbf{Answer length}. The prover only needs to answer the queries, so the answer consists of at most $3(\kappa + q(n))$ bits.

\textbf{Sampling time}. The sampling algorithm of the answer-reduced protocol needs to run the sampling algorithm of the $\PCPP$ verifier, which takes time
\begin{align*}
    t_{\PCPP, Q}(n) =  \poly(n+\ell_{V,Q}(n), \log_2(2^{\ell_{V,A}(n)}), \log_2(T(n))).
\end{align*}
The sampling algorithm must also run $\Alg_Q$, the embedding algorithm $\mu_{\ell_2(n)}$, and sample random indices in the encodings.
Hence, the sampling time is 
\begin{align*}
    t_{V^{AR}, Q}(n)   &\leq t_{V,Q}(n) + t_{\PCPP, Q}(n) + q(n)\cdot\ell_2(n) + 2 \kappa \ell_1(n) + \kappa \ell_2(n) \\
                    &=t_{V,Q}(n) + \poly(n+\ell_{V,Q}(n), \log_2(2^{\ell_{V,A}(n)}), \log_2(T(n))) + O(\ell_1(n) + \ell_2(n)),
\end{align*}
where $q(n) = O(1)$ is the query complexity of the $\PCPP$ verifier.

\textbf{Verification time}. Since the verification time of the code checks and consistency tests are $O(1)$, 
\begin{align*}
    t_{V^{AR}, A}(n) \leq t_{\PCPP,A}(n) + O(1) = \poly(n+\ell_{V,Q}(n), \ell_1(n), \log_2(T(n))).
\end{align*}

\textbf{Efficient computatbility.} This follows from the observation that the descriptions of $\Alg'_Q$ and $\Alg'_A$
contains both $\Alg_Q$ and $\Alg_A$ respectively with new instructions. 
Since the new instructions added to $\Alg_Q$ and $\Alg_A$ are independent of $\ipt$, $\Alg_Q$ and $\Alg_A$, the time to write them down are $O(1)$, and their sizes are also $O(1)$.

\textbf{Completeness}. This follows the same proof of the completeness part of \cite[Theorem 17.10]{neexp}.
If an honest prover gets one question $x_b$, the prover will compute its answer, encode its answer using the Hadamard code, and answer the queries accordingly.
If an honest prover gets both questions, the prover will compute its answers, compute a $\PCPP$ proof to certify these answers are correct, compute the Hadamard encodings of the answers and the proof, and answer the queries accordingly.

\textbf{Soundness}. The constant $K_1$ depends on the parameter $\kappa = \abs{\bI_0'}$, so we should set $\kappa$ to be a sufficiently large constant so that $1 - K_1 - K_2 \ep^{1/128}$ is greater than the soundness $s$ of $V$.
Operationally, the views are augmented by $\kappa$ uniformly randomly chosen coordinates. The purpose of this is to drive the distance of the Hadamard code up from $1/2$ to $1 - 1/2^\kappa$, which will be needed for \cref{lm:k_prod_distr}. 

Let $C_I = \abs{\bI_0} + \kappa = \abs{\bI_1} + \kappa$ and $C_J = \abs{\bJ} + \kappa$ be two constants.
Suppose $\ipt$ is not in $L$. Let $(\ket{\psi}, M)$ be a strategy that
passes with probability $1 - \ep$.
This strategy can pass each \textbf{Answer code check} with probability $1 - 10\ep$. Given values $c$ and $x_c$,
write $1 - \ep_{c,x_c}$ for the probability the code check passes conditioned on these values.
Then with probability at least $1 - 10 \ep^{1/2}$, $\ep_{c,x_c} \leq \ep^{1/2}$.
When this occurs, we can apply \cref{prop:subset_test} to $G_{\ell_1}(\bT_c)$ where the distribution of 
$\bT_c$ denoted by $D_{x_c}$ is determined by $c$ and $x_c$.
\Cref{prop:subset_test} implies that there exists Hilbert spaces $\calH_{x_c}$, $\ket{aux_{x_c}} \in \calH_{x_c} \x \calH_{x_c}$
and projective measurement $\set{ G^{x_c}_u }$ on $\calH^{x_c}$ such that
\begin{align*}
	\bigE_{T_c \sim D_{x_c}} \sum_{a \in \bF_2^{C_I}} \norm{  (M^{x_c, T_c}_a \x \id_{\calH_{A, x_c}} \x \id_B -\id_A \x G^{x_c}_{[w|_{T_c} =a]})\ket{\psi} \x \ket{aux_{x_c}}}^2
	\leq O(C_I^3\sqrt{\ep_{c,x_c}})
\end{align*}
where $\id_A = \id_{\calH_A \x \calH_{A, x_c}}$ and similar for $\id_B$.
When this does not occur, we can still assume such Hilbert spaces and projective measurements so that
\begin{align*}
	\bigE_{T_c \sim D_{x_c}} \sum_{a \in \bF_2^{C_I}} \norm{  (M^{x_c, T_c}_a \x \id_{\calH_{A, x_c}} \x \id_B -\id_A \x G^{x_c}_{[w|_{T_c} =a]})\ket{\psi} \x \ket{aux_{x_c}}}^2
	\leq O(1).
\end{align*}
When averaging over $c$ and $x_c$, 
\begin{align*}
	\bigE_{c,x_c}\bigE_{T_c \sim D_{x_c}} \sum_{a \in \bF_2^{C_I}} \norm{  (M^{x_c, T_c}_a \x \id_{A,x_c} \x \id_B -\id_A \x G^{x_c}_{[w|_{T_c} =a]})\ket{\psi} \x \ket{aux_{x_c}}}^2 \leq O(C_I^3\ep^{1/4}).
\end{align*}
Passing the \textbf{Proof code check} implies that there exists Hilbert spaces $\calH_{x_0,x_1}$, states $\ket{aux_{x_0,x_1}} \in \calH_{x_0,x_1} \x \calH_{x_0,x_1}$
and projective measurements $\set{H^{x_0,x_1}_w}$ on $\calH \x \calH_{x_0,x_1}$ such that 
\begin{align*}
	\bigE_{x_0,x_1} \bigE_{U \sim D_{(x_0,x_1)}} \sum_{a \in \bF_2^{C_J}} \norm{  (M^{x_0,x_1, U}_a \x \id_{\calH_{A, x_0,x_1}} \x \id_B -\id_A \x H^{x_0,x_1}_{[w|_{U} =a]})\ket{\psi} \x \ket{aux_{x_0,x_1}}}^2 \leq O(C_J^3\ep^{1/4}).
\end{align*}

The next step is ensuring the $G$ and $H$ measurements act on the same Hilbert space.
Let 
\begin{align*}
\ket{\tpsi} = \ket{\psi} \x (\x_{x} \ket{aux_{x}}) \x (\x_{x_0,x_1} \ket{aux_{x_0,x_1}})
\end{align*}
and
\begin{align*}
	&\tG^{x_c}_u = G^{x_c}_u \x (\x_{x \neq x_c} \id_{\calH_{x}}) \x (\x_{x_0,x_1} \id_{\calH_{x_0,x_1}}) \\
	&\tH^{x_0,x_1}_u = H^{x_0,x_1} \x (\x_x \id_{\calH_x}) \x (\x_{(z_0,z_1) \neq (x_0,x_1)} \id_{\calH_{z_0,z_1}}),
\end{align*}
and, let 
\begin{align*}
	&N^{x_c, T_c}_{a_c}  = M^{x_c, T_c}_{a_c} \x (\x_x \id_{\calH_x}) \x (\x_{x_0,x_1} \id_{\calH_{x_0,x_1}}) \\
	&N^{x_0, x_1, U} = M^{x_0, x_1, U}_{a_2}  \x (\x_x \id_{\calH_x}) \x (\x_{x_0,x_1} \id_{\calH_{x_0,x_1}}) \\
	&N^{x_0, x_1, T_0, T_1, U}_{a_0,a_1,a_2} = M^{x_0, x_1, T_0, T_1, U}_{a_0,a_1,a_2} \x  (\x_x \id_{\calH_x}) \x (\x_{x_0,x_1} \id_{\calH_{x_0,x_1}}). 
\end{align*}
Note that we omit the permutation of the Hilbert spaces in the definitions above.
Then for all $x_c$
\begin{align*}
	&\bigE_{T_c \sim D_{x_c}} \sum_{a \in \bF_2^{C_I}} \norm{  (N^{x_c, T_c}_a  \x \id_B -\id_A \x \tG^{x_c}_{[w|_{T_c} =a]})\ket{\tpsi}}^2\\
	&=
	\bigE_{T_c \sim D_{x_c}} \sum_{a \in \bF_2^{C_I}} \norm{  (M^{x_c, T_c}_a \x \id_{\calH_{A, x_c}} \x \id_B -\id_A \x G^{x_c}_{[w|_{T_c} =a]})\ket{\psi} \x \ket{aux_{x_c}}}^2.
\end{align*}
Thus
\begin{align}
	\label{eq:N_to_G}
	\bigE_{c,x_c}\bigE_{T_c \sim D_{x_c}} \sum_{a \in \bF_2^{C_I}} \norm{  (N^{x_c, T_c}_a \x \id_B -\id_A \x \tG^{x_c}_{[w|_{T_c} =a]})\ket{\tpsi}}^2 \leq O(C_I^3\ep^{1/4}),
\end{align}
and
\begin{align}
	\label{eq:N_to_H}
	\bigE_{x_0,x_1} \bigE_{U \sim D_{(x_0,x_1)}} \sum_{a \in \bF_2^{C_J}} \norm{  (N^{x_0,x_1, U}_a \x \id_B -\id_A \x \tH^{x_0,x_1}_{[w|_{U} =a]})\ket{\tpsi}}^2 \leq O(C_J^3\ep^{1/4}).
\end{align}
Note these relations also hold with the two systems flipped. 

Similarly, passing the \textbf{Cross Checks} implies that
\begin{align}
	&N^{x_0, x_1, T_0, T_1, U}_{a_0} \x \id_B \appd{O(\ep)} \id_A \x N^{x_0, T_0}_{a_0} \\
	&N^{x_0, x_1, T_0, T_1, U}_{a_1} \x \id_B \appd{O(\ep)} \id_A \x N^{x_1, T_1}_{a_1} \\
	\label{eq:Nx0_consist}
	&N^{x_0, T_0}_{a_0} \x \id \appd{O(\ep)} \id_A \x N^{x_0, T_0}_{a_0} \\
	\label{eq:Nx1_consist}
	&N^{x_1, T_1}_{a_1} \x \id \appd{O(\ep)} \id_A \x N^{x_1, T_1}_{a_1} \\
	&N^{x_0, x_1, T_0, T_1, U}_{a_2} \x \id_B \appd{O(\ep)} \id_A \x N^{x_0, x_1, U}_{a_2} \\  
	&N^{x_0, x_1, T_0, T_1, U}_{a_0,a_1,a_2} \x \id_B \appd{O(\ep)} \id_A \x N^{x_0, x_1, T_0, T_1, U}_{a_0,a_1,a_2},
\end{align}
with respect to $\ket{\tpsi}$ over the distribution $D$ of $x_0, x_1, T_0, T_1, U$.
Note that here we use the $\approx$ notation introduced at the beginning of \cref{sec:lemma_mip}
to make the dependence on the distribution implicit.
These equations combined with \cref{eq:N_to_G,eq:N_to_H} imply the 
measurements $\set{ N^{x_0,x_1,T_0,T_1,U}_{a_0, a_1, a_2} }$, $\set{ \tG^x_u }$ and $\set{\tH^{x_0,x_1}_w}$ satisfy
conditions of \cref{lm:k_prod_distr} with respect to $\ket{\tpsi}$ and distribution $D$.
Let 
\begin{align*}
	\set{ \Lambda^{x_0,x_1}_{u_0,u_1,w} := \tG^{x_0}_{u_0} \cdot \tG^{x_1}_{u_1} \cdot \tH^{x_0,x_1}_{w} \cdot \tG^{x_1}_{u_1} \cdot 
	\tG^{x_0}_{u_0}}
\end{align*}
be a POVM constructed following \cref{lm:k_prod_distr}. 
Recall that $T_c$ and $U$ has $\kappa$ independent coordinates, so two different codewords agree on $T_c$ or $U$ with a probability at most 
$\eta_H^\kappa = 1/2^\kappa$.
Letting $C_0 = \max\{C_I, C_J\}$,
Hence we can applying \cref{lm:k_prod_distr} to this POVM with $k = 3$, $\delta := C_0^3\ep^{1/4} $ and $\ep:=1/2^{\kappa}$, and get that 
\begin{align}
	\label{eq:N_to_Lambda}
	N^{x_0, x_1, T_0, T_1, U}_{a_0,a_1,a_2} \x \id_B \appd{O(C_0^{3/16}\ep^{1/64} + \frac{1}{2^{\kappa/8}})} \id_A \x \Lambda^{x_0,x_1}_{[u_0|_{T_0}, u_1|_{T_1}, w|_U =
	a_0,a_1,a_2]} 
\end{align}
with respect to $\ket{\tpsi}$ and $D$, where $[u_0|_{T_0}, u_1|_{T_1}, w|_U =
	a_0,a_1,a_2]$ means that $\Enc_{\ell_1}(u_0)|_{T_0} = a_0$ and etc..
Passing \textbf{Verify} with a probability at least $1 - 10\ep$ along with 
\Cref{eq:N_to_Lambda,lm:close} implies that $\set{\Lambda^{x_0,x_1}_{u_0,u_1,w} }$ can be used to pass the verify test with probability $1 - 10\ep - O(C_0^{1/8}\ep^{1/128} + \frac{1}{2^{\kappa/16}})$ where we upper bound $C_0^{3/16}$ by $C_0^{1/4}$. The player would measure $\id_A \x \Lambda$ on $\ket{\tpsi}$ and return the local views of the measurement outcomes according to the questions. 

Consider the measurements $\set{ \Lambda^{x_0, x_1}_{u_0, u_1} := \sum_w \Lambda^{x_0,x_1}_{u_0,u_1,w}}$
Let
\begin{align*}
	p &:= \bigE_{x_0,x_1} \sum_{u_0, u_1: V(\ipt, x_0,x_1,u_0,u_1)=1} \bra{\tpsi} \id_A \x \Lambda^{x_0,x_1}_{u_0,u_1} \ket{\tpsi},
\end{align*}
which is the probability that measuring with $\Lambda^{x_0, x_1}_{u_0, u_1}$ gives answers 
$u_0$ and $u_1$ accepted by the verifier $V$ when the questions are $x_0$ and $x_1$.
Then
\begin{align*}
	 p &= \bigE_{x_0,x_1} \sum_{\substack{u_0,u_1: \\ V(\ipt, x_0,x_1,u_0,u_1)=1}}
	 \sum_{w}
	   \bra{\tpsi} \id_A \x \Lambda^{x_0,x_1}_{u_0,u_1,w} \ket{\tpsi} \\
	&\geq \bigE_{x_0,x_1} \sum_{\substack{u_0,u_1: \\ V(\ipt, x_0,x_1,u_0,u_1)=1}}
	 \sum_{w} \bra{\tpsi} \id_A \x \Lambda^{x_0,x_1}_{u_0,u_1,w} \ket{\tpsi} \cdot 
	\Pr_R[V_{\PCPP}^{u_0,u_1,w}(\ipt,x_0,x_1,2\cdot 2^{\ell_1}; R)=1]\\
	&= \Pr[(\ket{\tpsi}, \Lambda) \text{ pass \textbf{verify} check }]  \\
	&\quad-  \sum_{\substack{u_0,u_1: \\V(\ipt, x_0,x_1,u_0,u_1)=0}} \sum_w\bra{\tpsi} \id_A \x \Lambda^{x_0,x_1}_{u_0,u_1,w} \ket{\tpsi} \cdot 
	\Pr_R[V^{u_0,u_1,w}_{\PCPP}(\ipt, x_0, x_1, 2\cdot 2^{\ell_1}; R)=1]\\
	&\geq 1 - 10 \ep - O(C_0^{1/8}\ep^{1/128} + \frac{1}{2^{\kappa/16}}) \\
	&\quad - \sum_{\substack{u_0,u_1:\\ V(\ipt, x_0,x_1,u_0,u_1)=0}} \sum_w \bra{\tpsi}  \id_A \x \Lambda^{x_0,x_1}_{u_0,u_1,w} \ket{\tpsi} \cdot 
	\Pr_R[V^{u_0,u_1,w}_{\PCPP}(\ipt,x_0,x_1,2\cdot 2^{\ell_1};R)=1] \\
	&\geq 1 - 10 \ep - O(C_0^{1/8}\ep^{1/128} + \frac{1}{2^{\kappa/16}}) - (1-p)s,
\end{align*}
where $s$ is the soundness of $V_{\PCPP}$.
In the derivation above, $\Pr_R[V^{u_0,u_1,w}_{\PCPP}(\ipt, x_0, x_1, 2\cdot 2^{\ell_1}; R)=1]$ is the probability that $V_{\PCPP}$
accepts $\ipt$.
For any $x_0,x_1,u_0,u_1$ not accepted by $V$, this probability is below $s$ by \cite[Proposition 17.8]{neexp}.
Hence
\begin{align*}
	p \geq \frac{ 1 - 10 \ep -O(C_0^{1/8}\ep^{1/128} + \frac{1}{2^{\kappa/16}}) - s }{1 - s} = 1 - \frac{10\ep+O(C_0^{1/8}\ep^{1/128} + \frac{1}{2^{\kappa/16}}) }{1 - s}.
\end{align*}

In the end, we use $(\set{ \tG^x_u }, \ket{\tpsi})$ as a strategy for $V$.
Applying \cref{lm:chain_rule} to \cref{eq:Nx0_consist,eq:Nx1_consist,eq:N_to_G}, we get that 
\begin{align*}
	\tG^{x_0}_{u|_{T_0}=a} \x \id \appd{O(C_0^3\ep^{1/4})} \id \x \tG^{x_0}_{u|_{T_0}= a}
\end{align*}
with respect to the distribution of $x_0$ and the distribution of $T_0$ determined by $x_0$ on the state $\ket{\tpsi}$.
Since $\set{ \tG^{x_0}_u }$ is a projective measurement, we know
\begin{align*}
	\bigE_{x_0} \bigE_{T_0 \sim D_{x_0}} \sum_a \bra{\tpsi} \tG^{x_0}_{u|_{T_0} = a} \x \tG^{x_0}_{u|_{T_0} = a} \ket{\tpsi} \geq 1 - O(C_0^3\ep^{1/4}).
\end{align*}
On the other hand
\begin{align*}
	&\bigE_{x_0} \bigE_{T_0 \sim D_{x_0}} \sum_a \bra{\tpsi} \tG^{x_0}_{u|_{T_0} = a} \x \tG^{x_0}_{u|_{T_0} = a} \ket{\tpsi} \\
	&= \bigE_{x_0} \sum_u \bra{\tpsi} \tG^{x_0}_{u} \x \tG^{x_0}_{u} \ket{\tpsi} 
	+ \bigE_{x_0} \bigE_{T_0\sim D_{x_0}} \sum_{u\neq u': u|_{T_0} = u'|_{T_0}} \bra{\tpsi} \tG^{x_0}_{u} \x \tG^{x_0}_{u'} \ket{\tpsi} \\
	&= \bigE_{x_0} \sum_u \bra{\tpsi} \tG^{x_0}_{u} \x \tG^{x_0}_{u} \ket{\tpsi} + 
	\bigE_{x_0} \bigE_{T_0\sim D_{x_0}} \sum_{u \neq u'} \id[u|_{T_0} = u'|_{T_0}] \bra{\tpsi} \tG^{x_0}_{u} \x \tG^{x_0}_{u'} \ket{\tpsi}.
\end{align*}
Since for all $x_0$ and $u \neq u'$, $\bigE_{T_0\sim D_{x_0}}\id[u|_{T_0} = u'|_{T_0}] \leq 1/2^\kappa$, we know
\begin{align*}
	 \bigE_{x_0} \sum_u \bra{\tpsi} \tG^{x_0}_{u} \x \tG^{x_0}_{u} \ket{\tpsi} \geq 1 - 1/2^\kappa - O(C_0^3\ep^{1/4}).
\end{align*}
Again, because $\set{\tG^{x_0}_u}$ is a projective measurement 
\begin{align*}
	\bigE_{x_0} \sum_u \norm{ (\tG^{x_0}_u \x \id - \id \x \tG^{x_0}_u) \ket{\tpsi}}^2 \leq \frac{1}{2^{\kappa-1}} + O(C_0^3\ep^{1/4}).
\end{align*}

Let $S(x_0,x_1) = \set{(a_0, a_1) \mid V(x_0, x_1,a_0,a_1) =1}$.
We can calculate
\begin{align*}
	&\abs{\bigE_{x_0,x_1} \sum_{(a_0, a_1) \in S} \bra{\tpsi} \tG^{x_0}_{a_0} \x \tG^{x_1}_{a_1} \ket{\tpsi}
	- \bra{\tpsi} \tG^{x_0}_{a_0}  \x \tG^{x_1}_{a_1}\tG^{x_0}_{a_0} \ket{\tpsi}} \\
	&\leq \sqrt{ \bigE_{x_0,x_1} \sum_{(a_0, a_1) \in S} \norm{ \tG^{x_0}_{a_0} \x \tG^{x_1}_{a_1} \ket{\tpsi} }^2} \\
	&\cdot \sqrt{  \bigE_{x_0,x_1} \sum_{(a_0, a_1) \in S} \bra{\tpsi} (\tG^{x_0}_{a_0} \x \id - \id \x \tG^{x_0}_{a_0}) 
	(\id \x \tG^{x_1}_{a_1} )  (\tG^{x_0}_{a_0} \x \id - \id \x \tG^{x_0}_{a_0}) \ket{\tpsi}} \\
	&\leq 1 \cdot \sqrt{  \bigE_{x_0} \sum_{a_0} \norm{  (\tG^{x_0}_{a_0} \x \id - \id \x \tG^{x_0}_{a_0}) \ket{\tpsi}}^2} \\
	&\leq O(\frac{1}{2^{\kappa/2}} + C_0^{3/2}\ep^{1/8}),
\end{align*}
and
\begin{align*}
	&\abs{\bigE_{x_0,x_1} \sum_{(a_0, a_1) \in S} \bra{\tpsi} \id \x \tG^{x_0}_{a_0} \tG^{x_1}_{a_1}\tG^{x_0}_{a_0}  \ket{\tpsi}
	- \bra{\tpsi} \tG^{x_0}_{a_0}  \x \tG^{x_1}_{a_1}\tG^{x_0}_{a_0} \ket{\tpsi}} \\
	&\leq \sqrt{ \bigE_{x_0,x_1} \sum_{(a_0, a_1) \in S} \norm{ \id \x \tG^{x_1}_{a_1}\tG^{x_0}_{a_0}  \ket{\tpsi} }^2} \\
	&\cdot \sqrt{  \bigE_{x_0,x_1} \sum_{(a_0, a_1) \in S} \bra{\tpsi} (\tG^{x_0}_{a_0} \x \id - \id \x \tG^{x_0}_{a_0}) 
	(\id \x \tG^{x_1}_{a_1} )  (\tG^{x_0}_{a_0} \x \id - \id \x \tG^{x_0}_{a_0}) \ket{\tpsi}} \\
	&\leq 1 \cdot \sqrt{  \bigE_{x_0} \sum_{a_0} \norm{  (\tG^{x_0}_{a_0} \x \id - \id \x \tG^{x_0}_{a_0}) \ket{\tpsi}}^2} \\
	&\leq O(\frac{1}{2^{\kappa/2}} + C_0^{3/2}\ep^{1/8}).
\end{align*}
Note that $\tG^{x_0}_{a_0} \tG^{x_1}_{a_1}\tG^{x_0}_{a_0} = \Lambda^{x_0,x_1}_{a_0,a_1}$.
Therefore, 
\begin{align*}
	\abs{\bigE_{x_0,x_1} \sum_{(a_0, a_1) \in S} \bra{\tpsi} (\tG^{x_0}_{a_0} \x \tG^{x_1}_{a_1} - \id \x \Lambda^{x_0,x_1}_{a_0,a_1}) \ket{\tpsi}}
	\leq O(\frac{1}{2^{\kappa/2}} + C_0^{3/2}\ep^{1/8}).
\end{align*}
On the other hand, we have shown
\begin{align*}
	\bigE_{x_0,x_1} \sum_{(a_0, a_1) \in S} \bra{\tpsi} \id \x \Lambda^{x_0,x_1}_{a_0,a_1}) \ket{\tpsi} = p \geq 1- O(C_0^{1/8}\ep^{1/128} + \frac{1}{2^{\kappa/16}}).
\end{align*}
Since the big-O notations in the derivations above hide constants that are independent of $\kappa$, 
the winning probability of the strategy $(\set{ \tG^x_u }, \ket{\tpsi})$ for $V$ 
is at least $1 -\frac{C_1}{2^{\kappa/16}} - C_2C_0^{3/2}\ep^{1/128}$ for some constants $C_1$ and $C_2$.
Hence, $K_1 = \frac{C_1}{2^{\kappa/16}}$ and $K_2 = C_2C_0^{3/2}$ in the soundness statement.
We need to pick $\kappa$ large enough such that $1 -\frac{C_1}{2^{\kappa/16}} > s$, then we can solve for 
\begin{align*}
    \ep_0 = \Big[\frac{1 - \frac{C_1}{2^{\kappa/16}} - s}{ C_2 C_0^{3/2}}\Big]^{128},
\end{align*}
such that $1 - \ep_0$ is the soundness of $V^{AR}$.
\end{proof}

\subsection{Putting everything together}
\begin{theorem}
    \label{thm:re}
    $\RE$ is contained in $\MIP^\ast[\poly, O(1)]$ with completeness $1$ and a constant soundness. 
\end{theorem}

\begin{proof}
    We know that there is an $\MIP^\ast[\poly,\poly]$ protocol $V = (\Alg_Q, \Alg_A)$ for any language in $\RE$. To prove the theorem, we construct an $\MIP^\ast[\poly, O(1)]$ protocol for the same language, by applying several answer reduction transformations to $V$. Specifically, we apply two iterations of an answer reduction scheme based on the low-degree test over finite fields to reduce the answer size to $O(\poly\log\log(n))$, followed by one iteration of answer reduction based on the Hadamard code over $\bF_2$ to further reduce the answer size to $O(1)$. The effect of each step is summarized in the following table.
    \begin{table}[h!]
\centering
\begin{tabular}{|c||c|c|c|c|c|}
\hline
 & \makecell{\textbf{Question} \\\textbf{size}} & \makecell{\textbf{Answer}\\ \textbf{size}} & \makecell{\textbf{Sampling}\\ \textbf{time}} & \makecell{\textbf{Verification}\\\textbf{time}} & \makecell{\textbf{Decision}\\\textbf{ complexity}}\\ \hline
Original protocol      & poly(n)      & poly(n)      & poly(n) & poly(n) & poly(n)\\ \hline
\makecell{After Step 1 \\ ( \cref{thm:newar-pr})}     & poly(n)     & polylog(n)   & poly(n)&   poly(n) & polylog(n) \\ \hline
\makecell{After Step 2\\ ( \cref{thm:newar-pr})}     & poly(n)     & polyloglog(n)   & poly(n)&   poly(n) & polyloglog(n) \\ \hline
\makecell{After Step 3 \\ ( \cref{thm:answer_reduce})}     & poly(n)     & $O(1)$   & poly(n)&   poly(n) & O(1) \\ \hline
\end{tabular}
\caption{Effect of each step of the proof}
\label{tab:sample}
\end{table}

    \textbf{Step 1.} We parallel repeat and oracularize $V$ to ensure its soundness is at most $1/2$, and apply the answer reduction technique summarized in \cref{thm:newar-pr}.
    Since parallel repetition and oracularization only introduce constant overhead in the question length, answer length, sampling time and verification time, we still use $V$ to denote the oracularized protocol.
    Denote the answer-reduced protocol by $V_1 = (\Alg_{Q_1},\Alg_{A_1})$.
    Since $d_{V,A}(n) = \poly(n)$ and $k(n)= \polylog(d_{V,A}(n)) = \polylog(n)$,
    by \cref{thm:newar-pr}, the new question length is
     \begin{align*}
         \ell_{V_1,Q}(n) = \polylog(d_{V,A}(n)) \cdot (2 \ell_{V,Q}(n) + \polylog(d_{V,A}(n))) = \poly(n).
     \end{align*}
     The new answer length and decision complexity are 
     \begin{align*}
         \ell_{V_1,A}(n) = d_{V_1, A}(n) = \polylog(d_{V,A}(n)) \cdot \polylog(d_{V,A}(n)) = \polylog(n).
     \end{align*}
     The new sampling time is
     \begin{align*}
         t_{V_1,Q}(n) = \polylog(d_{V,A}(n)) \cdot (t_{V,Q}(n) + \polylog(d_{V,A}(n))) = \poly(n).
     \end{align*}
     The new verification time is 
     \begin{align*}
         t_{V_1,A}(n) = \polylog(d_{V,A}(n)) \cdot (t_{V,Q}(n) + t_{V,A}(n) + \polylog(d_{V,A}(n))) = \poly(n).
     \end{align*}
     Lastly, $V$ has completeness 1 and soundness at most $1/2$, so is $V_1$.
    
    \textbf{Step 2.} We apply \cref{thm:newar-pr} again and denote the new protocol by $V_2 = (\Alg_{Q_2}, \Alg_{A_2})$.
    Since $d_{V_1,A}(n) = \polylog(n)$ and $k(n)= \polylog(d_{V_1,A}(n)) = \polyloglog(n)$,
    by \cref{thm:newar-pr}, the new question length is
     \begin{align*}
         \ell_{V_2,Q}(n) = \polylog(d_{V_1,A}(n)) \cdot (2 \ell_{V_1,Q}(n) + \polylog(d_{V_1,A}(n))) = \poly(n).
     \end{align*}
     The new answer length and decision complexity are 
     \begin{align*}
         \ell_{V_2,A}(n) = d_{V_2, A}(n) = \polylog(d_{V_1,A}(n)) \cdot \polylog(d_{V_1,A}(n)) = \polyloglog(n).
     \end{align*}
     The new sampling time is
     \begin{align*}
         t_{V_2,Q}(n) = \polylog(d_{V_1,A}(n)) \cdot (t_{V_1,Q}(n) + \polylog(d_{V_1,A}(n))) = \poly(n).
     \end{align*}
     The new verification time is 
     \begin{align*}
         t_{V_2,A}(n) = \polylog(d_{V_1,A}(n)) \cdot (t_{V_1,Q}(n) + t_{V_1,A}(n) + \polylog(d_{V_1,A}(n))) = \poly(n).
     \end{align*}
    Moreover, $V_2$ has perfect completeness and soundness at most $1/2$.
    
    \textbf{Step 3.}
    The protocol $V_2$ is oracularized again. 
    Again,  since oracularization only introduces constant overhead in the question length, answer length, sampling time and verification time, we still use $V_2$ to denote the oracularized protocol.
    Define the language $L_{\Enc}$ as in \cref{def:L_enc} for $V_2$.
    We can calculate the parameters in \cref{def:params}.
    First, $L_{\Enc} \in \mathrm{DTIME}(T(n))$ where 
    \begin{align*}
        T(n) = t_{V_2,A}(n) + 2^{\ell_{V_2,A}(n)} = \poly(n),
    \end{align*}
    where the decoder
    $\Dec_{\ell_{V_2,A}(n)}$ takes $O(2^{\polyloglog(n)})$ time.
    Moreover, the query lengths are
    \begin{align*}
        \ell_1(n) = \ell_{V_2,A}(n) = \polyloglog(n) && \ell_2(n) = \ell_\pi(n) = T(n)\log_2(T(n)) = \poly(n).
    \end{align*}
    Then the PCPP question sampling and verification times are
    \begin{align*}
        &t_{\PCPP,Q}(n) = t_{V_2,Q}(n) + \poly(n+\ell_{V_2,Q}(n), \ell_{V_2,A}(n), \log_2(T(n))) +O(\ell_1(n)+\ell_2(n)) = \poly(n), \\
        &t_{\PCPP,A}(n) = \poly(n+\ell_{V_2,Q}(n), \ell_{V_2,A}(n), \log_2(T(n))) = \poly(n). 
    \end{align*}
    Next, we apply the answer reduction technique in \cref{thm:answer_reduce} to get verifier $V^{AR}$.
    By \cref{thm:answer_reduce},
    \begin{align*}
        &\ell_{V^{AR},Q}(n) = O(\ell_{V_2,Q}(n) + \ell_1(n) + \ell_2(n)) = \poly(n), \\
        &\ell_{V^{AR},A}(n) = O(1), \\
        &t_{V^{AR},Q}(n) = O(t_{V_2,Q}(n) + \ell_1(n) + \ell_2(n) + t_{\PCPP,Q}(n)) = \poly(n), \\
        &t_{V^{AR},A}(n) = O(t_{\PCPP,A}(n)) = \poly(n).
    \end{align*} 
    Moreover, it is easy to see that the new decision complexity is $d_{V^{AR},A}(n) = O(1)$.
    The perfect completeness and constant soundness of $V^{AR}$ follow from \cref{thm:answer_reduce}.

Alternatively, we can apply the answer reduction technique of \cref{thm:newar-pr} iteratively until the answer size is constant.
The proof follows the same line of argument in the proof of \cite[Theorem 54]{10.1145/3564246.3585208}, so we only sketch the proof here.
The sampler $\Alg_Q$ and decider $\Alg_{A}$ both start by calculating the description of $V_0 = (\Alg_{Q_0}, \Alg_{A_0})$, which is an $\MIP^*[\poly,\poly]$ protocol for $\RE$, then repeatedly applying the answer reduction procedure from \cref{thm:newar-pr} followed by parallel repetition and oracularization to calculate the description of $V_{i+1}$
from the description of $V_i$ for $i \geq 0$ until $V_m$ has answer size $O(1)$.
Then $\Alg_Q$ executes $\Alg_{Q_m}$ to sample the questions.
When the answers are returned, the decider $\Alg_{A}$ executes $\Alg_{A_m}$ to check the answers.

Following the same analysis, we can get $m = O(\log\log\log(\ell_{V,A}(n)))$.
Besides the $O(1)$ answer size and decision complexity, the question size, sampling time, and verification time of $V_m$ are:

\textbf{Question size.} The question size follows the recursive relation 
\begin{align*}
    \ell_{V_{i+1},Q}(n) =  \polylog(d_{V_i,A}(n))(2 \ell_{V_i, Q}(n) +  \polylog(d_{V_i,A}(n))).
\end{align*}
Since $d_{V_{i+1},A}(n) = \polylog(d_{V_i,A}(n))$, we can bound 
\begin{align*}
    \ell_{V_m, Q}(n) &= \polylog(d_{V_{m-1},A}(n))(2 \ell_{V_{m-1},Q}(n) + \polylog(d_{V_{m-1},A}(n))) \\
    &= 2^m \cdot \prod_{i=0}^{m-1} [\polylog(d_{V_i,A}(n))] \cdot \ell_{V_0, Q}(n) + \sum_{i=0}^{m-1} 2^{m-1-i} \prod_{j=i}^{m-1} [\polylog(d_{V_j,A}(n))] \polylog(d_{V_i,A}(n)) \\
    &\leq 2^m \polylog(n)\ell_{V_0, Q}(n) + 2^m m \polylog(n) \polylog(d_{V_0,A}(n))) \\
    &= O(\polylog(n) \poly(n) + \polylog(n)) = O(\poly(n)),
\end{align*}
where we upper bound $\prod_{i=0}^{m-1} [\polylog(d_{V_i,A}(n))] = \polylog(n) \polyloglog(n) \ldots O(1)$ by
$\polylog(n) \cdot \polyloglog(n)^m = \polylog(n)$.

\textbf{Sampling time.} It takes $m$ iterations for the sampler to calculate $V_m$. 
The $(i+1)$th iteration takes time $O( \abs{\Alg_{Q_i}}+ \abs{\Alg_{A_i}}) = O(\abs{\Alg_{Q}} + \abs{\Alg_{A}} +i)$.
    Hence, the total computation time is $O( m(\abs{\Alg_{Q}} + \abs{\Alg_{A}}) + m^2)$.
The running time of $\Alg_{Q_m}$ follows the relation
\begin{align*}
    t_{V_m,Q}(n) &= \polylog(d_{V_{m-1},A}(n)) \cdot (t_{V_{m-1},Q}(n) + \polylog(d_{V_{m-1},A}(n))) \\
    &= \ldots \\
    &= \prod_{i=0}^{m-1} [\polylog(d_{V_i,A}(n))] \cdot t_{V_0, Q}(n) + \sum_{i=0}^{m-1} \prod_{j=i}^{m-1} [\polylog(d_{V_j,A}(n))] \cdot 
    \polylog(d_{V_{i},A}(n)) \\
    &\leq \polylog(n) t_{V_0, Q}(n) + m \polylog(n) \polylog(d_{V_0,A}(n)) = \poly(n).
\end{align*}
Hence, the total sampling time of $V_m$ is $O( m(\abs{\Alg_{Q}} + \abs{\Alg_{A}}) + m^2 + \poly(n)) = \poly(n)$.

\textbf{Verification time.} Similar to the previous case, the time to calculate $V_m$ is
$O( m(\abs{\Alg_{Q}} + \abs{\Alg_{A}}) + m^2)$.
The verification time of $V_m$ is
\begin{align*}
    t_{V_m,A}(n) &= \polylog(d_{V_{m-1},A}(n))(t_{V_{m-1},Q}(n) + t_{V_{m-1},A}(n) + \polylog(d_{V_{m-1},A}(n))) \\
    & = \ldots \\
    & = \prod_{i=0}^{m-1} [\polylog(d_{V_i,A}(n))] \cdot (t_{V_0, Q}(n) + t_{V_0,A}(n)) + \sum_{i=0}^{m-1} \prod_{j=i}^{m-1} [\polylog(d_{V_j,A}(n))] \cdot 
    \polylog(d_{V_{i},A}(n)) \\
    & \leq \polylog(n) (t_{V_0, Q}(n) + t_{V_0,A}(n)) + m \polylog(n) \polylog(d_{V_0,A}(n)) = \poly(n).
\end{align*}
Hence the total verification time is $O( m(\abs{\Alg_{Q}} + \abs{\Alg_{A}}) + m^2 + \poly(n)) = \poly(n)$.
Lastly, the protocol $V_m$ has completeness 1 and soundness at most $1/2$.

\end{proof}

\begin{remark}
We have just shown how to use our new answer reduction techniques to get very small answer sizes, without a large overhead in question length. A natural question that arises is whether this can be applied to the protocol of \cite{ 10.1145/3564246.3585208}, which has \emph{constant} question length but polylogarithmic answer length, in order to obtain a protocol with total communication that scales as $O(\poly\log\log(n))$. This would contradict the lower bound in \cite{ 10.1145/3564246.3585208}, which shows that $\RE$ (or indeed $\mathrm{EEXP}$) cannot be decided by $\MIP^*$ protocols with total communication smaller than $\log(n)$. Indeed, our tighter answer reduction fails to give such a result when applied to the constant-question-size protocol of \cite{ 10.1145/3564246.3585208}. This is because of the phenomenon described in \Cref{rem:qr-decision-complexity}: an application of question reduction resets the decision complexity to be $\poly(n)$, so in particular, the protocol from that work has decision complexity $\poly(n)$, and applying answer reduction to it would blow up both the question size and answer size to $\poly\log(n)$.
\end{remark}

\bibliographystyle{alpha}
\bibliography{references}

\appendix

\section{Lemmas for Noisy \texorpdfstring{$\mathrm{MIP}^\ast$}{MIP*}}
\paragraph*{Smoothing.} The following lemma reduces the degrees of the POVMs of an $\MIP^\ast$ strategy.
\begin{lemma}\label{lem:smoothing of strategies}\cite[Lemma 6.1]{qin2021nonlocal}\footnote{The statement is slightly different from that in \cite[Lemma 6.1]{qin2021nonlocal}. The difference arises due to our relocation of the truncating step, which was in \cite[Lemma 10.5]{qin2021nonlocal}.}
	Given parameters $0\leq\rho<1$, $0<\delta<1$, $n,m\in\posint$, $m\geq2$, and an $m$-dimensional noisy MES $\psi_{AB}$  with the quantum maximal correlation $\rho=\rho\br{\psi_{AB}}$, there exists $d=d\br{\rho,\delta}$ and a map $f:\H_m^{\otimes n}\rightarrow \H_m^{\otimes n},$ such that for any positive semi-definite matrices $P, Q\in\H_m^{\otimes n}$ satisfying $\nnorm{P}_2\leq1$ and $\nnorm{Q}_2\leq1$. The matrices $P^{(1)}=f\br{P}$ and $Q^{(1)}=f\br{Q}$ satisfy that		
		\begin{enumerate}
						\item $P^{(1)}$ and $Q^{(1)}$ are of degree at most $d$.
			\item $\nnorm{P^{(1)}}_2\leq1~\mbox{and}~\nnorm{Q^{(1)}}_2\leq1.$

			\item $\ab{\Tr\br{\br{P^{(1)}\otimes Q^{(1)}}\psi^{\otimes n}_{AB}}-\Tr\br{\br{P\otimes Q}\psi^{\otimes n}_{AB}}}\leq\delta.$
		\item $\frac{1}{m^n}\Tr~\zeta(P^{(1)})\leq\delta $ and $\frac{1}{m^n}\Tr~\zeta(Q^{(1)})\leq\delta. $
            \item the map $f$ is linear and unital.
		\end{enumerate}
		In particular, we can take $d=\frac{C\log^2\frac{1}{\delta}}{\delta(1-\rho)}$ for some absolute constant $C$.
\end{lemma}
\begin{remark}
  It is easily verified that for the above lemma,
  for each $\sigma\in[m^2]^n_{\ge0}$, we have
  $$\abs{\widehat{P}^{(1)}(\sigma)}\le\abs{\widehat{P}(\sigma)}~\mbox{and}~\abs{\widehat{Q}^{(1)}(\sigma)}\le\abs{\widehat{Q}(\sigma)}.$$
  This is because in fact $f$ applies depolarizing noise on $P$ and then eliminates the high degree parts. So the Fourier coefficients are non-increasing in absolute value.
\end{remark}

\paragraph*{Regularization.} The following lemma allows us to identify high-influence registers,
and the number of such registers can be upper-bounded.

\begin{lemma}\label{lem:regular}\cite[Lemma 7.4]{qin2021nonlocal}
	Given $0<\tau<1$, $d,n,m\in\posint$, $m\geq2$, and a degree-$d$ matrix $P\in\H_m^{\otimes n}$ satisfying $\nnorm{P}_2\leq1$, there exists a subset $H\subseteq[n]$ of size $h=\abs{H}\leq\frac{d}{\tau}$ such that for any $i\notin H$,
$$\infi{P^{\leq d}}\leq\tau.$$ 
\end{lemma}

\paragraph*{Rounding.} The following lemma shows that we can round a given set of matrices that sum up to $\id$
to a close-by POVM.
\begin{lemma}\label{lem:closedelta}
	Given $\vec{X}\in\br{\H_m^{\otimes n}}^t$ satisfying that $\sum_{i=1}^tX_i=\id$, define
	$$\R\br{\vec{X}}=\arg\min\set{\nnorm{\vec{X}-\vec{P}}_2^2:\vec{P}~\mbox{is a POVM}}$$
	
	It holds that
	
	$$\nnorm{\R\br{\vec{X}}-\vec{X}}_2^2\leq\frac{3(t+1)}{m^n}\sum_{i=1}^t\Tr~\zeta(X_i)+6\br{\frac{t}{m^n}\sum_{i=1}^t\Tr~\zeta(X_i)}^{1/2}.$$
\end{lemma}

\paragraph*{Miscellaneous Lemmas.} The following lemmas are used throughout \cref{sec:upperbound}.

	\begin{fact}\label{fac:cauchyschwartz}\cite[Fact 2.1]{qin2021nonlocal}
		Given registers $A, B$, operators $P\in\H\br{A}, Q\in\H\br{B}$ and a bipartite state $\psi_{AB}$, it holds that
		\[\abs{\Tr\br{\br{P\otimes Q}\psi_{AB}}}\leq\br{\Tr \,P^2\psi_A}^{1/2}\cdot\br{\Tr \,Q^2\psi_B}^{1/2}.\]
	\end{fact}

\begin{lemma}\label{lem:maincauchy}
Let $\set{\Pss}^{x\in\X}_{a\in\A},\set{\Qss}^{y\in\Y}_{b\in\B}$, $\set{\Pwss}^{x\in\X}_{a\in\A},\set{\Qwss}^{y\in\Y}_{b\in\B}\subseteq\herspace{n}$ be four sets of matrices. If for all \fourtuples, 
\[\abs{\Tr\br{\br{\Pss\otimes \Qss}\psi_{AB}^{\otimes n}}-\Tr\br{\br{\Pwss\otimes \Qwss}\psi_{AB}^{\otimes n}}}\leq\delta\nnorm{\Pss}_2\nnorm{\Qss}_2\]
for some $\delta>0$. 
Then
\[\ab{\gval{n}{\Pss}{\Qss}-\gval{n}{\Pwss}{\Qwss}}\leq\delta t\br{\sum_{x,a}\mu_A(x)\nnorm{\Pss}_2^2}^{1/2}\br{\sum_{y,b}\mu_B(y)\nnorm{\Qss}_2^2}^{1/2}.\]
\end{lemma}

\begin{proof}
\begin{align*}
&\ab{\gval{n}{\Pss}{\Qss}-\gval{n}{\Pwss}{\Qwss}}\\
\leq~&\sum_{x,y,a,b}\mu(x,y)\abs{\Tr\br{\br{\Pss\otimes \Qss}\psi_{AB}^{\otimes n}}-\Tr\br{\br{\Pwss\otimes \Qwss}\psi_{AB}^{\otimes n}}}\\
\leq~&\delta\sum_{x,y,a,b}\mu(x,y)\nnorm{\Pss}_2\nnorm{\Qss}_2\\
\leq~&\delta\br{\sum_{x,y,a,b}\mu(x,y)\nnorm{\Pss}_2^2}^{1/2}\br{\sum_{x,y,a,b}\mu(x,y)\nnorm{\Qss}_2^2}^{1/2}\quad\mbox{(Cauchy Schwarz)}\\
=~&\delta t\br{\sum_{x,a}\mu_A(x)\nnorm{\Pss}_2^2}^{1/2}\br{\sum_{y,b}\mu_B(y)\nnorm{\Qss}_2^2}^{1/2}.
\end{align*}
\end{proof}

\section{Deferred Proofs of Section \ref{sec:lemma_mip}}
\label{sec:ans}
\begin{proof}[Proof of \cref{lm:close}]
    We assume $\set{A^x_a}$ is projective. Then 
	\begin{align*}
		\bigE_x \sum_a \bra{\psi} \id \x B^x_a \ket{\psi} \geq \bigE_x \sum_a \bra{\psi} \id \x (B^x_a)^2  \ket{\psi} \geq 0,
	\end{align*}
	which implies that 
	\begin{align*}
		\abs{ \bigE_x \sum_a \bra{\psi} A^x_a \x \id \ket{\psi} -\bra{\psi} \id\x B^x_a \ket{\psi}} \leq 
		\abs{ \bigE_x \sum_a \bra{\psi} A^x_a \x \id \ket{\psi} -\bra{\psi} \id\x (B^x_a)^2 \ket{\psi}}.
	\end{align*}
	We can bound the second quantity in two steps.
	\begin{align*}
		&\abs{ \bigE_x \sum_a \bra{\psi} A^x_a\x \id \ket{\psi} - \bra{\psi} A^x_a \x B^x_a \ket{\psi}} \\
		&\leq \sqrt{\bigE_x \sum_a \norm{ A^x_a\ket{\psi}}^2} \sqrt{ \bigE_x \sum_a \norm{ (A^x_a \x \id - \id\x B^x_a)\ket{\psi}}^2} \leq \sqrt{\delta},
	\end{align*}
	and similarly
	\begin{align*}
		\abs{ \bigE_x \sum_a \bra{\psi} A^x_a\x B^x_a \ket{\psi} - \bra{\psi} \id \x (B^x_a)^2 \ket{\psi}} \leq \sqrt{\delta}.
	\end{align*}
        By the triangle inequality, the second quantity is at most $2\sqrt{\delta}$. So is the first one.
\end{proof}

\begin{proof}[Proof of \cref{lm:k_prod}]
	We start with 
	\begin{align*}
		A^x_{a_1, \ldots, a_k} = A^x_{a_k} \cdots A^x_{a_2} A^x_{a_1} A^x_{a_2} \cdots A^x_{a_k}.
	\end{align*}
	Because $A^x_{a_k} \x \id \appd{\delta} \id \x (B_k)^x_{a_k}$, 
	To apply \cref{lm:prod_meas}, 
	we can set $C^x_{a,b} = 
	A^x_{a_k} \cdots A^x_{a_2} A^x_{a_1} A^x_{a_2} \cdots A^x_{a_{k-1}} \x \id$ 
	with $a = a_k$ and $b = (a_1, \ldots, a_{k-1})$.
	Then $\sum_b (C^x_{a,b})\ct C^x_{a,b} \leq \id$.
	Hence by \cref{lm:prod_meas}
	\begin{align*}
		A^x_{a_1, \ldots, a_k} \x \id \appd{\delta} A^x_{a_k} \cdots A^x_{a_2} A^x_{a_1} A^x_{a_2} \cdots A^x_{a_{k-1}} \x (B_k)^x_{a_k}
	\end{align*}
	We can apply \cref{lm:prod_meas} again with $C^x_{a,b} = A^x_{a_k} \cdots A^x_{a_2} A^x_{a_1} A^x_{a_2} \cdots A^x_{a_{k-2}} \x B_k^{(a_k)}$ with $a= a_{k-1}$ and $b = (a_1,\ldots, a_{k-2}, a_k)$.
	Because $A^x_{a_{k-1}} \x \id \appd{\delta} \id \x (B_{k-1})^{(a_{k-1})}$, we can get that
	\begin{align*}
		A^x_{a_k} \cdots A^x_{a_2} A^x_{a_1} A^x_{a_2} \cdots A^x_{a_{k-1}} \x (B_k)^x_{a_k}
		\appd{\delta} A^x_{a_k} \cdots A^x_{a_2} A^x_{a_1} A^x_{a_2} \cdots A^x_{a_{k-2}}  \x (B_k)^x_{a_k}(B_{k-1})^x_{a_{k-1}}.
	\end{align*}
	Continuing similarly, we can get that
	\begin{align*}
	A^x_{a_k} \cdots A^x_{a_2} A^x_{a_1} \x  (B_k)^x_{a_k} \cdots (B_2)^x_{a_2}
	 \appd{\delta} 
	A^x_{a_k} \cdots A^x_{a_2}  \x (B_k)^x_{a_k} \cdots (B_1)^x_{a_1}.
	\end{align*}
	With another $(k-2)$ steps we can get that 
	\begin{align*}
		A^x_{a_k} \x (B_k)^x_{a_k} \cdots (B_2)^x_{a_2} (B_1)^x_{a_1} (B_2)^x_{a_2} \cdot  (B_{k-1})^x_{a_{k-1}}\appd{\delta}
		\id \x (B_k)^x_{a_k} \cdots (B_2)^x_{a_2} (B_1)^x_{a_1} (B_2)^x_{a_2} \cdot  (B_{k})^x_{a_{k}}. 
	\end{align*}
	Combining all the steps above with \cref{lm:chain_rule}
	\begin{align*}
		A^x_{a_1, \ldots, a_k} \x \id  \appd{(2k-1)^2\delta} \id \x
		(B_k)^x_{a_k} \cdots (B_2)^x_{a_2} (B_1)^x_{a_1} (B_2)^x_{a_2} \cdot  (B_{k})^x_{a_{k}},
	\end{align*}
	which completes the proof.
\end{proof}

\begin{proof}[Proof of \cref{lm:k_prod_distr}, the original proof]
	We first show the $k=2$ case. Notice that
	\begin{align*}
		J^{x,y_1,y_2}_{[g_1(y_1),g_2(y_2) = a_1, a_2]} = 
		\sum_{g_2: g_2(y_2)=a_2} (G_2)^x_{g_2} \left(\sum_{g_1: g_1(y_1) = a_1} (G_1)^x_{g_1}\right) (G_2)^x_{g_2}.
	\end{align*}
	Our goal is to bound
	\begin{align*}
		&\bigE_{x,y_1,y_2} \sum_{a_1, a_2} \bra{\psi} A^{x,y_1,y_2}_{a_1,a_2} \x J^{x,y_1,y_2}_{[g_1(y_1),g_2(y_2) = a_1, a_2]} \ket{\psi} \\
		&= \bigE_{x,y_1,y_2} \sum_{a_1, a_2} \bra{\psi} A^{x,y_1,y_2}_{a_1,a_2} \x 
		\sum_{g_2: g_2(y_2)=a_2} (G_2)^x_{g_2} (G_1)^x_{[g_1(y_1) = a_1]}(G_2)^x_{g_2} \ket{\psi} \\
		&=\bigE_{x,y_1,y_2} \sum_{a_1, g_2 }\bra{\psi} A^{x,y_1,y_2}_{a_1,g_2(y_2) } \x 
		 (G_2)^x_{g_2} (G_1)^x_{[g_1(y_1) = a_1]}(G_2)^x_{g_2} \ket{\psi}.
	\end{align*}
	First notice that 
	\begin{align*}
		\bigE_{x,y_1,y_2} \sum_{a_1, g_2 }\bra{\psi} A^{x,y_1,y_2}_{a_1 , g_2(y_2) } \x 
		 (G_2)^x_{g_2} (G_1)^x_{[g_1(y_1) = a_1]} \ket{\psi}
		 \appd{2\sqrt{2\delta}} 
		 \bigE_{x,y_1,y_2} \sum_{a_1, a_2} \bra{\psi} A^{x,y_1,y_2}_{a_1,a_2} \x \id \ket{\psi} = 1.
	\end{align*}
	This is because
	\begin{align*}
		&\abs{ \bigE_{x,y_1,y_2} \sum_{a_1, g_2 }\bra{\psi} A^{x,y_1,y_2}_{a_1, g_2(y_2) }  \x 
		 (G_2)^x_{g_2}  \ket{\psi} -
		  \bra{\psi} A^{x,y_1,y_2}_{a_1, g_2(y_2) } \x
		 (G_2)^x_{g_2} (G_1)^x_{[g_1(y_1) = a_1]} \ket{\psi}} \\
		& =  \abs{\bigE_{x,y_1,y_2}  \sum_{a_1, g_2} \bra{\psi} A^{x,y_1,y_2}_{a_1, g_2(y_2) } \x (G_2)^x_{g_2} 
		(A^{x,y_1,y_2}_{a_1} \x\id -\id\x(G_1)^x_{[g_1(y_1) = a_1]})\ket{\psi} }\\
		& \leq \sqrt{ \bigE_{x,y_1,y_2} \sum_{a_1, g_2} \norm{ A^{x,y_1,y_2}_{a_1, g_2(y_2) } \x (G_2)^x_{g_2} \ket{\psi}}^2} \cdot \\
		&\quad 
		\sqrt{\bigE_{x,y_1,y_2} \sum_{a_1, g_2} \bra{\psi} (A^{x,y_1,y_2}_{a_1} \x\id -\id\x(G_1)^x_{[g_1(y_1) = a_1]})  A^{x,y_1,y_2}_{a_1, g_2(y_2) }  (A^{x,y_1,y_2}_{a_1} \x\id -\id\x(G_1)^x_{[g_1(y_1) = a_1]})\ket{\psi}} \\
		& \leq \sqrt{ \bigE_{x,y_1,y_2} \sum_{a_1,g_2} \norm{ A^{x,y_1,y_2}_{a_1, g_2(y_2) } \x (G_2)^x_{g_2} \ket{\psi}}^2} \cdot \\
		&
		\sqrt{ \bigE_{x,y_1,y_2} \sum_{a_1} \bra{\psi} (A^{x,y_1,y_2}_{a_1} \x\id -\id\x(G_1)^x_{[g_1(y_1) = a_1]}) \sum_{g_2} A^{x,y_1,y_2}_{a_1, g_2(y_2) }
		(A^{x,y_1,y_2}_{a_1} \x\id -\id\x(G_1)^x_{[g_1(y_1) = a_1]})\ket{\psi}}	\\
		&\leq 1 \cdot \sqrt{ \bigE_{x,y_1,y_2} \sum_{a_1} \norm{(A^{x,y_1,y_2}_{a_1} \x\id -\id\x(G_1)^x_{[g_1(y_1) = a_1]})\ket{\psi}}^2} \\
		&\leq \sqrt{2\delta}
	\end{align*}
	and
	\begin{align*}
		&\abs{\bigE_{x,y_1,y_2} \sum_{a_1, g_2 }\bra{\psi} A^{x,y_1,y_2}_{a_1, g_2(y_2) }  \x 
		 (G_2)^x_{g_2}  \ket{\psi} - \bra{\psi} A^{x,y_1,y_2}_{a_1,g_2(y_2)} \x \id \ket{\psi}} \\
		 &= \abs{ \bigE_{x,y_1,y_2} \sum_{a_1, a_2 } \bra{\psi} A^{x,y_1,y_2}_{a_1, a_2 } \cdot
		 (\id \x (G_2)^x_{[g_2(y_2)=a_2]} - A^{x,y_1,y_2}_{a_2} \x \id) \ket{\psi}} \\
		 &\leq \sqrt{\bigE_{x,y_1,y_2} \sum_{a_1,a_2} \norm{ A^{x,y_1,y_2}_{a_1, a_2 } \ket{\psi}}^2} 
		 \cdot \\
		 & \sqrt{\bigE_{x,y_1,y_2} \sum_{a_1,a_2}\bra{\psi}  (\id \x (G_2)^x_{[g_2(y_2)=a_2]} - A^{x,y_1,y_2}_{a_2} \x \id) 
		 A^{x,y_1,y_2}_{a_1, a_2 } (\id \x (G_2)^x_{[g_2(y_2)=a_2]} - A^{x,y_1,y_2}_{a_2} \x \id) \ket{\psi}} \\
		 & \leq \sqrt{ \bigE_{x,y_1,y_2} \sum_{a_1,a_2} \norm{ A^{x,y_1,y_2}_{a_1, a_2 } \ket{\psi}}^2} \cdot \\
		 & \sqrt{  \bigE_{x,y_1,y_2} \sum_{a_2} \bra{\psi}  (\id \x (G_2)^x_{[g_2(y_2)=a_2]} - A^{x,y_1,y_2}_{a_2} \x \id) 
		 \sum_{a_1} A^{x,y_1,y_2}_{a_1, a_2 } (\id \x (G_2)^x_{[g_2(y_2)=a_2]} - A^{x,y_1,y_2}_{a_2} \x \id) \ket{\psi}} \\
		 & \leq 1 \cdot \sqrt{ \bigE_{x,y_1,y_2} \sum_{a_2} \norm{ (\id \x (G_2)^x_{[g_2(y_2)=a_2]} - A^{x,y_1,y_2}_{a_2} \x \id) \ket{\psi}}^2} \\
		 & \leq \sqrt{2\delta},
	\end{align*}
	Hence, we focus on proving 
	\begin{align}
        \label{eq:g_1g_2}
		\bigE_{x,y_1,y_2} \sum_{a_1, g_2} \norm{ \id \x \left((G_1)^x_{[g_1(y_1) = a_1]}(G_2)^x_{g_2} - (G_2)^x_{g_2}(G_1)^x_{[g_1(y_1) = a_1]}\right) \ket{\psi}}^2 \leq C_1\sqrt{\delta} +C_2 \ep
	\end{align}
	for some constants $C_1$ and $C_2$, which will imply that 
	\begin{align*}
		& \abs{ \bigE_{x,y_1,y_2} \sum_{a_1, g_2} 
		\bra{\psi} A^{x,y_1,y_2}_{a_1,g_2(y_2) } \x 
		(G_2)^x_{g_2}  \left(  (G_1)^x_{[g_1(y_1) = a_1]}(G_2)^x_{g_2} - (G_2)^x_{g_2}(G_1)^x_{[g_1(y_1) = a_1]} \right)  \ket{\psi} } \\
	& \leq \sqrt{  \bigE_{x,y_1,y_2} \sum_{a_1, g_2} \norm{  A^{x,y_1,y_2}_{a_1,g_2(y_2) } \x 
		(G_2)^x_{g_2} \ket{\psi}}^2} \cdot \\
		&\sqrt{ \bigE_{x,y_1,y_2} \sum_{a_1, g_2} \norm{ \bra{\psi} \id\x \left(  (G_1)^x_{[g_1(y_1) = a_1]}(G_2)^x_{g_2} - (G_2)^x_{g_2}(G_1)^x_{[g_1(y_1) = a_1]} \right)  \ket{\psi}}^2} \\
		&\leq \sqrt{ C_1 \sqrt{\delta} + C_2\ep}
	\end{align*}
	and
	\begin{align*}
		\abs{	\bigE_{x,y_1,y_2} \sum_{a_1, a_2} \bra{\psi} A^{x,y_1,y_2}_{a_1,a_2} \x J^{x,y_1,y_2}_{[g_1(y_1),g_2(y_2) = a_1, a_2]} \ket{\psi}-1}
		\leq 2\sqrt{2\delta} + \sqrt{ C_1 \sqrt{\delta} + C_2\ep}.
	\end{align*}
	
	To prove \cref{eq:g_1g_2}, we start with \cref{eq:A_G_consist}
	\begin{align*}
		\bigE_{x,y_1,y_2} \sum_{a_i} \norm{ (A^{x,y_1,y_2}_{a_i} \x\id - \id\x (G_i)^x_{[g_i(y_i)=a_i]})\ket{\psi}}^2 \leq 2\delta
	\end{align*}
	for $i= 1,2$.
	Then by \cref{lm:prod_meas}
	\begin{align*}
		&\id \x (G_1)^x_{[g_1(y_1)=a_1]} (G_2)^x_{[g_2(y_2)=a_2]}  \ket{\psi} \\
		&\appd{2\delta} A^{x,y_1,y_2}_{a_2} \x  (G_1)^x_{[g_1(y_1)=a_1]}  \ket{\psi} \\
		&\appd{2\delta}  A^{x,y_1,y_2}_{a_2} A^{x,y_1,y_2}_{a_1} \x \id \ket{\psi} \\
		&= A^{x,y_1,y_2}_{a_1} A^{x,y_1,y_2}_{a_1} \x \id \ket{\psi} \\
		&\appd{2\delta}  A^{x,y_1,y_2}_{a_2} \x (G_2)^x_{[g_2(y_2)=a_2]} \ket{\psi} \\
		&\appd{2\delta} \id \x  (G_2)^x_{[g_2(y_2)=a_2]} (G_1)^x_{[g_1(y_1)=a_1]} \ket{\psi}.
	\end{align*}
	Chaining the inequalities together using \cref{lm:chain_rule} gives
	\begin{align*}
		\bigE_{x,y_1,y_2} \sum_{a_1,a_2} \norm{ \id \x  \left( (G_1)^x_{[g_1(y_1)=a_1]} (G_2)^x_{[g_2(y_2)=a_2]} - (G_1)^x_{[g_1(y_1)=a_1]} (G_2)^x_{[g_2(y_2)=a_2]} \right)  \ket{\psi}}^2 \leq 32 \delta.
	\end{align*}
	Let 
	\begin{align*}
	S_1 &= \bigE_{x,y_1,y_2} \sum_{a_1, g_2} \norm{ \id \x \left((G_1)^x_{[g_1(y_1) = a_1]}(G_2)^x_{g_2} - (G_2)^x_{g_2}(G_1)^x_{[g_1(y_1) = a_1]}\right) \ket{\psi}}^2 \\
	S_2 &= \bigE_{x,y_1,y_2} \sum_{a_1,a_2} \norm{ \id \x \left( (G_1)^x_{[g_1(y_1)=a_1]} (G_2)^x_{[g_2(y_2)=a_2]} - (G_1)^x_{[g_1(y_1)=a_1]} (G_2)^x_{[g_2(y_2)=a_2]} \right)  \ket{\psi}}^2.
	\end{align*}
	We are going to show that $S_1$ is close to $S_2$. 
	Expanding $S_1 - S_2$, we get $\abs{ S_1 - S_2 } \leq \Delta_1 + \Delta_2 + \Delta_3 + \Delta_4$,
	where
	\begin{align*}
		\Delta_1 &=\abs{ \bigE_{x,y_1,y_2} \sum_{a_1, g_2}
		\bra{\psi} \id \x  (G_2)^x_{g_2} (G_1)^x_{[g_1(y_1) = a_1]} (G_1)^x_{[g_1(y_1) = a_1]} (G_2)^x_{g_2}  \ket{\psi} \\ 
		&\quad -
		 \sum_{a_1, a_2} \bra{\psi}  \id \x (G_2)^x_{[g_2(y_2) = a_2]} 
		(G_1)^x_{[g_1(y_1) = a_1]} (G_1)^x_{[g_1(y_1) = a_1]} (G_2)^x_{[g_2(y_2) = a_2]}  \ket{\psi}} \\
		\Delta_2 &= \abs{ 
		\bigE_{x,y_1,y_2} \sum_{a_1, g_2}
		\bra{\psi} \id \x (G_1)^x_{[g_1(y_1) = a_1]} (G_2)^x_{g_2}  (G_2)^x_{g_2}  (G_1)^x_{[g_1(y_1) = a_1]}  \ket{\psi} \\ 
		&\quad -
		 \sum_{a_1, a_2} \bra{\psi}  
		 \id \x (G_1)^x_{[g_1(y_1) = a_1]}(G_2)^x_{[g_2(y_2) = a_2]} (G_2)^x_{[g_2(y_2) = a_2]} (G_1)^x_{[g_1(y_1) = a_1]}  \ket{\psi}} \\
		\Delta_3 &=\abs{\bigE_{x,y_1,y_2} \sum_{a_1, g_2}
		\bra{\psi}  \id \x (G_2)^x_{g_2} (G_1)^x_{[g_1(y_1) = a_1]}   (G_2)^x_{g_2}  (G_1)^x_{[g_1(y_1) = a_1]}  \ket{\psi} \\ 
		&\quad -
		 \sum_{a_1, a_2} \bra{\psi}  \id \x
		(G_2)^x_{[g_2(y_2) = a_2]} (G_1)^x_{[g_1(y_1) = a_1]}  (G_2)^x_{[g_2(y_2) = a_2]} (G_1)^x_{[g_1(y_1) = a_1]}  \ket{\psi}} \\
		\Delta_4 &=\abs{\bigE_{x,y_1,y_2} \sum_{a_1, g_2}
		\bra{\psi}  \id \x (G_1)^x_{[g_1(y_1) = a_1]}   (G_2)^x_{g_2}  (G_1)^x_{[g_1(y_1) = a_1]} (G_2)^x_{g_2} \ket{\psi} \\ 
		&\quad -
		 \sum_{a_1, a_2} \bra{\psi}  \id \x
		 (G_1)^x_{[g_1(y_1) = a_1]}  (G_2)^x_{[g_2(y_2) = a_2]} (G_1)^x_{[g_1(y_1) = a_1]} (G_2)^x_{[g_2(y_2) = a_2]}  \ket{\psi}}	.
	\end{align*}
	First of all
	\begin{align*}
		\Delta_1 = \abs{ 1 - \bigE_{x,y_1,y_2} \sum_{a_1,a_2}
		\bra{\psi}  \id \x (G_2)^x_{[g_2(y_2) = a_2]} 
		(G_1)^x_{[g_1(y_1) = a_1]} (G_2)^x_{[g_2(y_2) = a_2]}  \ket{\psi}}.
	\end{align*}
	By \cref{eq:G_A_consist},
	\begin{align*}
		\id \x (G_2)^x_{[g_2(y_2) = a_2]} 
		(G_1)^x_{[g_1(y_1) = a_1]} (G_2)^x_{[g_2(y_2) = a_2]}  \ket{\psi}
		\appd{ 18 \delta}  A^{x,y_1,y_2}_{a_1,a_2} \x \id \ket{\psi},
	\end{align*}
	then \cref{lm:close} implies that 
	\begin{align*}
		\abs{ \bigE_{x,y_1,y_2} \sum_{a_1, a_2} \bra{\psi}  A^{x,y_1,y_2}_{a_1,a_2} \x \id \ket{\psi} - \bra{\psi} \id \x (G_2)^x_{[g_2(y_2) = a_2]} 
		(G_1)^x_{[g_1(y_1) = a_1]} (G_2)^x_{[g_2(y_2) = a_2]} \ket{\psi} } \leq 6\sqrt{2\delta}.
	\end{align*}
	Since $\bigE_{x,y_1,y_2} \sum_{a_1, a_2} \bra{\psi} \id \x A^{x,y_1,y_2}_{a_1,a_2} \ket{\psi} = 1$, 
	$\Delta_1 \leq 6\sqrt{2\delta}$.
	Next, observe that $\Delta_2 = 0$ as $(G_2)^x_{g_2}$ and $(G_2)^x_{[g_2(y_2)= a_2]}$ are projective measurements.
	Lastly, observe that $\Delta_3 = \Delta_4$, so we focus on bounding $\Delta_3$.
	First notice that 
	\begin{align*}
		&\bigE_{x,y_1,y_2} \sum_{a_1, g_2} \bra{\psi} 
		\id \x (G_2)^x_{g_2} (G_1)^x_{[g_1(y_1) = a_1]}   (G_2)^x_{g_2}  (G_1)^x_{[g_1(y_1) = a_1]}  \ket{\psi} \\
		 &\appd{ 3\sqrt{2\delta}}
		\bigE_{x,y_1,y_2} \sum_{a_1, g_2} \bra{\psi}(G_1)^x_{[g_1(y_1) = a_1]} \x  (G_2)^x_{g_2} (G_1)^x_{[g_1(y_1) = a_1]}   (G_2)^x_{g_2}  \ket{\psi} \\
		&\bigE_{x,y_1,y_2} \sum_{a_1, a_2} \bra{\psi}  \id \x 
		(G_2)^x_{[g_2(y_2) = a_2]} (G_1)^x_{[g_1(y_1) = a_1]}  (G_2)^x_{[g_2(y_2) = a_2]} (G_1)^x_{[g_1(y_1) = a_1]} \ket{\psi} \\
		& \appd{3\sqrt{2\delta}}  \bigE_{x,y_1,y_2} \sum_{a_1, a_2} \bra{\psi}  (G_1)^x_{[g_1(y_1) = a_1]} \x 
		(G_2)^x_{[g_2(y_2) = a_2]} (G_1)^x_{[g_1(y_1) = a_1]}  (G_2)^x_{[g_2(y_2) = a_2]}  \ket{\psi}
	\end{align*}
	The reason why $\id \x (G_1)^x_{[g_1(y_1) = a_1]} \appd{18\delta}  (G_1)^x_{[g_1(y_1) = a_1]} \x \id$ is the following.
	Applying \cref{lm:consist_to_approx} to \cref{eq:A_G_consist,eq:G_A_consist} we get
	\begin{align*}
		&\bigE_{x,y_1,y_2} \sum_{a_i} \norm{ (A^{x,y_1,y_2}_{a_i} \x \id - \id \x (G_i)^x_{[g_i(y_i)=a_i]})\ket{\psi}}^2 \leq 2\delta \\
		&\bigE_{x,y_1,y_2} \sum_{a_i} \norm{ ((G_i)^x_{[g_i(y_i)=a_i]} \x \id - \id \x  A^{x,y_1, y_2}_{a_i} )\ket{\psi}}^2 \leq 2\delta.
	\end{align*}
	Notice that for any $i \in [2]$, 
	\begin{align*}
		&\bigE_{x,y_1,y_2} \sum_{a_i}  \bra{\psi} A^{x,y_1, y_2}_{a_i} \x A^{x,y_1,y_2}_{a_i} \ket{\psi} \geq
		\bigE_{x,y_1,y_2} \sum_{a_1,a_2} \bra{\psi} A^{x,y_1, y_2}_{a_1,a_2} \x A^{x,y_1, y_2}_{a_1,a_2} \ket{\psi}
		\geq 1 - \delta 
	\end{align*}
	because $A^{x,y_1, y_2}_{a_1,a_2} \x A^{x,y_1, y_2}_{b_1,b_2} \geq 0$ for any $a_1, a_2,b_1,b_2$.
	Then \cref{lm:consist_to_approx} also implies that 
	\begin{align*}
		\bigE_{x,y_1,y_2} \sum_{a_1} \norm{ (A^{x,y_1, y_2}_{a_i} \x \id - \id \x  A^{x,y_1, y_2}_{a_i})
		\ket{\psi}}^2 \leq 2 \delta.
	\end{align*}
	Hence, \cref{lm:chain_rule} implies that for all $i \in [2]$.
	\begin{align*}
		\bigE_{x,y_1,y_2} \sum_{a_i} \norm{ \Big( (G_i)^x_{[g_i(y_i)=a_i]} \x \id - \id \x (G_i)^x_{[g_i(y_i) = a_i]} \Big) \ket{\psi}}^2 \leq 18\delta.
	\end{align*}	
	Also, notice that
	\begin{align*}
		&\abs{ \bigE_{x,y_1,y_2} \sum_{a_1, a_2} \bra{\psi}  
		(G_2)^x_{[g_2(y_2) = a_2]} (G_1)^x_{[g_1(y_1) = a_1]}  (G_2)^x_{[g_2(y_2) = a_2]}  \x  (G_1)^x_{[g_1(y_1) = a_1]} \ket{\psi} \\
		&- \bigE_{x,y_1,y_2} \sum_{a_1, g_2} \bra{\psi}  
		(G_2)^x_{g_2} (G_1)^x_{[g_1(y_1) = a_1]}  (G_2)^x_{g_2}  \x  (G_1)^x_{[g_1(y_1) = a_1]} \ket{\psi}} \\
		&= \abs{  \bigE_{x,y_1,y_2} \sum_{a_1} \sum_{g_2, g_2'} \bra{\psi} (G_2)^x_{g_2} (G_1)^x_{[g_1(y_1) = a_1]}  (G_2)^x_{g_2'}  \x  (G_1)^x_{[g_1(y_1) = a_1]} \ket{\psi} \id[g_2(y_2) = g_2'(y_2)]}  \\
		&\leq \ep \abs{ \bigE_{x,y_1} \sum_{a_1} \bra{\psi} (G_1)^x_{[g_1(y_1) = a_1]} \x (G_1)^x_{[g_1(y_1) = a_1]} \ket{\psi}} \\
		&\leq \ep.
	\end{align*}
	Therefore, $\Delta_3 = \Delta_4 \leq 6\sqrt{2\delta}  + \ep$, and
	\begin{align*}
		\abs{ S_1 - S_2 } \leq \sum_{j = 1}^4 \Delta_j \leq 18\sqrt{2 \delta} + 2 \ep,
	\end{align*}
	and
	\begin{align*}
		S_1 \leq 32\delta + 18\sqrt{2\delta} + 2\ep.
	\end{align*}
	In conclusion,
	\begin{align*}
		&\abs{ \bigE_{x,y_1,y_2} \sum_{a_1, a_2} \bra{\psi} A^{x,y_1,y_2}_{a_1,a_2} \x J^{x,y_1,y_2}_{[g_1(y_1),g_2(y_2) = a_1, a_2]} \ket{\psi}  - 1} \\
		 &\leq 2\sqrt{2\delta} + \sqrt{ 32\delta + 18\sqrt{2\delta} + 2 \ep} \leq 11 \delta^{1/4} + 2 \sqrt{\ep},
	\end{align*}
	and equivalently
	\begin{align*}
		A^{x,y_1,y_2}_{a_1,a_2} \x \id \appd{22 \delta^{1/4} + 4\se} \id \x J^{x,y_1,y_2}_{[g_1(y_1),g_2(y_2) = a_1, a_2]}.
	\end{align*}
	Switching the roles of Alice and Bob, the same proof gives us that 
	\begin{align*}
		 J^{x,y_1,y_2}_{[g_1(y_1),g_2(y_2) = a_1, a_2]} \x \id \appd{22 \delta^{1/4} + 4\se} \id \x A^{x,y_1,y_2}_{a_1,a_2}.
	\end{align*}
	
	For the general case, assume 
	\begin{align*}
		& A^{x,y_1,\ldots,y_i}_{a_1,\ldots,a_i} \x \id \appd{f(i,\delta,\ep)}  \id \x J^x_{[g_1(y_1), \ldots , g_i(y_i) 
		= a_1, \ldots, a_i]} \text{ and }\\
		&\id \x A^{x,y_1,\ldots,y_i}_{a_1,\ldots,a_i} \appd{f(i,\delta,\ep)}   J^x_{[g_1(y_1), \ldots , g_i(y_i) 
		= a_1, \ldots, a_i]} \x \id,
	\end{align*}
	which imply that 
	\begin{align*}
		 \id \x J^x_{[g_1(y_1), \ldots , g_i(y_i)} \appd{3( 2 \delta + 2 f(i,\delta, \ep))}  J^x_{[g_1(y_1), \ldots , g_i(y_i)} \x \id.
	\end{align*}
	Since $\delta$ and $\ep$ are fixed, we write $f(i,\delta, \ep)$ as $f(i)$ in the rest of the proof and proceed to the $i+1$ case.
	As in the base case, our goal is to bound
	\begin{align*}
		&\bigE_{x,y_1,\ldots,y_{i+1}} \sum_{a_1,\ldots, a_{i+1}} \bra{\psi} A^{x,y_1,\ldots,y_{i+1}}_{a_1,\ldots,a_{i+1}} \x J^{x,y_1,\ldots,y_{i+1}}_{[g_1(y_1),\ldots,g_{i+1}(y_{i+1}) = a_1, \ldots,a_{i+1}]} \ket{\psi} \\
		&=\bigE_{x,y_1,\ldots, y_{i+1}} \sum_{a_1, \ldots, a_i, g_{i+1} }\bra{\psi} A^{x,y_1,\ldots,y_{i+1}}_{a_1,\ldots, a_i, g_{i+1}(y_{i+1}) } \x 
		 (G_{i+1})^x_{g_{i+1}} J_{[g_1(y_1),\ldots,g_{i}(y_{i}) = a_1, \ldots,a_{i}]}^{x,y_1,\ldots,y_i}   (G_{i+1})^x_{g_{i+1}} \ket{\psi}.
	\end{align*}
	by relating it to
	\begin{align*}
		&\bigE_{x,y_1,\ldots,y_{i+1}} \sum_{a_1, \ldots, a_i, g_{i+1} }\bra{\psi} A^{x,y_1,\ldots, y_{i+1}}_{a_1,\ldots,a_i, g_{i+1}(y_{i+1}) } \x 
		 (G_{i+1})^x_{g_{i+1}}  J_{[g_1(y_1),\ldots,g_{i}(y_{i}) = a_1, \ldots,a_{i}]}^{x,y_1,\ldots,y_i} \ket{\psi}\\
		 & \appd{\sqrt{2\delta} + \sqrt{f(i)} } 
		 \bigE_{x,y_1,\ldots, y_{i+1}} \sum_{a_1, \ldots, a_{i+1}} \bra{\psi} A^{x,y_1,\ldots,y_{i+1}}_{a_1,\ldots, a_{i+1}} \x \id \ket{\psi} = 1.
	\end{align*}
	So the central step is bounding 
	\begin{align*}
		\bigE_{x,y_1,\ldots,y_{i+1}} \sum_{a_1,\ldots,a_i, g_{i+1}} \norm{ \id \x \left( J_{[g_1(y_1),\ldots,g_{i}(y_{i}) = a_1, \ldots,a_{i}]}^{x,y_1,\ldots,y_i}(G_{i+1})^x_{g_{i+1}} - (G_{i+1})^x_{g_{i+1}} J_{[g_1(y_1),\ldots,g_{i}(y_{i}) = a_1, \ldots,a_{i}]}^{x,y_1,\ldots,y_i} \right) \ket{\psi}}^2.
	\end{align*}
	As in the base case, we can use similar arguments to show
	\begin{align*}
		\bigE_{x,y_1,\ldots,y_{i+1}} \sum_{a_1,\ldots,a_i, g_{i+1}} \norm{ \id \x &\Big( J_{[g_1(y_1),\ldots,g_{i}(y_{i}) = a_1, \ldots,a_{i}]}^{x,y_1,\ldots,y_i}(G_{i+1})^x_{[g_{i+1}(y_{i+1}) = a_{i+1}]} \\
		&- (G_{i+1})^x_{[g_{i+1}(y_{i+1})=a_{i+1}]}   J_{[g_1(y_1),\ldots,g_{i}(y_{i}) = a_1, \ldots,a_{i}]}^{x,y_1,\ldots,y_i} \Big) \ket{\psi}}^2
		\leq 4(2f(i) + 4 \delta),
	\end{align*}
	and 
	\begin{align*}
		\abs{ \bigE_{x,y_1,\ldots,y_{i+1}} \sum_{a_1,\ldots,a_i, g_{i+1}} \norm{ \id \x &\Big( J_{[g_1(y_1),\ldots,g_{i}(y_{i}) = a_1, \ldots,a_{i}]}^{x,y_1,\ldots,y_i}(G_{i+1})^x_{g_{i+1}} \\
		 &- (G_{i+1})^x_{g_{i+1}} J_{[g_1(y_1),\ldots,g_{i}(y_{i}) = a_1, \ldots,a_{i}]}^{x,y_1,\ldots,y_i} \Big) \ket{\psi}}^2
		-  \\
		\bigE_{x,y_1,\ldots,y_{i+1}} \sum_{a_1,\ldots,a_i, g_{i+1}} \norm{ \id \x &\Big( J_{[g_1(y_1),\ldots,g_{i}(y_{i}) = a_1, \ldots,a_{i}]}^{x,y_1,\ldots,y_i}(G_{i+1})^x_{[g_{i+1}(y_{i+1}) = a_{i+1}]} \\
		&- (G_{i+1})^x_{[g_{i+1}(y_{i+1})=a_{i+1}]}   J_{[g_1(y_1),\ldots,g_{i}(y_{i}) = a_1, \ldots,a_{i}]}^{x,y_1,\ldots,y_i} \Big) \ket{\psi}}^2} \\
		\leq 2 \sqrt{2f(i) + 4 \delta} + &2 \sqrt{6f(i) + 4\delta} + 2\ep.
	\end{align*}
	Therefore,
	\begin{align*}
		&\bigE_{x,y_1,\ldots,y_{i+1}} \sum_{a_1,\ldots,a_i, g_{i+1}} \norm{ \id \x \Big( J_{[g_1(y_1),\ldots,g_{i}(y_{i}) = a_1, \ldots,a_{i}]}^{x,y_1,\ldots,y_i}(G_{i+1})^x_{g_{i+1}} - (G_{i+1})^x_{g_{i+1}} J_{[g_1(y_1),\ldots,g_{i}(y_{i}) = a_1, \ldots,a_{i}]}^{x,y_1,\ldots,y_i} \Big) \ket{\psi}}^2 \\
		&\leq 4(2f(i) + 4 \delta) + 2 \sqrt{2f(i) + 4 \delta} + 2 \sqrt{6f(i) + 4\delta} + 2\ep,
	\end{align*}
	and
	\begin{align*}
		&\abs{ \bigE_{x,y_1,\ldots,y_{i+1}} \sum_{a_1,\ldots, a_{i+1}} \bra{\psi} A^{x,y_1,\ldots,y_{i+1}}_{a_1,\ldots,a_{i+1}} \x J^{x,y_1,\ldots,y_{i+1}}_{[g_1(y_1),\ldots,g_{i+1}(y_{i+1}) = a_1, \ldots,a_{i+1}]} \ket{\psi} -1} \\
		&\leq \sqrt{2\delta} + \sqrt{f(i)} + \sqrt{16\sqrt{f(i)} + 24 \sqrt{\delta} + 2\ep }
	\end{align*}
	That is $f(i+1) = 5f(i)^{1/4}  + 7 \delta^{1/4} + \sqrt{2\ep}$. Then the lemma follows.
\end{proof}
\section{Upper Bound on the Number of Noisy MES's for Nonlocal Games}
\label{sec:upperbound}

The proof follows closely to that of~\cite{qin2021nonlocal}. The major difference is that in the proof of ~\cite{qin2021nonlocal}, each pair of questions $(x,y)$ is treated independently. Then, a union bound is applied to all possible questions. To improve the upper bound, we take into account the distribution of the questions, combined with a better Gaussian dimension reduction in \cite{qin_et_al:LIPIcs.ICALP.2023.97}.
Then
our new upper bound below only depends polynomially on the size of the question set
whereas the previous one has an exponential dependence.

\subsection{Gaussian Dimension Reduction}

The following lemma is a simplified version of \cite[Lemma 5.13]{qin_et_al:LIPIcs.ICALP.2023.97}, with the questions and answers being classical.
In the proof of \cref{thm:upperbound},
we will use this lemma, after we replace the low-influence registers by Gaussian random variables, to further reduce the dimension of the Gaussian space.
The only difference is in Item 3 of \cref{lem:expectationdimensionreduction}, where we preserve the expectation of the $\zeta$ function value over the random variable $\randM$.
In the previous version (Item 2 of \cite[Lemma 5.13]{qin_et_al:LIPIcs.ICALP.2023.97}), we used Markov's inequality on the expectation value.
As the notations are considerably different, we include a new proof for completeness.

\begin{lemma}\cite[Lemma 5.13]{qin_et_al:LIPIcs.ICALP.2023.97}\label{lem:expectationdimensionreduction}
Given parameters $\rho\in\Br{0,1}$, $\delta>0$, $d,n,h\in\posint$, $m\geq2$, an $m$-dimensional noisy MES $\psi_{AB}$ with the quantum maximal correlation $\rho=\rho(\psi_{AB})$, and degree-d multilinear joint random matrices 
\[
\br{P(\randg),Q(\randh)}=\br{\sum_{S\subseteq[n]}\mathbf{g}_S P_S,\sum_{S\subseteq[n]}\mathbf{h}_S Q_S}_{\br{\mathbf{g},\mathbf{h}}\sim\G_\rho^{\otimes n}},
\]
where $\mathbf{g}_S=\prod_{i\in S}\mathbf{g}_i, \mathbf{h}_S=\prod_{i\in S}\mathbf{h}_i$ and $ P_S, Q_S\in\H_m^{\otimes h}$ for all $S\subseteq[n]$, satisfying 
\[\expec{\randg}{\nnorm{P(\randg)}_2^2}\leq1~\mbox{and}~\expec{\randh}{\nnorm{Q(\randh)}_2^2}\leq1.\] 
Let $L^2\br{\herspace{h},\gamma_{n}}$ be the space of random operators whose Fourier coefficients are square-integrable with respect to the measure $\gamma_n$. Then there exists an explicitly computable $n_0=n_0(d,\delta)$ and maps $f_M,g_M: L^2\br{\herspace{h},\gamma_{n}}\rightarrow L^2\br{\herspace{h},\gamma_{n}}$ for $M\in\reals^{n\times n_0}$ and joint random operators $\br{P(M\tilde{\randx}),Q(M\tilde{\randy})}=\br{f_M(P(\randg)),g_M(Q(\randh))}$:
 \[
\br{P(M\tilde{\randx}),Q(M\tilde{\randy})}=\br{\sum_{S\subseteq[n]}\randu_S P_S,\sum_{S\subseteq[n]}\randv_S Q_S}_{\br{\mathbf{x},\mathbf{y}}\sim\G_\rho^{\otimes n_0}},
\]
where $\tilde{\randx}=\randx/\norm{\randx}_2$, $\tilde{\randy}=\randy/\norm{\randy}_2$, $\mathbf\randu_S=\prod_{i\in S}\innerproduct{m_i}{\tilde{\randx}}$, $\mathbf\randv_S=\prod_{i\in S}\innerproduct{m_i}{\tilde{\randy}}$, $\innerproduct{\cdot}{\cdot}$ denotes the standard inner product over $\reals^{n_0}$ and $m_i$ denotes the $i$'th row of $M$, such that if we sample $\mathbf{M}\sim\gamma_{n\times n_0}$, then the following hold:
\begin{enumerate}
\item 
With probability at least $1-2\delta$, we have 
\[\expec{\randx}{\nnorm{P(\randM\tilde{\randx})}_2^2}\leq1+\delta\quad\mbox{and}\quad\expec{\randy}{\nnorm{Q(\randM\tilde{\randy})}_2^2}\leq1+\delta.\]

\item With probability at least $1-\delta$, we have 

\[\ab{\expec{\mathbf{x},\mathbf{y}}{\Tr\br{\br{P(\randM\tilde{\randx})\otimes Q(\randM\tilde{\randy})}\psi_{AB}^{\otimes h}}}-\expec{\mathbf{g},\mathbf{h}}{\Tr\br{\br{P(\randg)\otimes Q(\randh)}\psi_{AB}^{\otimes h}}}}\leq\delta.\]

\item \[\expec{\randg}{\Tr~\zeta\br{P(\randg)}}=\expec{\randM,\randx}{\Tr~\zeta\br{P(\randM\tilde{\randx})}}\quad\mbox{and}\quad\expec{\randh}{\Tr~\zeta\br{Q(\randh)}}=\expec{\randM,\randy}{\Tr~\zeta\br{Q(\randM\tilde{\randy})}}.\]

\item the maps $f_M, g_M$ are linear and unital for any nonzero $M\in\reals^{n\times n_0}$.
\end{enumerate}

In particular, one may take $n_0=\frac{d^{O(d)}}{\delta^6}$.
\end{lemma}

For $M\in\reals^{n\times n_0},$ denote $F(M)=\expec{\mathbf{x},\mathbf{y}}{\Tr\br{\br{P(M\tilde{\randx})\otimes Q(M\tilde{\randy})}\psi_{AB}^{\otimes h}}}$. 
To prove \cref{lem:expectationdimensionreduction} item 2, we need the following lemma.

\begin{lemma}\label{lem:meanvar}
In the setting of \cref{lem:expectationdimensionreduction}, given $d\in\posint,$ $\delta>0,$ there exists $n_0=\frac{d^{O(d)}}{\delta^2}$ such that the following holds: For $\mathbf{M}\sim\gamma_{n\times n_0}$,
\begin{align*}
\ab{\expec{}{F(\mathbf{M})}-\expec{\mathbf{g},\mathbf{h}}{\Tr\br{\br{P(\randg)\otimes Q(\randh)}\psi_{AB}^{\otimes h}}}}\leq~&\delta,\\
\var{F(\mathbf{M})}\leq~&\delta.\\
\end{align*}
\end{lemma}

We use the following lemma to prove \cref{lem:meanvar}.
\begin{lemma}\cite[Lemma A.8,A.9]{Ghazi:2018:DRP:3235586.3235614}\label{lem:intermediate}
Given parameters $d$ and $\delta$, there exists an explicitly computable $n_0(d,\delta)$ such that the followings hold: 
\begin{itemize}
\item For any subsets $S,T\subseteq[n]$ satisfying $\ab{S},\ab{T}\leq d$, it holds that
\begin{align*}
&\text{if }S\ne T:\quad\expec{\mathbf{M},\mathbf{x},\mathbf{y}}{\mathbf\randu_S\mathbf\randv_T}=0,\\
&\text{if }S= T:\quad\ab{\expec{\mathbf{M},\mathbf{x},\mathbf{y}}{\mathbf\randu_S\mathbf\randv_T}-\rho^{\ab{S}}}\leq\delta.\\
\end{align*}

\item Let $(\randx',\randy')\sim\G_\rho^{\otimes n_0}$ be independent of $(\randx,\randy)$, and let $\mathbf\randu'_S=\prod_{i\in S}\innerproduct{m_i}{\frac{\randx'}{\twonorm{\randx'}}}$, $\mathbf\randv'_S=\prod_{i\in S}\innerproduct{m_i}{\frac{\randy'}{\twonorm{\randy'}}}$. For any subsets $S,T,S',T'\subseteq[n]$ satisfying $\ab{S},\ab{T},\ab{S'},\ab{T'}\leq d$, it holds that
\begin{multline*}
\text{if } S\bigtriangleup T\bigtriangleup S'\bigtriangleup T'\ne\emptyset:\\\ab{\expec{\mathbf{M},\mathbf{x},\mathbf{y},\mathbf{x'},\mathbf{y'}}{\mathbf\randu_S\mathbf\randv_T\mathbf{u'}_{S'}\mathbf{v'}_{T'}}-\br{\expec{\mathbf{M},\mathbf{x},\mathbf{y}}{\mathbf\randu_S\mathbf\randv_T}}\br{\expec{\mathbf{M},\mathbf{x'},\mathbf{y'}}{\mathbf{u'}_{S'}\mathbf{v'}_{T'}}}}=0,
\end{multline*}

\begin{multline*}
\text{if } S\bigtriangleup T\bigtriangleup S'\bigtriangleup T'=\emptyset:\\\ab{\expec{\mathbf{M},\mathbf{x},\mathbf{y},\mathbf{x'},\mathbf{y'}}{\mathbf\randu_S\mathbf\randv_T\mathbf{u'}_{S'}\mathbf{v'}_{T'}}-\br{\expec{\mathbf{M},\mathbf{x},\mathbf{y}}{\mathbf\randu_S\mathbf\randv_T}}\br{\expec{\mathbf{M},\mathbf{x'},\mathbf{y'}}{\mathbf{u'}_{S'}\mathbf{v'}_{T'}}}}\leq\delta.
\end{multline*}

Here, $S\bigtriangleup T\bigtriangleup S'\bigtriangleup T'$ is the symmetric difference of the sets $S,T,S',T'$, equivalently, the set of all $i\in[n]$ which appear an odd number of times in the multiset $S\sqcup T\sqcup S'\sqcup T'$.
\end{itemize}

In particular, one may take $n_0=\frac{d^{O(d)}}{\delta^2}$.
\end{lemma}

\begin{proof}[Proof of \cref{lem:meanvar}]
Use \cref{lem:intermediate} with parameters $d$ and $\delta$, we have 
\begin{align*}
&\ab{\expec{\mathbf{M}}{F(\mathbf{M})}-\expec{\mathbf{g},\mathbf{h}}{\Tr\br{\br{P(\randg)\otimes Q(\randh)}\psi_{AB}^{\otimes h}}}}\\
=~&\ab{\sum_{S,T\subseteq[n]}\br{\expec{\mathbf{M},\mathbf{x},\mathbf{y}}{\mathbf\randu_S\mathbf\randv_T}-\expec{\mathbf{g},\mathbf{h}}{\mathbf{g}_S\mathbf{h}_T}}\Tr\br{\br{ P_S\otimes Q_T}\psi_{AB}^{\otimes h}}}\\
=~&\ab{\sum_{S\subseteq[n]}\br{\expec{\mathbf{M},\mathbf{x},\mathbf{y}}{\mathbf\randu_S\mathbf\randv_S}-\rho^{\ab{S}}}\Tr\br{\br{ P_S\otimes Q_S}\psi_{AB}^{\otimes h}}}\\
\leq~&\delta\sum_{S\subseteq[n]}\ab{\Tr\br{\br{ P_S\otimes Q_S}\psi_{AB}^{\otimes h}}}\quad\mbox{(\cref{lem:intermediate})}\\
\leq~&\delta\sum_{S\subseteq[n]}\nnorm{ P_S}_2\nnorm{ Q_S}_2\quad\mbox{(\cref{fac:cauchyschwartz})}\\
\leq~&\delta\sqrt{\sum_{S\subseteq[n]}\nnorm{ P_S}_2^2\cdot\sum_{S\subseteq[n]}\nnorm{ Q_S}_2^2}\\
=~&\delta \br{\expec{\randg}{\nnorm{P(\randg)}_2^2}\expec{\randg}{\nnorm{Q(\randh)}_2^2}}^{1/2}\quad\leq\delta.\\
\end{align*}

Use \cref{lem:intermediate} with parameters $d$ and $\delta\leftarrow\delta/9^d$, we have 

\begin{align*}
&\var{F(\mathbf{M})}\\
=~&\expec{\mathbf{M}}{F(\mathbf{M})^2}-\br{\expec{\mathbf{M}}{F(\mathbf{M})}}^2\\
\leq~&\sum_{S,T,S',T'\subseteq[n]}\ab{\expec{\mathbf{M},\mathbf{x},\mathbf{y},\mathbf{x'},\mathbf{y'}}{\mathbf\randu_S\mathbf\randv_T\mathbf{u'}_{S'}\mathbf{v'}_{T'}}-\br{\expec{\mathbf{M},\mathbf{x},\mathbf{y}}{\mathbf\randu_S\mathbf\randv_T}}\br{\expec{\mathbf{M},\mathbf{x'},\mathbf{y'}}{\mathbf{u'}_{S'}\mathbf{v'}_{T'}}}}\\
&\ab{\Tr\br{\br{ P_S\otimes Q_S}\psi_{AB}^{\otimes h}}\Tr\br{\br{ P_{S'}\otimes Q_{S'}}\psi_{AB}^{\otimes h}}}\\
\leq~&\frac{\delta}{9^d}\sum_{\substack{S,T,S',T'\subseteq[n]\\ S\bigtriangleup T\bigtriangleup S'\bigtriangleup T'=\emptyset}}\nnorm{ P_S}_2\nnorm{ Q_T}_2\nnorm{ P_{S'}}_2\nnorm{ Q_{T'}}_2
\end{align*}

To finish the proof, we will show that,
\[\sum_{\substack{S,T,S',T'\subseteq[n]\\ S\bigtriangleup T\bigtriangleup S'\bigtriangleup T'=\emptyset}}\nnorm{ P_S}_2\nnorm{ Q_T}_2\nnorm{ P_{S'}}_2\nnorm{ Q_{T'}}_2\leq9^d\expec{\randg}{\nnorm{P(\randg)}_2^2}\expec{\randg}{\nnorm{Q(\randh)}_2^2}\]

Define functions $f,g:\set{1,-1}^n\rightarrow\reals$ over the boolean hypercube as,
\[f(x)=\sum_{\substack{S\subseteq[n]\\\ab{S}\leq d}}\nnorm{ P_S}_2\chi_S(x)\quad\text{and}\quad g(x)=\sum_{\substack{T\subseteq[n]\\\ab{T}\leq d}}\nnorm{ Q_T}_2\chi_T(x)\]

By the hypercontractivity inequality over the boolean hypercube \cite[Page 240]{Odonnell08}
\[\expec{x}{f(x)^4}\leq9^d\br{\expec{x}{f(x)^2}}^2\quad\text{and}\quad\expec{x}{g(x)^4}\leq9^d\br{\expec{x}{g(x)^2}}^2,\]

we have
\begin{align*}
&\sum_{\substack{S,T,S',T'\subseteq[n]\\ S\bigtriangleup T\bigtriangleup S'\bigtriangleup T'=\emptyset}}\nnorm{ P_S}_2\nnorm{ Q_T}_2\nnorm{ P_{S'}}_2\nnorm{ Q_{T'}}_2\\
=~&\expec{x}{f(x)^2g(x)^2}\\
\leq~&\sqrt{\expec{x}{f(x)^4}\expec{x}{g(x)^4}}\\
\leq~&9^d\expec{x}{f(x)^2}\expec{x}{g(x)^2}\\
=~&9^d\sum_{S\subseteq[n]}\nnorm{ P_S}_2^2\sum_{S\subseteq[n]}\nnorm{ Q_S}_2^2\\
=~&9^d\expec{\randg}{\nnorm{P(\randg)}_2^2}\expec{\randg}{\nnorm{Q(\randh)}_2^2}\quad \leq9^d.
\end{align*}

Thus $\var{F(\mathbf{M})}\leq\delta.$
\end{proof}

To prove \cref{lem:expectationdimensionreduction} Item 1, we need the following lemma whose proof is similar to that of \cref{lem:meanvar}. We omit the proof here.

\begin{lemma}\label{lem:normmeanvar}
In the setting of \cref{lem:expectationdimensionreduction}, given $d\in\posint,$ $\delta>0,$ there exists $n_0=\frac{d^{O(d)}}{\delta^2}$ such that the following holds: For $\mathbf{M}\sim\gamma_{n\times n_0}$,
\begin{align*}
\ab{\expec{\randM,\randx}{\nnorm{P(\randM\tilde{\randx})}_2^2}-\expec{\randg}{\nnorm{P(\randg)}_2^2}}\leq~&\delta,\\
\var{\expec{\randx}{\nnorm{P(\randM\tilde{\randx})}_2^2}}\leq~&\delta,\\
\ab{\expec{\randM,\randy}{\nnorm{Q(\randM\tilde{\randy})}_2^2}-\expec{\randh}{\nnorm{Q(\randh)}_2^2}}\leq~&\delta,\\
\var{\expec{\randy}{\nnorm{Q(\randM\tilde{\randy})}_2^2}}\leq~&\delta.
\end{align*}
\end{lemma}

\begin{proof}[Proof of \cref{lem:expectationdimensionreduction}]

For item 2, we invoke \cref{lem:meanvar} with parameters $d$ and $\delta\leftarrow\delta^3/2$. 
 Using Chebyshev's inequality, we have that for any $\eta>0$,
\[\Pr_{\mathbf{M}}\Br{\ab{F(\mathbf{M})-\expec{\mathbf{M}}{F(\mathbf{M})}}>\eta}\leq\frac{\delta^3}{2\eta^2}.\]

Using the triangle inequality, we get
\begin{align*}
&\Pr_{\mathbf{M}}\Br{\ab{F(\mathbf{M})-\expec{\mathbf{g},\mathbf{h}}{\Tr\br{\br{P(\randg)\otimes Q(\randh)}\psi_{AB}^{\otimes h}}}}>\delta}\\
\leq~&\Pr_{\mathbf{M}}\Br{\ab{F(\mathbf{M})-\expec{\mathbf{M}}{F(\mathbf{M})}}+\ab{\expec{\mathbf{M}}{F(\mathbf{M})}-\expec{\mathbf{g},\mathbf{h}}{\Tr\br{\br{P(\randg)\otimes Q(\randh)}\psi_{AB}^{\otimes h}}}}>\delta}\\
\leq~&\Pr_{\mathbf{M}}\Br{\ab{F(\mathbf{M})-\expec{\mathbf{M}}{F(\mathbf{M})}}>\delta-\delta^3/2}\leq\delta.
\end{align*}

By \cref{lem:normmeanvar}, we can similarly argue for item 1. For item 3, note that for any fixed $x\in\reals^{n_0}$, the distribution of $\randM x/\norm{x}_2$ is identical to $\gamma_{n}$.
	It is easy to verify Item 4.
\end{proof}

\subsection{Upper Bound}
We are now ready to prove \cref{thm:upperbound}.
\begin{proof}[Proof of \cref{thm:upperbound}]
The proof follows that in \cite{qin2021nonlocal} step by step, except that the Gaussian dimension reduction step in the original proof is replaced by \cref{lem:expectationdimensionreduction}. 
Here, we include the proof for completeness.

Suppose the players use the strategy $\br{\set{\Pnn{0}}^{x\in\calX}_{a\in\calA},\set{\Qnn{0}}^{y\in\calY}_{b\in\calB}}$ to achieve the highest winning probability when sharing $n$ copies of $\psi_{AB}$, where $\Pnn{0}$ is the POVM element of Alice corresponding to the answer $a$ upon receiving the question $x$, and $\Qnn{0}$ is the POVM element of Bob corresponding to the answer $b$ upon receiving the question $y$. Then for all $\br{x,y,a,b}\in\calX\times\calY\times\calA\times\calB$, $\Pnn{0}\geq0$, $\Qnn{0}\geq0$, $\sum_a\Pnn{0}=\id$, $\sum_b\Qnn{0}=\id$, and $\omega_n(\mathfrak{G},\psi_{AB})=\gval{n}{\Pnn{0}}{\Qnn{0}}$.

Let $\delta,\tau$ be parameters which are chosen later. The proof is composed of several steps.

\begin{itemize}
\item \textbf{Smoothing.} This step allows us to restrict ourselves to strategies with low-degree POVMs.

 More specifically, for any \fourtuples, we apply the map $f^{(1)}$ implied by \cref{lem:smoothing of strategies} to $\Pnn{0}$ and $\Qnn{0}$ to get $\Pnn{1}$ and $\Qnn{1}$, respectively\footnote{Specifically, we apply a depolarizing channel $\Delta_\gamma$ for some $\gamma\in(0,1)$ to $\Pnn{0}$ and $\Qnn{0}$, and then truncate it to be of degree $d$ to get $\Pnn{1}$. Readers may refer to \cite{qin2021nonlocal} for details.}. Note that for all $x,y,a,b,$ $\nnorm{\Pnn{0}}_2^2\leq1$ and $\nnorm{\Qnn{0}}_2^2\leq1.$ Let $d=\frac{C\log^2\frac{1}{\delta}}{\delta(1-\rho)}$, by \cref{lem:smoothing of strategies} Item 3 and Item 4, 

			\[\ab{\Tr\br{\br{\Pnn{1}\otimes \Qnn{1}}\psi^{\otimes n}_{AB}}-\Tr\br{\br{\Pnn{0}\otimes \Qnn{0}}\psi^{\otimes n}_{AB}}}\leq\delta\]
			and
		\[\frac{1}{m^n}\Tr~\zeta(\Pnn{1})\leq\delta,\quad\frac{1}{m^n}\Tr~\zeta(\Qnn{1})\leq\delta.\]
     By \cref{lem:maincauchy} and \cref{lem:smoothing of strategies} items 1, 2 and 5, the following hold.
     
     \begin{enumerate}
						\item For any $x,y,a,b$, $\Pnn{1}$ and $\Qnn{1}$ are of degree at most $d$.
\item For any $x,y,a,b$, $\displaystyle\nnorm{\Pnn{1}}_2\leq1~\mbox{and}~\nnorm{\Qnn{1}}_2\leq1.$
			\item $\ab{\gval{n}{\Pnn{1}}{\Qnn{1}}-\gval{n}{\Pnn{0}}{\Qnn{0}}}\leq\delta t^2,$
\item $\displaystyle\frac{1}{m^n}\sum_{x,a}\mu_A(x)\Tr~\zeta\br{\Pnn{1}}\leq\delta t $ and $\displaystyle\frac{1}{m^n}\sum_{y,b}\mu_B(y)\Tr~\zeta\br{\Qnn{1}}\leq\delta t. $
            \item For any $x,y$, $\displaystyle\sum_{a\in\calA}\Pnn{1}=\sum_{b\in\calB}\Qnn{1}=\id$.
		\end{enumerate}
     
\item \textbf{Regularization.}
In this step, we identify the set $H$ of high-influence registers for all POVM elements.

      For any \fourtuples, we apply \cref{lem:regular} to $\Pnn{1}$ and $\Qnn{1}$ to get sets $H_{x,a}$ and $H_{y,b}$ of size at most $d/\tau$, respectively, such that 
      \[\br{\forall i\notin H_{x,a}}~\infi{\Pnn{1}}\leq\tau\quad\mbox{and}\quad\br{\forall i\notin H_{y,b}}~\infi{\Qnn{1}}\leq\tau.\]
      Set $H=\br{\bigcup_{x,a}H_{x,a}}\cup\br{\bigcup_{y,b}H_{y,b}}$, then 
      $h=\ab{H}\leq\frac{2std}{\tau}$, and 
      \[\br{\forall i\notin H}~\infi{\Pnn{1}}\leq\tau\quad\mbox{and}\quad\infi{\Qnn{1}}\leq\tau.\]
      
\item \textbf{Invariance from \herspace{n} to $L^2\br{\herspace{h},\gamma_{(m^2-1)(n-h)}}$.}
In this step, we only keep the quantum registers in $H$ and replace 
the rest of the quantum registers by Gaussian random variables.
Hence, 
the number of quantum registers is reduced from $n$ to $h = \abs{H} = d/\tau$.

      For any \fourtuples, applying \cite[Lemma 10.5]{qin2021nonlocal} to $\Pnn{1}$, $\Qnn{1}$ and $H$, we obtain joint random matrices
      \[\br{\Pnn{2}(\randg),\Qnn{2}(\randh)}\in L^2\br{\herspace{h},\gamma_{2(m^2-1)(n-h)}}\times L^2\br{\herspace{h},\gamma_{2(m^2-1)(n-h)}},\]
      where $\br{\mathbf{g},\mathbf{h}}\sim\G_{\rho}^{\otimes2(m^2-1)(n-h)}$, such that the following hold.
      \begin{enumerate}
		\item For any $x,y,a,b,$ $\displaystyle\expec{\randg}{\nnorm{\Pnn{2}(\randg)}_2^2}\leq1$ and 		
		$\displaystyle\expec{\randh}{\nnorm{\Qnn{2}(\randh)}_2^2}\leq1$.
		\item $\expec{\randg,\randh}{\gval{h}{\Pnn{2}(\randg)}{\Qnn{2}(\randg)}}=\gval{n}{\Pnn{1}}{\Qnn{1}}.$
		\item $\displaystyle\sum_{x,a}\mu_A(x)\ab{\frac{1}{m^h}\expec{}{\Tr~\zeta\br{\Pnn{2}(\randg)}}-\frac{1}{m^n}\Tr~\zeta\br{\Pnn{1}}}\leq O\br{t\br{3^dm^{d/2}\sqrt{\tau}d}^{2/3}}$ and

$\displaystyle\sum_{y,b}\mu_B(y)\ab{\frac{1}{m^h}\expec{}{\Tr~\zeta\br{{\Qnn{2}(\randh)}}}-\frac{1}{m^n}\Tr~\zeta\br{\Qnn{1}}}\leq O\br{t\br{3^dm^{d/2}\sqrt{\tau}d}^{2/3}}$.
\item For any $x,y$, $\sum_{a\in\calA}\Pnn{2}(\randg)=\sum_{b\in\calB}\Qnn{2}(\randh)=\id.$
	\end{enumerate}
	
\item \textbf{Gaussian dimension reduction.} 
In this step, we apply \cref{lem:expectationdimensionreduction} to further reduce the number of Gaussian random variables.
This is the only part different from the proof in \cite{qin2021nonlocal}.

	Let $n_0$ be determined later. 
	For any \fourtuples~and $M\in\reals^{n\times n_0}$, applying \cref{lem:expectationdimensionreduction} to $\Pnn{2}(\randg)$ and $\Qnn{2}(\randh)$ with $\delta\gets\delta/\br{2s^2t^2}$, $d\gets d$, $n\gets 2(m^2-1)(n-h)$, we get joint random matrices $\Pnn{3}(M\tilde{\randx})$ and $\Qnn{3}(M\tilde{\randy})$. If we sample $\randM\sim\gamma_{n\times n_0}$, by \cref{lem:expectationdimensionreduction} item 3 we have
\[\sum_{x,a}\mu_A(x)\expec{\randM,\randx}{\Tr~\zeta\br{\Pnn{3}(\randM\tilde{\randx})}}=\sum_{x,a}\mu_A(x)\expec{\randg}{\Tr~\zeta\br{\Pnn{2}(\randg)}}\]and\[\sum_{y,b}\mu_B(y)\expec{\randM,\randy}{\Tr~\zeta\br{\Qnn{3}(\randM\tilde{\randy})}}=\sum_{y,b}\mu_B(y)\expec{\randh}{\Tr~\zeta\br{\Qnn{2}(\randh)}}.\]

Then by Markov's inequality, with probability each at most $1/6$, 
\[\sum_{x,a}\mu_A(x)\expec{ \randx}{\Tr~\zeta\br{\Pnn{3}(\randM\tilde{\randx})}}>6\sum_{x,a}\mu_A(x)\expec{\randg}{\Tr~\zeta\br{\Pnn{2}(\randg)}}\]and\[\sum_{y,b}\mu_B(y)\expec{ \randy}{\Tr~\zeta\br{\Qnn{3}(\randM\tilde{\randy})}}>6\sum_{y,b}\mu_B(y)\expec{\randh}{\Tr~\zeta\br{\Qnn{2}(\randh)}}.\]

By \cref{lem:expectationdimensionreduction} item 1, 2, and using a union bound, with probability at least $2/3-\delta$ the following hold:

\begin{enumerate}
\item For any $x,y,a,b$, $\displaystyle\expec{ \randx}{\nnorm{\Pnn{3}(M\tilde{\randx})}_2^2}\leq2$ and
$\displaystyle\expec{ \randy}{\nnorm{\Qnn{3}(M\tilde{\randy})}_2^2}\leq2.$

\item $\displaystyle\ab{\expec{\randx,\randy}{\gval{h}{\Pnn{3}(M\tilde{\randx})}{\Qnn{3}(M\tilde{\randy})}}-\expec{\randg,\randh}{\gval{h}{\Pnn{2}(\randg)}{\Qnn{2}(\randg)}}}\leq\delta t^2.$

\item $\displaystyle\sum_{x,a}\mu_A(x)\expec{ \randx}{\Tr~\zeta\br{\Pnn{3}(M\tilde{\randx})}}\leq6\sum_{x,a}\mu_A(x)\expec{\randg}{\Tr~\zeta\br{\Pnn{2}(\randg)}}$ \quad and

$\displaystyle\sum_{y,b}\mu_B(y)\expec{ \randy}{\Tr~\zeta\br{\Qnn{3}(M\tilde{\randy})}}\leq6\sum_{y,b}\mu_B(y)\expec{\randh}{\Tr~\zeta\br{\Qnn{2}(\randh)}}.$

\item For any $x,y$, $\displaystyle\sum_{a\in\calA}\Pnn{3}(M\tilde{\randx})=\sum_{b\in\calB}\Qnn{3}(M\tilde{\randy})=\id.$
\end{enumerate}
Here $n_0=\frac{d^{O(d)}s^{12}t^{12}}{\delta^6}$. Therefore, there must exist  an $M$ such that all the above four requirements hold. We will use this fixed M throughout the rest of the proof. 

\item \textbf{Smoothing random matrices.} 
In this step, we reduce $\deg(\Pnn{3})$ and $\deg(\Qnn{3})$ for any \fourtuples. We apply \cite[Lemma 12.1]{qin2021nonlocal} to $\Pnn{3}(M\tilde{\randx})$ and $\Qnn{3}(M\tilde{\randy})$ with $\delta\gets\delta$, $h\gets h$, $n\gets n_0$ and obtain joint random matrices $\Pnn{4}(\randx),\Qnn{4}(\randy)\in L^2\br{\herspace{h},\gamma_{n_0}}$ such that the following holds.
\begin{enumerate}
\item For any $x,y,a,b$, the entries of $\Pnn{4}(\randx)$ and $\Qnn{4}(\randy)$ are polynomials of degree at most $d$.
						
\item For any \fourtuples, $\displaystyle\expec{\randx}{\nnorm{\Pnn{4}(\randx)}_2^2}\leq2$ and
$\displaystyle\expec{\randy}{\nnorm{\Qnn{4}(\randy)}_2^2}\leq2.$

\item $\displaystyle\ab{\expec{\randx,\randy}{\gval{h}{\Pnn{4}(\randx)}{\Qnn{4}(\randx)}}-\expec{\randx,\randy}{\gval{h}{\Pnn{3}(M\tilde{\randx})}{\Qnn{3}(M\tilde{\randy})}}}\leq \delta t^2.$

\item $\displaystyle\ab{\sum_{x,a}\mu_A(x)\expec{\randx}{\Tr~\zeta\br{\Pnn{4}(\randx)}}-\sum_{x,a}\mu_A(x)\expec{ \randx}{\Tr~\zeta\br{\Pnn{3}(M\tilde{\randx})}}}\leq\delta t$ \quad and

$\displaystyle\ab{\sum_{y,b}\mu_B(y)\expec{\randy}{\Tr~\zeta\br{\Qnn{4}(\randy)}}-\sum_{y,b}\mu_B(y)\expec{ \randy}{\Tr~\zeta\br{\Qnn{3}(M\tilde{\randy})}}}\leq\delta t.$

\item For any $x,y$, $\sum_{a\in\calA}\Pnn{4}(\randx)=\sum_{b\in\calB}\Qnn{4}(\randy)=\id.$
\end{enumerate}

\item \textbf{Multilinearization.}
For any \fourtuples, we apply \cite[Lemma 13.1]{qin2021nonlocal} to $\Pnn{4}(\randx)$ and $\Qnn{4}(\randy)$ with $d\gets d$, $\delta\gets\tau$, $h\gets h$, $n\gets n_0$ and obtain joint random matrices $\Pnn{5}(\randx),\Qnn{5}(\randy)\in L^2\br{\herspace{h},\gamma_{n_0n_1}}$ such that the following holds.
\begin{enumerate}
\item For any $x,y,a,b$, the entries of $\Pnn{5}(\randx)$ and $\Qnn{5}(\randy)$ are multilinear polynomials of degree at most $d$, and every variable in $\Pnn{5}(\randx)$ and $\Qnn{5}(\randx)$ has influence at most $\tau$.
						
\item For any $x,y,a,b$, $\displaystyle\expec{\randx}{\nnorm{\Pnn{5}(\randx)}_2^2}\leq2$ and
$\displaystyle\expec{\randy}{\nnorm{\Qnn{5}(\randy)}_2^2}\leq2.$

\item $\displaystyle\ab{\expec{\randx,\randy}{\gval{h}{\Pnn{5}(\randx)}{\Qnn{5}(\randx)}}-\expec{\randx,\randy}{\gval{h}{\Pnn{4}(\randx)}{\Qnn{4}(\randy)}}}\leq\tau t^2.$

\item $\displaystyle\ab{\sum_{x,a}\mu_A(x)\expec{\randx}{\Tr~\zeta\br{\Pnn{5}(\randx)}}-\sum_{x,a}\mu_A(x)\expec{ \randx}{\Tr~\zeta\br{\Pnn{4}(\randx)}}}\leq\tau t$ \quad and

$\displaystyle\ab{\sum_{y,b}\mu_B(y)\expec{\randy}{\Tr~\zeta\br{\Qnn{5}(\randy)}}-\sum_{y,b}\mu_B(y)\expec{ \randy}{\Tr~\zeta\br{\Qnn{4}(\randy)}}}\leq\tau t.$

\item For any $x,y$, $\sum_{a\in\calA}\Pnn{5}(\randx)=\sum_{b\in\calB}\Qnn{5}(\randy)=\id.$
\end{enumerate}
Here $n_1=O\br{\frac{d^2}{\tau^2}}.$

\item \textbf{Invariance from $L^2\br{\herspace{h},\gamma_{n_0n_1}}$ to $\herspace{h+n_0n_1}$.}
In this step, we transform all the random matrices from the previous step to matrices without any classical randomness.
In particular, we replace all the Gaussian random variables with $n_0n_1$ quantum registers, so after this step, the number of quantum registers is $h + n_0n_1$.

For any \fourtuples, applying \cite[Lemma 10.11]{qin2021nonlocal} to $\Pnn{5}(\randx)$, $\Qnn{5}(\randy)$ with $n\gets n_0n_1$, $h\gets h$, $d\gets 2d$, $\tau\gets\tau$ to get $\Pnn{6},\Qnn{6}\in\herspace{h+n_0n_1}$ satisfying the following.
      
   \begin{enumerate}
		\item For any $x,y,a,b$, $\displaystyle\nnorm{\Pnn{6}}_2^2\leq2$ and 		
		$\displaystyle\nnorm{\Qnn{6}}_2^2\leq2.$
		\item $\gval{h+n_0n_1}{\Pnn{6}}{\Qnn{6}}=\expec{\randx,\randy}{\gval{h}{\Pnn{5}(\randx)}{\Qnn{5}(\randy)}}.$
		\item $\displaystyle\sum_{x,a}\mu_A(x)\ab{\frac{1}{m^{h+n_0n_1}}\Tr~\zeta\br{\Pnn{6}}-\frac{1}{m^h}\expec{}{\Tr~\zeta\br{\Pnn{5}(\randx)}}}\leq O\br{t\br{9^dm^{d}\sqrt{\tau}d}^{2/3}}$ and

$\displaystyle\sum_{y,b}\mu_B(y)\ab{\frac{1}{m^{h+n_0n_1}}\Tr~\zeta\br{\Qnn{6}}-\frac{1}{m^h}\expec{}{\Tr~\zeta\br{{\Qnn{5}(\randy)}}}}\leq O\br{t\br{9^dm^{d}\sqrt{\tau}d}^{2/3}}$.
\item For any $x,y$, $\sum_{a\in\calA}\Pnn{6}=\sum_{b\in\calB}\Qnn{6}=\id.$
	\end{enumerate}

\item \textbf{Rounding.} Note that the matrices from the previous step may not form valid POVMs, so in this step we round them to close POVMs. In this step, the number of quantum registers remains the same as $h+n_0n_1$.

By \cref{lem:closedelta} there exist operators $\set{\Pnn{7}}$ and $\set{\Qnn{7}}$ satisfying for all $x$
\begin{align}
\sum_a \nnorm{\Pnn{7}-\Pnn{6}}_2^2&\le\frac{3(t+1)}{m^D}\sum_{a}\Tr~\zeta\br{\Pnn{6}}+6\sqrt{t}\br{\frac{1}{m^D}\sum_{a}\Tr~\zeta\br{\Pnn{6}}}^{1/2} \nonumber\\
&\le10t\br{\frac{1}{m^D}\sum_{a}\Tr~\zeta\br{\Pnn{6}}}^{1/2}\label{eqn:rounding1}.
\end{align}
Similarly, for all $y$, we have
\begin{equation}\label{eqn:rounding2}
\sum_a \nnorm{\Qnn{7}-\Qnn{6}}_2^2\le10t\br{\frac{1}{m^D}\sum_{b}\Tr~\zeta\br{\Qnn{6}}}^{1/2}.
\end{equation}
Then
\begin{align*}
&\ab{\gval{D}{\Pnn{7}}{\Qnn{7}}-\gval{D}{\Pnn{6}}{\Qnn{6}}}\\
\leq~&\ab{\gval{D}{\Pnn{7}-\Pnn{6}}{\Qnn{7}}}+\ab{\gval{D}{\Pnn{6}}{\Qnn{7}-\Qnn{6}}}\\
\leq~&\sum_{x,y,a,b}\mu(x,y)\br{\nnorm{\Pnn{7}-\Pnn{6}}_2\nnorm{\Qnn{7}}_2+\nnorm{\Pnn{6}}_2\nnorm{\Qnn{7}-\Qnn{6}}_2}\\
\leq~&\br{\sum_b\sum_{x,a}\mu_A(x)\nnorm{\Pnn{7}-\Pnn{6}}_2^2}^{1/2}\br{\sum_a\sum_{y,b}\mu_B(y)\nnorm{\Qnn{7}}_2^2}^{1/2}\\
&+\br{\sum_b\sum_{x,a}\mu_A(x)\nnorm{\Pnn{6}}_2^2}^{1/2}\br{\sum_a\sum_{y,b}\mu_B(y)\nnorm{\Qnn{7}-\Qnn{6}}_2^2}^{1/2}\quad\mbox{(Cauchy Schwarz)}\\
\leq~&\sqrt{10}t^2\br{\sum_{x}\mu_A(x)\br{\frac{1}{m^{D}}\sum_{a}\Tr~\zeta\br{\Pnn{6}}}^{1/2}}^{1/2}+2\sqrt{5}t^2\br{\sum_{y}\mu_B(y)\br{\frac{1}{m^{D}}\sum_{b}\Tr~\zeta\br{\Qnn{6}}}^{1/2}}^{1/2}\\
\leq~&\sqrt{10}t^2\br{\frac{1}{m^{D}}\sum_{x, a}\mu_A(x)\Tr~\zeta\br{\Pnn{6}}}^{1/4}+2\sqrt{5}t^2\br{\frac{1}{m^{D}}\sum_{y, b}\mu_B(y)\Tr~\zeta\br{\Qnn{6}}}^{1/4},
\end{align*}
\end{itemize}
where in the second last inequality, we use $\nnorm{\Pnn{6}} \leq 2$, $\nnorm{\Qnn{7}} \leq 1$, and \cref{eqn:rounding1,eqn:rounding2}.
The last inequality follows from concavity of the function $x\mapsto \sqrt{x}$.

Keeping track of the parameters in the construction, we can upper bound $\frac{1}{m^{D}}\sum_{x,a}\mu_A(x)\Tr~\zeta\br{\Pnn{6}}$ and $\frac{1}{m^{D}}\sum_{y,b}\mu_B(y)\Tr~\zeta\br{\Qnn{6}}$. 
We choose
\begin{equation}\label{eqn:paramters}
  \delta=\frac{\epsilon^4}{300t^9}, \tau=\frac{\epsilon^{12}}{t^{27}}\exp\br{-\frac{300t^{9}\log m}{\epsilon^4(1-\rho)}\log^2\br{\frac{t}{\epsilon}}}
\end{equation}
such that the difference in the game value at the final step matches that of the previous steps, remaining on the order of $O(\delta t^2)$. We conclude that the number of quantum registers is
 \begin{align*}
     D = h + n_0n_1 = \frac{d}{\tau} + \frac{d^{O(d)}s^{12}t^{12}}{\delta^6}\cdot O\br{\frac{d^2}{\tau^2}} = O\br{\frac{s^{12}t^{120}}{\epsilon^{48}}\exp\br{\frac{600t^{9}\log m}{\epsilon^4(1-\rho)}\log^2\br{\frac{t}{\epsilon(1-\rho)}}}},
 \end{align*}
 which completes the proof.
\end{proof}
\section{Truncation}
\begin{lemma}[Truncation]\label{lem:truncating_low_degrees}
  Let $\set{\Pss}, \set{\Qss}$ be two sets of operators satisfying
  \begin{enumerate}
    \item For all $x, y$, $\sum_{a}\Pss=\sum_{b}\Qss=\id$.
    \item For all $x, a, y, b, \sigma$, $\ab{\Pff}\le1$ and $\ab{\Qff}\le 1$.
  \end{enumerate}
  Let $\wireal = D\log m + \log\br{\frac{2}{\delta}}$.
  Then there exist operators $\set{\Pnn{2}}, \set{\Qnn{2}}$ satisfying
  \begin{enumerate}
    \item For each $x, y, a, b, \sigma$, the Fourier coefficients of $\Pnn{2}$ and $\Qnn{2}$ consists of at most $\wireal$ bits.
    \item For all $x, y$, $\sum_{a}\Pnn{2}=\sum_{b}\Qnn{2}=\id$.
    \item For all $x, y, a, b$, $\nnorm{\Pnn{2}}_2\le1$ and $\nnorm{\Qnn{2}}_2\le1$.
    \item For all $x, y, a, b$, $\ab{\Tr\br{\br{\Pnn{2}\otimes \Qnn{2}}\psi^{\otimes n}_{AB}}-\Tr\br{\br{\Pss\otimes \Qss}\psi^{\otimes n}_{AB}}}\leq\delta$.
	\item For all $x, y, a, b$, \[\ab{\frac{1}{m^\dlbo}\Tr~\zeta\br{\Pnn{2}}-\frac{1}{m^\dlbo}\Tr~\zeta\br{\Pss}}\leq\delta~\mbox{and}~\ab{\frac{1}{m^\dlbo}\Tr~\zeta\br{\Qnn{2}}-\frac{1}{m^\dlbo}\Tr~\zeta\br{\Qss}}\leq\delta. \]
  \end{enumerate}
\end{lemma}
\begin{proof}
    \newcommand{\tsms} {\alpha}
    Let $\tsms=2^{-\wireal}=\delta/(2m^D)$.
    For each $x, y, \sigma$, define $\Pfn{1} = \lfloor\Pff/\tsms\rfloor\tsms$.
    For each $x, \sigma\neq0^\dlbo$, define integer $k_{x, \sigma}$ as
    $$-\sum_{a}\Pfn{1} = k_{x, \sigma}\cdot\tsms$$
    and for $\sigma=0^D$, define
    $$1 - \sum_{a}\Pfn{1} = k_{x, 0^\dlbo}\cdot\tsms.$$
    Let $t_{x,\sigma}=\ab{\set{a\in\A: \Pfn{1}\neq \Pff}}$,
    we can see that $0\le k_{x, \sigma}< t_{x,\sigma}$ always holds because
    $\sum_{a}\Pss=\id$
    and by the fact that $\Pfn{1}>\Pff-\alpha$.
    Let $S_{x, \sigma}$ be an arbitrary subset of $\set{a\in\A: \Pfn{1}\neq \Pff}$ of size $k_{x,\sigma}$.
    Define $\Pnn{2}$ as
    $$\Pfn{2}=\begin{cases}
        \Pfn{1} &\text{ if } a \not\in S_{x,\sigma} \\
        \Pfn{1} + \tsms&\text{ if } a \in S_{x,\sigma} \\
    \end{cases}$$
    Then item 1 and item 2 hold for $\Pnn{2}$.
    Also, since for $a\in S_{x,\sigma}$ we have $\Pfn{1}< \Pff\le 1$, we have $\Pfn{1}\le1-\alpha$.
    So, it can be verified that $\ab{\Pfn{2}}\le1$ always holds, which implies that item 3 also holds.
    To prove the remaining items, we need
    $$\nnorm{\Pss-\Pnn{2}}_2 = \sqrt{\sum_{\sigma}\br{\Pff-\Pfn{2}}^2}
                                   <\sqrt{\sum_{\sigma}\tsms^2}\le m^D\tsms.$$
    We can apply the same operations to $\set{\Qss}$ and get $\set{\Qnn{2}}$.
    Then for all $x, y, a, b$,
    \begin{align*}
     &\ab{\Tr\br{\br{\Pnn{2}\otimes \Qnn{2}}\psi^{\otimes n}_{AB}}-\Tr\br{\br{\Pss\otimes \Qss}\psi^{\otimes n}_{AB}}} \\
  \le&\ab{\Tr\br{\br{\Pnn{2}\otimes \Qnn{2}}\psi^{\otimes n}_{AB}}-\Tr\br{\br{\Pnn{2}\otimes \Qss}\psi^{\otimes n}_{AB}}}\\ 
    &+\ab{\Tr\br{\br{\Pnn{2}\otimes \Qss}\psi^{\otimes n}_{AB}}-\Tr\br{\br{\Pss\otimes \Qss}\psi^{\otimes n}_{AB}}} \\
    =&\ab{\Tr\br{\br{\Pnn{2}\otimes \br{\Qnn{2}-\Qss}}\psi^{\otimes n}_{AB}}} +
      \ab{\Tr\br{\br{\br{\Pnn{2}-\Pss}\otimes \Qss}\psi^{\otimes n}_{AB}}} \\
 \leq&\nnorm{\Pnn{2}}_2\nnorm{\Qnn{2}-\Qss}_2 + \nnorm{\Pnn{2}-\Pss}_2\nnorm{\Qss}_2
 \leq2m^D\tsms=\delta,
    \end{align*}
    and item 4 follows.
    Then item 5 follows from \cref{lem:zetaadditivity}.
\end{proof}

\end{document}